\documentclass[preprints,article,submit,oneauthor,pdftex]{Definitions/mdpi}
\firstpage{1}
\pubvolume{xx}
\issuenum{1}
\articlenumber{5}
\pubyear{2020}
\copyrightyear{2020}
\history{Received: date; Accepted: date; Published: date}

\usepackage{inputenc}
\Title{Generalized Description of Intermittency in Turbulence via Stochastic Methods}

\Author{Jan Friedrich $^{1,2,*}$ \orcidA{},Rainer Grauer $^{2}$ \orcidB{}}
\AuthorNames{Jan Friedrich, Rainer Grauer}

\address{%
$^{1}$ \quad Univ. Lyon, ENS de Lyon, Univ. Claude Bernard, CNRS, Laboratoire de Physique,
F-69342, Lyon, France; jan-friedrich1@ens-lyon.fr,\\
$^{2}$ \quad Institute for Theoretical Physics I, Ruhr-University Bochum, Universit\"{a}tsstr. 150,
D-44780 Bochum, Germany; grauer@tp1.rub.de}
\corres{Correspondence: jan-friedrich1@ens-lyon.fr}

\abstract{We present a generalized picture of intermittency in turbulence that is based on the theory of stochastic processes. To this end, we rely on the experimentally and numerically verified finding by R.~Friedrich and J.~Peinke [Phys. Rev. Lett. 78, 863 (1997)]
that allows for an interpretation of the turbulent energy cascade as a Markov process of velocity increments in scale. It is explicitly shown that all known phenomenological models of turbulence can be reproduced by the Kramers-Moyal expansion of the  velocity increment probability density function that is associated to a Markov process. We compare the different sets of Kramers-Moyal coefficients of each phenomenology and deduce that an accurate description of intermittency should take into account an infinite number of
coefficients. This is demonstrated in more detail for the case of Burgers turbulence that exhibits pronounced intermittency effects. Moreover, the influence of nonlocality on Kramers-Moyal coefficients is investigated by direct numerical simulations of a generalized Burgers equation. Depending
on the balance between nonlinearity and nonlocality, we encounter different intermittency behavior that ranges from self-similarity (purely nonlocal case) to intermittent behavior (intermediate case that agrees with Yakhot's mean field theory
[Phys. Rev. E 63 026307 (2001)]) to shock-like behavior (purely nonlinear Burgers case).}


\begin{document}

\section{Introduction}
%description of the problem, how to efficiently decrease the numbers of freedom,
%different approaches: operator product expansion, instantons, stochastic approach
The phenomenon of homogeneous and isotropic
turbulence can still be considered as one of
the main unsolved problems in classical physics~\cite{Nelkin1991,monin}.
An adequate treatment of the underlying Navier-Stokes
equation should make an assertion about the
small-scale fluctuations of the longitudinal velocity increments
\begin{equation}
 \delta_{r} v(\mathbf{x},t) = \left(\mathbf{u}(\mathbf{x}+\mathbf{r},t) - \mathbf{u}(\mathbf{x},t) \right)
 \cdot \frac{\mathbf{r}}{r}\;,
 \label{eq:vel_inc}
\end{equation}
in a statistical sense.
Here, deviations from Kolmogorov's mean field theory~\cite{Kolmogorov1941} that predicts
$\langle (\delta_{r} v)^n \rangle \sim \langle \varepsilon \rangle^{n/3} |r|^{n/3}$
are commonly attributed to the intermittent fluctuations of the local energy dissipation rate
$\varepsilon$ and  manifest themselves by a non-self-similar
probability density function (PDF)
of the velocity increments. In turn, this implies a nonlinear order dependence for the
scaling exponents $\zeta_n$ of the moments $\langle (\delta_{r} v)^n \rangle \sim |r|^{\zeta_n}$.
In this context,
considerable efforts have been devoted to the development of phenomenological
models of turbulence that all try to account for
the intermittent character of
the local energy dissipation rate such as the log-normal model
~\cite{Kolmogorov1962,Oboukhov1962}
or the popular model by She and Leveque~\cite{She1994} (we also refer the reader to
the monograph by  Frisch \cite{frisch:1995} for further discussion). Despite
their success in fitting experimental observations
of structure function scaling, these phenomenological models
are not obtained from ``first principles'', i.e., they are not
directly derived from the Navier-Stokes equation.

In this paper we follow a different phenomenological approach~\cite{Friedrich1997} that interprets the concept of the turbulent energy cascade, i.e., the transport of energy from large to small scales, as a Markov process of velocity increments \emph{in scale}. The vigor of this phenomenology is its capability to reproduce the entire multi-scale velocity
increment statistics from the integral length scale down to a scale
where the Markov property is violated~\cite{Luck2006}.
The experimentally and numerically verified Markov property of the velocity
increments in the inertial range of scales, however,
implies that the increment PDF as well as the transition PDF
are governed by the same partial differential equation in scale, the so-called
Kramers-Moyal expansion. As it is discussed in Section~\ref{sec:markov} of the present paper, the Kramers-Moyal approach
allows for a general description of anomalous scaling. Consequently, it is
able to reproduce all known phenomenological models of turbulence by the proper
choice of the Kramers-Moyal coefficients that enter the Kramers-Moyal expansion.

The result of this paper is that, in order to obtain
an accurate description of intermittency effects, higher order Kramers-Moyal
coefficients have to be small but non-vanishing. Therefore,
the truncation of the Kramers-Moyal expansion  as it is done
in the usual Fokker-Planck approach~\cite{Friedrich1997,Renner2002a,Friedrich2011a} might result in an inaccurate description of the tails of the PDFs. To this end, we investigate
the asymptotics of higher-order
Kramers-Moyal coefficients of the corresponding phenomenologies in
Section~\ref{sec:markov}.

Section~\ref{sec:num} substantiates the existence
of higher order coefficients by direct numerical simulations of the Burgers equation. Due to its advantageous properties in comparison to the Navier-Stokes equation (no nonlocal pressure contributions, integrability via the Hopf-Cole transformation~\cite{Hopf1950,Cole1951}), the Burgers equation has been widely used as a model system for turbulence~\cite{Polyakov1995,E1999,bouchaud1996velocity,Eule2006,friedrich2018multiscale} and exhibits pronounced intermittency effects due to strong negative velocity gradients occurring in shocks
(we also refer the reader to~\cite{Bec2007a} for further references). Further applications of Burgers equation range from astrophysical problems~\cite{zel1970gravitational,hussain2011korteweg,dreher2005axisymmetric} to solid surface growth by vapor deposition via
the equivalent Kardar-Parisi-Zhang equation~\cite{kardar1986dynamic}.
Moreover, the inclusion of an additional nonlocal term allows for the incorporation
of intermittency effects similar to the ones that are encountered in Navier-Stokes turbulence~\cite{Zikanov1997}. Therefore, we will explicitly investigate the influence of the balance between nonlinearity and nonlocality and its consequences for the Kramers-Moyal coefficients. Furthermore, we will give an outlook on the extension of this analysis to ordinary Navier-Stokes turbulence.
\section{Interpretation of the Turbulent Energy Cascade as a Markov Process
of Velocity Increments in Scale}
\label{sec:markov}
A key quantity in the statistical description of turbulence~\cite{monin} is the $n$-increment PDF of longitudinal velocity
increments (\ref{eq:vel_inc}) defined according to
\begin{equation}
f_n(v_{n}, r_n; \ldots; v_{1},r_{1};\mathbf{x},t) = \prod_{i=1}^n \left \langle \delta(v_i-\delta_{r_i} v(\mathbf{x},t)) \right \rangle\;,
\label{eq:n-inc}
\end{equation}
where the brackets indicate ensemble averaging.
The $n$-increment PDF is a high-dimensional object whose determination from first principles is inaccessible due to the hierarchical ordering that is inherent in turbulent flows. In the following, we will focus on the spatial properties of the $n$-increment PDF at different scales $r_i$, i.e., we will assume stationarity. Due to the left-bounded velocity increment definition (\ref{eq:vel_inc}), we can further assume homogeneity with respect to the point of reference $\mathbf{x}$. In stochastic processes~\cite{risken}, the $n$-increment PDF can be expressed as a product of the $n-1$-increment PDF and a conditional probability through Bayes' theorem
\begin{equation}
p(v_{n},r_n | v_{{n-1}}, r_{n-1};\ldots; v_{1}, r_1)=
\frac{f_n(v_{n}, r_n; \ldots; v_{1},r_{1})}{f_{n-1}(v_{{n-1}}, r_{n-1}; \ldots; v_{1},r_{1})}\;.
\label{eq:bayes}
\end{equation}
In their seminal work, Friedrich and Peinke~\cite{Friedrich1997} investigated the multi-scale
velocity increment statistics in a
free jet experiment. They could show that longitudinal
velocity increments (\ref{eq:vel_inc}) possess a Markov property \emph{in scale},
namely
\begin{equation}
 p(v_3,r_3|v_2,r_2;v_1,r_1) = p(v_3, r_3|v_2,r_2) \quad \textrm{for} \quad 0 \le r_3 \le r_2 \le r_1\;,
 \label{eq:markov_prop}
\end{equation}
or more general
\begin{equation}
p(v_{n},r_n | v_{{n-1}}, r_{n-1};\ldots; v_{1}, r_1)=
p(v_{n},r_n | v_{{n-1}}, r_{n-1}1) \quad
\textrm{for} \quad 0 \le r_n \le r_{n-1} \le \ldots \le r_1\;.
\label{eq:markov_prop_gen}
\end{equation}
Further experiments~\cite{Luck2006} revealed that the Markov property
(\ref{eq:markov_prop}) is valid in the inertial range and is only broken at small scale separations
$r_2-r_3 < \lambda_{ME}$, where $\lambda_{ME}$ is termed the Markov-Einstein length. An important consequence of
the Markov property is that the $n$-increment PDF (\ref{eq:n-inc})
can be factorized into products containing only transition probabilities
\begin{equation}
  f_n(v_n,r_n;v_{n-1},r_{n-1};\ldots; v_1,r_1)
  =p(v_n, r_n|v_{n-1},r_{n-1})\ldots p(v_2, r_2|v_{1},r_{1})f_1(v_1,r_1)\;,
 \label{eq:n-inc-fact}
\end{equation}
for all $ r_{i-1}- r_i > \lambda_{ME}$ and $r_n \le r_{n-1} \le \cdots \le r_2 \le r_1.$
This means a considerable reduction of the complexity of the problem, since the knowledge
of the transition probabilities $p(v_i, r_i|v_{i-1},r_{i-1})$
is sufficient for the determination of the $n$-increment PDF ($f_1(v_1,r_1)$
is presumed to be known at large scales).
Moreover,
a central notion of a Markov process is that the one-increment
PDF and the transition
PDF follow the same Kramers-Moyal expansion in scale~\cite{risken}
\begin{align}
 -\frac{\partial}{\partial r_1}  f_1(v_1,r_1) =& \hat L_{KM}(v_1,r_1)  f_1(v_1,r_1),
  \label{eq:km_exp}\\
 -\frac{\partial}{\partial r_2}  p(v_2,r_2|v_1,r_1) =& \hat L_{KM}(v_2,r_2)  p(v_2,r_2|v_1,r_1)\;,
    \label{eq:km_exp_trans}
\end{align}
where $\hat L_{KM}$ is the Kramers-Moyal operator
\begin{equation}
 \hat L_{KM}(v,r) =  \sum_{k=1}^{\infty}(-1)^k \frac{\partial^k}{\partial v^k}  D^{(k)}(v,r)\;,
\end{equation}
and $D^{(k)}(v,r)$ are the Kramers-Moyal coefficients
\begin{equation}
 D^{(k)}(v_1,r_1)=\frac{1}{k!} \lim_{r_2 \rightarrow r_1} \frac{1}{r_1-r_2} \int \textrm{d} v_2 (v_2-v_1)^k
 p(v_2,r_2|v_1,r_1)\;.
 \label{eq:km_coeff}
\end{equation}
Here, the minus signs in Eqs.
(\ref{eq:km_exp} - \ref{eq:km_exp_trans}) indicate that the process runs from large to small
scales. In the following we want to relate the Kramers-Moyal expansion to
the scaling solutions of the different phenomenologies
of turbulence. To this
end, we take the moments $ \langle (\delta_r v)^n \rangle = \int \textrm{d}v v^n f_1(v,r)$ of the one-increment PDF in Equation (\ref{eq:km_exp})
\begin{align} \nonumber
- \frac{\partial }{\partial r} \langle (\delta_r v)^n \rangle=&
 -\frac{\partial}{\partial r} \int_{-\infty}^{\infty}\textrm{d}v v^n
 f_1(v,r)\\ \nonumber
 =& \sum_{k=1}^{\infty}(-1)^k \int_{-\infty}^{\infty}\textrm{d}v v^n
 \frac{\partial^k}{\partial v^k}  D^{(k)}(v,r) f_1(v,r)\\
  =& \sum_{k=1}^{n} \frac{n!}{(n-k)!} \int_{-\infty}^{\infty}\textrm{d}v v^{n-k}
   D^{(k)}(v,r) f_1(v,r)\;.
\end{align}
In order to match powers of $v$, we choose
$D^{(k)}(v,r)= \tilde D^{(k)}(r) v^k$ and obtain
\begin{equation}
\frac{\partial }{\partial r}\ln  \langle (\delta_r v)^n \rangle = -\sum_{k=1}^{n} \frac{n!}{(n-k)!}
\tilde D^{(k)}(r)\;.
\end{equation}
Scaling solutions $\langle (\delta_r v)^n \rangle \sim r^{\zeta_n}$ are recovered by the particular choice $\tilde D^{(k)}(r)=\frac{(-1)^k K_k}{k!}\frac{1}{r}$
\begin{equation}
  \frac{\partial }{\partial r}\ln \langle (\delta_r v)^n \rangle = \sum_{k=1}^{n} (-1)^{1+k} {n \choose k} K_k \frac{1}{r} \;,
\end{equation}
and integrating this equation from integral scale $L$ to small scales $r$
\begin{equation}
  \ln \left[\frac{\langle (\delta_r v)^n \rangle}{\langle (\delta_L v)^n \rangle}\right] = \sum_{k=1}^{n} (-1)^{k+1} {n \choose k} \ln \left[\frac{r}{L} \right]\;.
\end{equation}
Hence, in this alternative formulation of universality in turbulence~\cite{friedrich2018multiscale}, scaling exponents $\zeta_n$ of structure functions
\begin{equation}
   \langle (\delta_r v)^n \rangle = \langle (\delta_L v)^n \rangle r^{\sum_{k=1}^{n} (-1)^{k+1} {n \choose k}K_k} \;.
\end{equation}
are related to the sequence of so-called reduced Kramers-Moyal coefficients $K_n$
by a binomial transform $T$
\begin{equation}
 \zeta_n =-\sum_{k=1}^n (-1)^{k} {n \choose k} K_k.
 \label{eq:zeta_final}
\end{equation}
The binomial transform is an involution $TT=\mathbf{1}$, and, hence, the sequence of reduced Kramers-Moyal coefficients $K_n$ can be associated with the scaling exponents $\zeta_n$ of each phenomenological model of turbulence according to
\begin{equation}
  K_n =-\sum_{k=1}^n (-1)^{k} {n \choose k} \zeta_k.
  \label{eq:km_final}
\end{equation}
It should be noted that the binomial transform is usually defined to start from $k=0$, which can readily be included in Equations (\ref{eq:zeta_final}) and (\ref{eq:km_final}) since $\zeta_0=K_0=0$. Furthermore,
the fact that the reduced Kramers-Moyal coefficients $K_n$ are determined
by the scaling exponents $\zeta_n$
shows that the Kramers-Moyal expansion (\ref{eq:km_exp}) with specific Kramers-Moyal coefficients
\begin{equation}
  D^{(n)}(v,r)= \frac{(-1)^n K_n}{n!} \frac{v^n}{r}\;,
  \label{eq:km_scale}
\end{equation}
is general enough to
capture the essence of anomalous scaling. In other words, \emph{all} currently known phenomenological models of turbulence - characterized by their corresponding sets of scaling exponents $\zeta_n$ - can be reproduced by the Kramers-Moyal expansion (\ref{eq:km_exp}) with Kramers-Moyal coefficients (\ref{eq:km_scale}) where reduced Kramers-Moyal coefficients $K_n$ are related to the scaling exponents $\zeta_n$ by Equation (\ref{eq:km_final}). In the next subsections, we will describe in
detail how these different phenomenological models can be mapped onto the Kramers-Moyal coefficients.\\
~\\
\emph{i.) Kolmogorov's theory K41:}\\
The monofractal K41 phenomenology \cite{Kolmogorov1941} states that
$\langle (\delta_r v)^n \rangle = C_n \langle \varepsilon\rangle ^{n/3} r^{n/3}$
and an evaluation of the reduced Kramers-Moyal coefficients (\ref{eq:km_final})
suggests that it can be reproduced by just
a single Kramers-Moyal coefficient
\begin{equation}
 K_n = \bigg \{\begin{array}{ll}
		1/3 & \textrm{ for } n \le 1, \\
		 0 & \textrm{ for } n > 1.
	   \end{array}
\end{equation}
\emph{ii.) Kolmogorov-Oboukhov theory K62:}\\
A first intermittency model which assumes a log-normal distribution of
the local rate of energy dissipation $\varepsilon$
has been proposed by Kolmogorov~\cite{Kolmogorov1962}
and Oboukhov~\cite{Oboukhov1962}.
\begin{figure}
\centering
\includegraphics[angle=270, width=0.64 \textwidth]{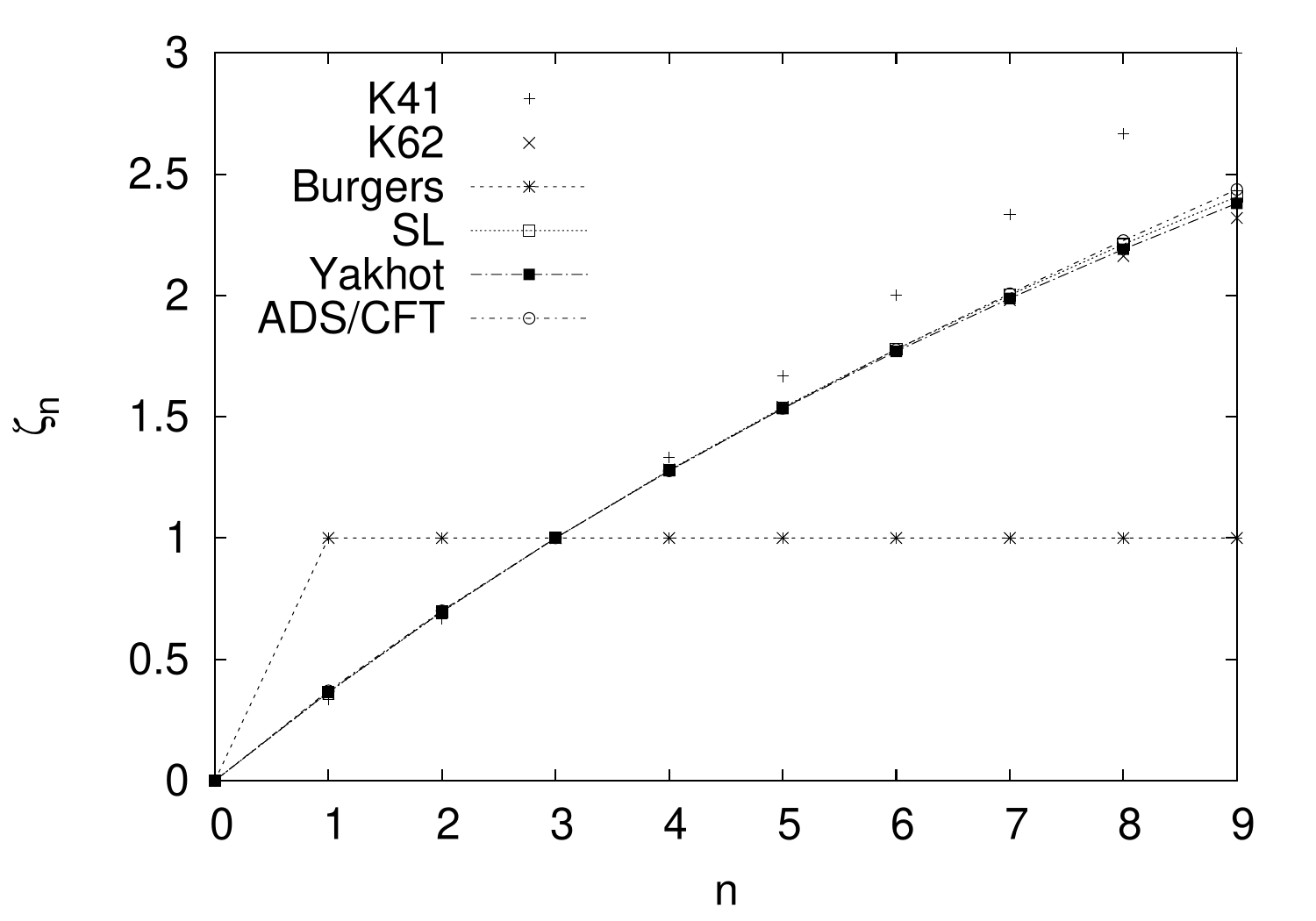}
\caption{Scaling exponents $\zeta_n$ of velocity structure functions
for the different phenomenologies discussed in \emph{i.)-vi.)}. The crosses that are arranged
on the straight $n/3$-line correspond to the self-similar K41 phenomenology \emph{i.)}.
Burgers phenomenology \emph{iii.)}
exhibits the strongest intermittency behavior whereas the other phenomenologies
can only be distinguished for higher orders $n$. Note that the K62 phenomenology \emph{ii.)}
has a parabolic form that violates the structure function convexity condition
\cite{frisch:1995} for $n \ge \frac{3}{2} + \frac{3}{\mu}$ (not observable in the figure).}
\label{fig:scale_exp}
\end{figure}
It predicts the scaling of the structure functions according to
$\langle (\delta_r v)^n \rangle = C_n \langle \varepsilon \rangle^{\frac{n}{3}} r^{\frac{n}{3}} \left(\frac{r}{L}\right)^{-\frac{n(n-3)\mu}{18}}$
where $L$ is the integral length scale and $\mu$ is the so-called intermittency
coefficient which is of the order $\mu \approx 0.227$. As it has been discussed
by Friedrich and Peinke~\cite{Friedrich1997}, this reduces the Kramers-Moyal expansion to a Fokker-Planck
equation with drift and diffusion coefficient
\begin{equation}
	K_1 = \frac{3+\mu}{9} \quad \textrm {and}\quad
	K_2= \frac{\mu}{9} ,
\end{equation}
and implies the vanishing of all higher-order coefficients.\\
~\\
\emph{iii.) Burgers scaling:}\\
The velocity structure functions in Burgers turbulence ~\cite{Bec2007a}
follow the extreme scaling
\begin{equation}
 \langle (\delta_r v)^n \rangle=\Bigg \{\begin{array}{ll}
		C_n \frac{\langle \varepsilon^{n/2}\rangle }{\nu^{n/2}} r^n  & \textrm{ for } n < 1, \\[0.5em]
		C_n  \langle \varepsilon \rangle^{\frac{n}{3}} L^{\frac{n}{3}-1} r & \textrm{ for } n \geq 1.
	   \end{array}
 \label{v_scaling}
\end{equation}
Here, the first scaling is due to smooth positive velocity increments in the ramps,
whereas the latter scaling corresponds to negative velocity increments
dominated by shocks that form due to the compressibility of the velocity field in the vicinity of
the viscosity $\nu \rightarrow 0$.
\begin{figure}
\centering
\includegraphics[angle=270, width=0.68 \textwidth]{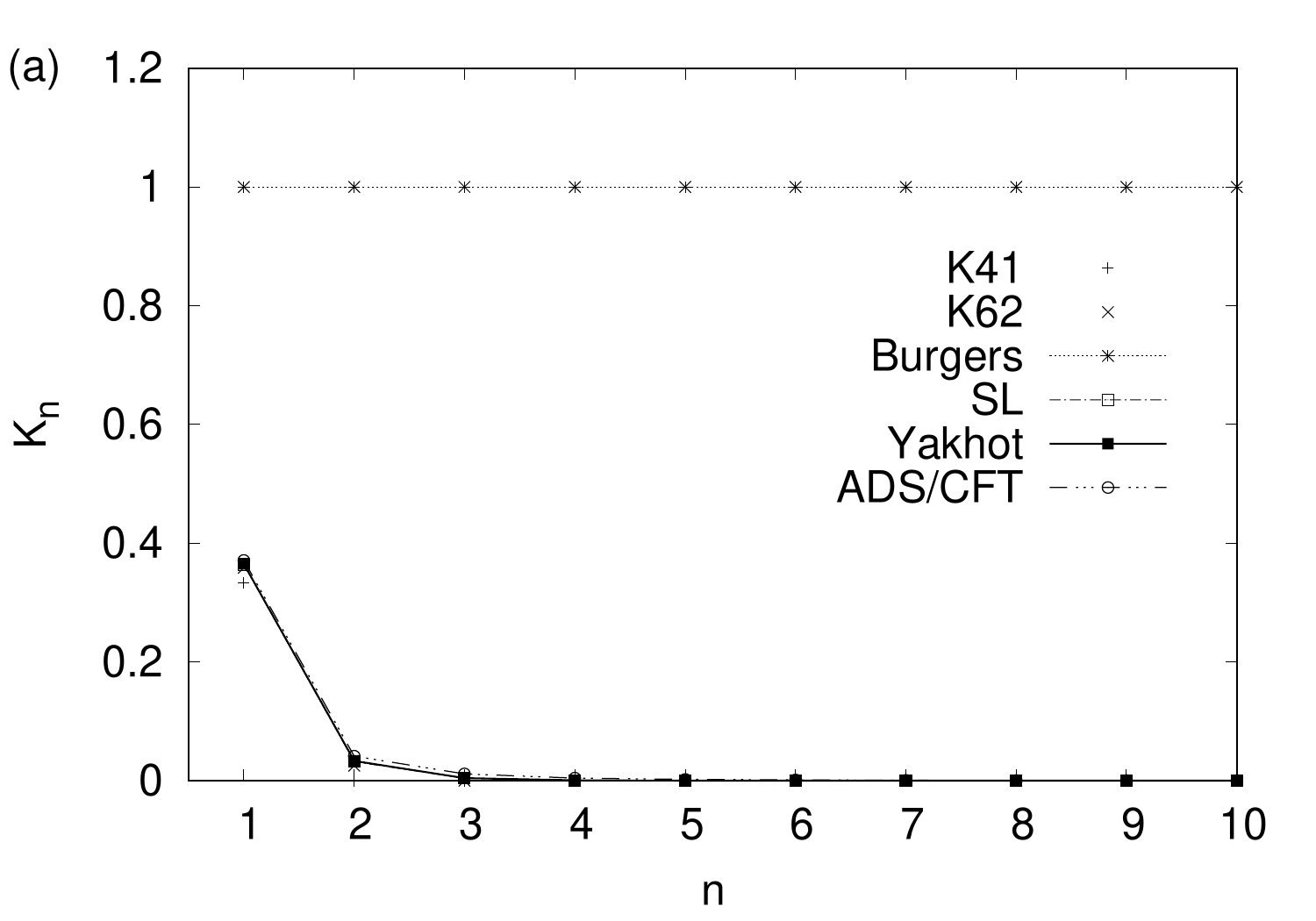}
  \includegraphics[angle=270, width=0.73 \textwidth]{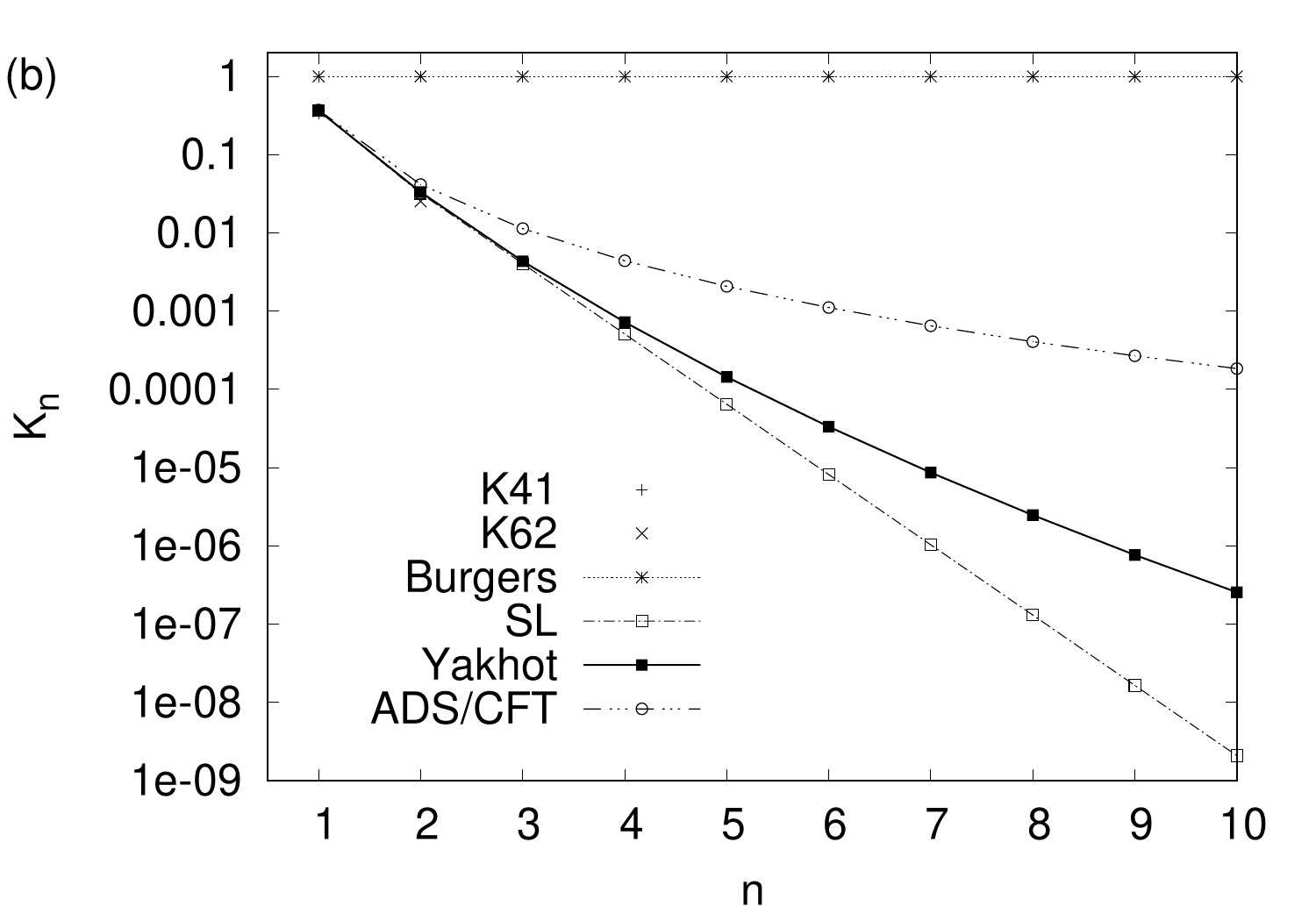}~~~~~~
\caption{(a) Reduced Kramers-Moyal coefficients from Equation (\ref{eq:km_final})
for different phenomenological models of turbulence
up to the order $n=10$. Coefficients for $n > 2$ seem to tend towards zero.
(b) Semi-logarithmic plot of the reduced Kramers-Moyal
coefficients. All phenomenological models
except for K41 and K62 show an asymptotic behavior.
Note that the She-Leveque model possesses a nearly linear slope in the semi-logarithmic
representation.}
\label{fig:km}
\end{figure}
The smooth solutions correspond to a single Kramers-Moyal coefficient, whereas
the shock solutions can only be reproduced by an infinite number of Kramers-Moyal
coefficients and we obtain
\begin{equation}
  \begin{array}{ll}
    K_1 = 1, \;\;\;
   	K_n = 0 \;\textrm{ for }\; n > 1 \;,\;\; & \textrm{for positive increments} . \\
    K_n = 1 , \;\;\; \forall ~n               &\textrm{for negative increments} .
  \end{array}
 %\label{eq:km_ramps}
 \label{eq:km_shock}
\end{equation}
\emph{iv.) She-Leveque model:}\\
The She-Leveque model~\cite{She1994} for 3D Navier-Stokes turbulence predicts
scaling exponents
$\zeta_n= \frac{n}{9} + 2\left( 1 - \left( \frac{2}{3}\right)^{n/3} \right)$
that are in very good agreement with both
experimental and numerical data. This yields an infinite set of coefficients~\cite{Nickelsen2015} and the reduced Kramers-Moyal coefficients read
\begin{equation}
	K_n =  \frac{n}{9} \, _1F_0(1-n;;1)+2
   \left(1-\sqrt[3]{\frac{2}{3}}\right)^n\;,
\end{equation}
where $_{\nu}F_{n}(a;b;z)$ is the generalized hypergeometric function. It has been shown recently~\cite{Nickelsen_2017} that this particular model can be apprehended as a jump process of a stochastic process for the velocity increments in scale.\\
\\
\emph{v.) Yakhot model:}\\
Yakhot~\cite{yakhot:2001,yakhot:2006} introduced a model for structure function exponents
$\zeta_{2n} = \frac{2(1+3\beta)n}{3(1+2\beta n)}$
based on a mean-field approximation. Similar scaling exponents were first derived by Novikov~\cite{Novikov1994} and subsequently by Castaing~\cite{Castaing1996}.
With the choice of $\beta=0.05$, structure functions agree equally
well with experimental data as the popular She-Leveque model. The
translation to the Krames-Moyal coefficients is given by
\begin{equation}
	K_n = \frac{\Gamma[n+1]}{\Gamma\left[n+1 + \frac{1}{\beta}\right]}
   \textstyle \left(\Gamma \left [1+\frac{1}{\beta}\right] +
   \frac{1}{3 \beta^2} \Gamma \left[\frac{1}{\beta}\right] \right) .
\end{equation}
\\
\emph{vi.) ADS/CFT random geometry model:}\\
Eling and Oz~\cite{Eling2015} introduced a structure function scaling model
which is motivated by a gravitational Knizhnik- Polyakov-Zamolodchikov
(KPZ)-type relation. For 3D
Navier-Stokes turbulence, they derive
\begin{equation}
	\zeta_n = \frac{\left( (1+\gamma^2)^2 + 4 \gamma^2(\frac{n}{3} -1)\right)^\frac{1}{2} + \gamma^2 -1}
	{2\gamma^2},
\end{equation}
where experimental data suggests the value
$\gamma^2=0.161$.
Unfortunately, we could not obtain an analytical formula for the coefficients of
this particular model and have restricted ourselves to a numerical evaluation of Equation (\ref{eq:km_final}).

We have plotted the reduced Kramers-Moyal coefficients $K_n$
for the different models up to the order $n=10$ in Figure~\ref{fig:km} (a).
As one can see, all models besides K41 and Burgers can hardly be distinguished from one another
and the reduced Kramers-Moyal coefficients seem to tend towards zero very quickly.
According to a theorem due to Pawula~\cite{Pawula1967}
(see also \cite{risken}), the vanishing of the
fourth-order Kramers-Moyal coefficient implies that all higher coefficients are
zero as well and the Kramers-Moyal expansion (\ref{eq:km_exp}) reduces to an ordinary Fokker-Planck equation. The latter is particularly suitable for
modeling approaches via its corresponding Langevin equation as well as
the undemanding determination of statistical quantities via the
exact short-scale propagator of the Fokker-Planck equation~\cite{risken}.

In the original work~\cite{Friedrich1997} and also
all subsequent works~\cite{Luck2006,Renner2001,Renner2002a}
it was argued in favor of Pawula's theorem since the
experimentally determined Kramers-Moyal coefficient of order four was very close to zero. Figure~\ref{fig:km} (a) seems to agree qualitatively with this finding. However, in order to demonstrate that this can be
misleading, we  show a semi-logarithmic plot of  Figure~\ref{fig:km} (a) in
Figure~\ref{fig:km} (b). It can be seen that the
models \emph{iv.)-vi.)} tend asymptotically towards zero
and higher-order Kramers-Moyal are rather small but strictly non-zero.
At this point, we want to emphasize that since $K_4 \approx 10^{-3}$,
the significant detection of these higher-order coefficients
in the experiment might be quite challenging due to the presence of measurement noise or insufficient statistics. Nevertheless, since
the models \emph{iv.)-vi.)} agree quite well with experimental data
an accurate determination of the higher-order coefficients should be
within the reach of a spatially and temporarily well-resolved high-Reynolds number experiment. Moreover, Pawula's theorem directly reduces the velocity increment statistics to families of the K62 phenomenology \emph{ii.)}. It
should therefore be noted that the latter is only valid for moments $\langle v^n \rangle$ that do not exceed the order $n \ge \frac{3}{2} + \frac{3}{\mu}$, due to the convexity condition for $\zeta_{2n}$ (see also \cite{frisch:1995} for further discussion). Consequently, one should bear in mind that whilst modeling or other purposes of the Fokker-Planck approach, the tails of the PDFs might not be described accurately, although - admittedly - this effect should be rather small.
In the following section, we want to quantify the small-scale intermittency behavior on the basis of the reduced Kramers-Moyal cofficients (\ref{eq:km_final}) at the example of a generalized Burgers equation with an additional nonlocality.
\begin{table}[h]
\centering
\begin{tabular}{c c c c}
  run  & \#1 ($\alpha =1$) & \#2 ($\alpha=0$) & \#3 ($\alpha=0.15)$ \\[3pt]
  \hline
  $u_{rms}$ &  0.0079 & 0.0058    & 0.0026\\
  $\nu$ & 0.000014 &     0.00001  & $1 \times 10^{-6}$\\
  $\langle \varepsilon \rangle$ & $5.45 \times 10^{-7}$ & $1.38 \times 10^{-7}$  & $6.23 \times 10^{-8}$\\
  dx & 0.002    & 0.0015  & 0.0015\\
  $\eta$ & 0.0084 & 0.0092 & 0.002 \\
  $\lambda$ & 0.0401 & 0.04895 & 0.0104\\
  Re$_{\lambda}$ & 22.71   & 28.1668 & 27.36\\
  $L$ & 0.9119 & 1.379  & 0.286\\
  $T$ in $T_L$ & 7441  & 1057 &2299\\
  $N$ & 3072& 4096 & 4096\\
  cut-off & 3052 & 4066 & 4066\\
\end{tabular}
\caption{Characteristic parameters of the
numerical simulations: root mean square velocity $u_{rms}= \sqrt{ \langle u^2 \rangle}$, viscosity $\nu$, averaged rate of local energy dissipation
$\langle \varepsilon \rangle= 2
\nu \left\langle \left(\frac{\partial u}{\partial x} \right)^2 \right\rangle$,
grid spacing dx,  dissipation length $\eta=\left(\frac{\nu^3}{\langle \varepsilon \rangle} \right)^{1/4}$,
Taylor length $\lambda = u_{rms} \sqrt{\frac{\nu}{\langle \varepsilon \rangle }}$,
Taylor-Reynolds number $\textrm{Re}_{\lambda}=
\frac{u_{rms} \lambda}{\nu}$,
integral length scale $L= \frac{u_{rms}^3}{\langle \varepsilon \rangle}$,
large-eddy turn-over time
$T_L=\frac{L}{u_{rms}}$, number of grid points $N$ and cut-off
of the power law forcing. The intermediate case ($\alpha=0.15$) included a damping term of the form $-\gamma u(x,t)$ with $\gamma=0.002$ on the r.h.s.
of Equation (\ref{eq:gen_burgers}).}
\label{tab:1}
\end{table}
\section{Direct Numerical Simulations of a Generalized Burgers Equation}
\label{sec:num}
We consider the generalized Burgers equation
\begin{equation}
 \frac{\partial }{\partial t} u(x,t) + \alpha u(x,t) \frac{\partial}{\partial x} u(x,t)
 +\frac{1-\alpha}{\pi} \textrm{p.v.}\int \textrm{d} x' \frac{u(x',t)}{x-x'} \frac{\partial}{\partial x} u(x,t)
 = \nu \frac{\partial^2}{\partial x^2}u(x,t) +F(x,t)\;,
 \label{eq:gen_burgers}
\end{equation}
with forcing that is assumed to be white noise in time
$\langle F(x,t) F(x',t) \rangle = \chi(x-x') \delta(t-t')$. Here, the spatial correlation
of the forcing follows a power law in Fourier space~\cite{Chekhlov1995}, namely
\begin{equation}
 \langle \hat F(k,t) \hat F(k',t) \rangle \sim k^{-1} \delta(k-k') \delta(t-t')\;.
 \label{eq:forcing}
\end{equation}
Moreover, $\alpha = 1$ in Equation (\ref{eq:gen_burgers})
corresponds to the case of Burgers turbulence, whereas $\alpha = 0$ corresponds
to a purely nonlocal case that is dominated by self-similar behavior~\cite{Zikanov1997}.
The intermediate case $\alpha=0.15$, however, exhibits several similarities to the intermittency behavior of 3D hydrodynamic turbulence. The latter
manifests itself by a skewed velocity gradient PDF, as well as by nonlinear
scaling exponents $\zeta_n$ of the velocity structure functions $\langle (\delta_r v)^n \rangle \sim |r|^{\zeta_n}$ which has been reported by Zikanov, Thess and Grauer~\cite{Zikanov1997}.
Apparently, the balance between  the nonlinear and the nonlocal term results in particular dissipative structures that have considerable influence on the intermittency behavior of the system.

Equation (\ref{eq:gen_burgers}) has been solved with the help of a standard pseudo-spectral code.
A third order Runge-Kutta method was used for the time stepping due to its vigor of capturing the temporal evolution of shock-like structures~\cite{Shu1988}. Furthermore, aliasing errors that occur
due to an insufficient resolution of the quadratic nonlinearities were treated
with the help of a filter in Fourier space~\cite{Hou2007}. %The advantages of the latter method in comparison to the more frequently used $2/3$-rule~\cite{Orszag1971}, is its treatment of
%higher wavenumbers which occur during the computation of the quadratic nonlinear terms: The $2/3$-rule simply cuts off wave-numbers that exceed the wavenumber $2/3 ~k_{\max}$ of the largest wavenumber $k_{\max}$ that can be resolved by the grid. Especially for the case of Burgers turbulence, the smoother Fourier filter~\cite{Hou2007} leads to a more efficient suppression of oscillations, caused by nearly singular shock structures. In both methods the nonlinearities that are calculated
%in real space and then transformed into Fourier space are multiplied with the filter function
%$S(k)$, where the different filters read
%
%\begin{equation}
 %S_{Orzsag}(k) =  \bigg \{\begin{array}{ll}
%		1 & \textrm{ for } k < 2k_{\max}/3\;, \\
%		 0 & \textrm{ for } k \geq 2 k_{\max}/3\;,
%	   \end{array}  \qquad \textrm{and} \qquad
% S_{Hou}(k) = \exp \left[-36 \left(\frac{k}{k_{\max}}\right)^{36} \right]\;.
%\end{equation}
%
Concerning the realization of the forcing determined by Equation (\ref{eq:forcing}),
we assured the white-noise in time condition with a numerical
method for the Langevin equation discussed by Higham~\cite{Higham2001}.
Cheklov and Yakhot~\cite{Chekhlov1995} reported that a power law correlation function $\sim k^{-1}$ results in a Kolmogorov-type spectrum which can also be determined on the basis of a one-loop renormalization group consideration. Here, the nonlinearity directly
balances forcing contributions that are acting on all scales. Therefore,
the cut-off of the forcing is located in the neighborhood of the dissipation range in all simulations.

Table~\ref{tab:1} shows the characteristic parameters of the DNS. The attained Reynolds numbers in the simulations are not as high as the Reynolds numbers attained in references~\cite{Chekhlov1995,Zikanov1997}.
This is due to the fact that the latter results were obtained with a so-called
hyperviscous term, i.e., replacing $\frac{\partial^2}{\partial x^2}$
by $\frac{\partial^{2\alpha}}{\partial x^{2\alpha}}$, which leads to an efficient damping of the higher wavenumber (small-scale) components of the velocity field. By this method, higher Reynolds numbers can be attained, which leads to an increased inertial range. Apparently, the concept
of hyperviscosity has no considerable influence on the inertial range behavior~\cite{Chekhlov1995}, therefore it can be considered as an efficient method to attain higher Reynolds numbers. Nonetheless, in this work we intend to investigate the original dissipative effects, and thus only the Laplacian viscous term has been considered. Moreover, we chose the resolution $N$ in
a way which allows for the production of a sufficient amount of data. These restrictions lead to the moderate Reynolds numbers in Table~\ref{tab:1}.
\begin{figure}
 \centering
\includegraphics[width=0.49 \textwidth]{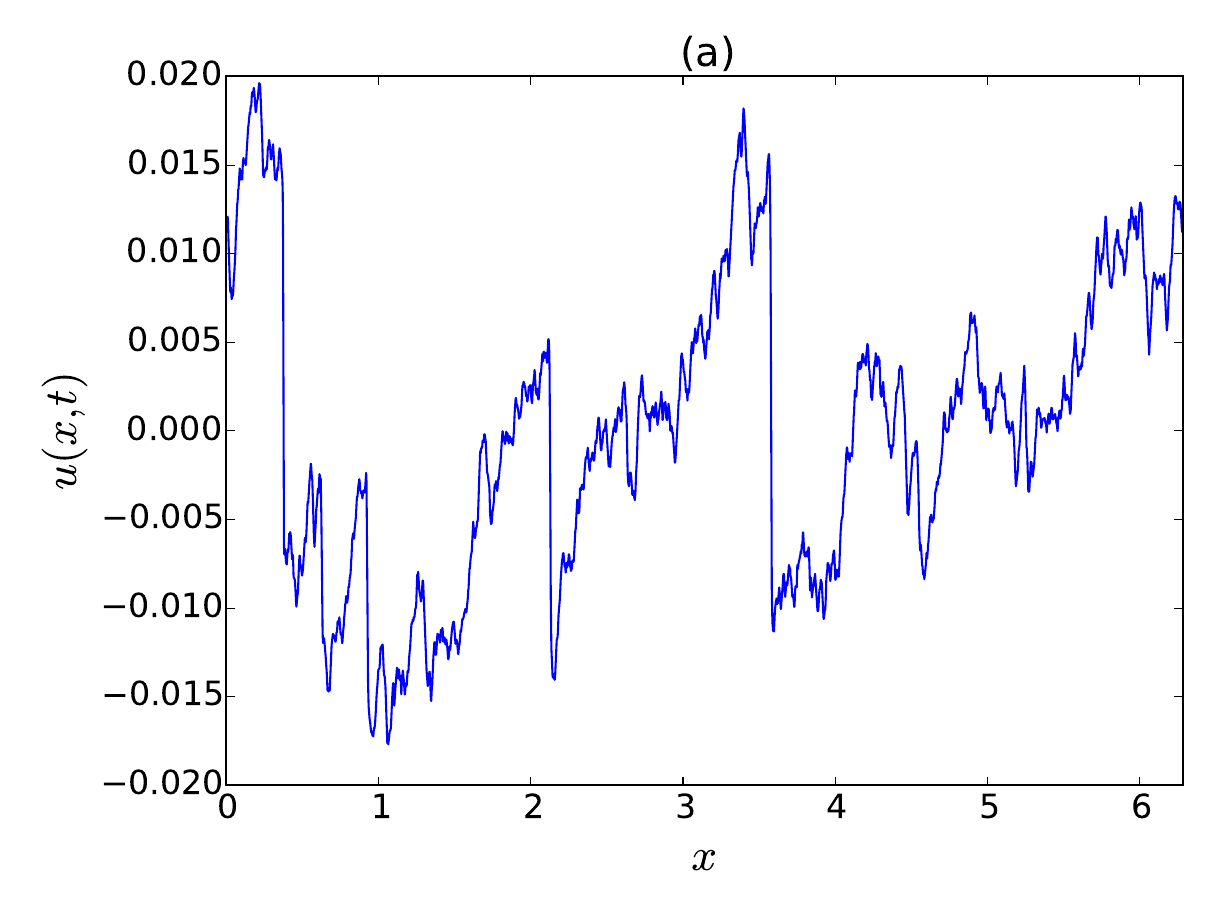}
\includegraphics[width=0.49 \textwidth]{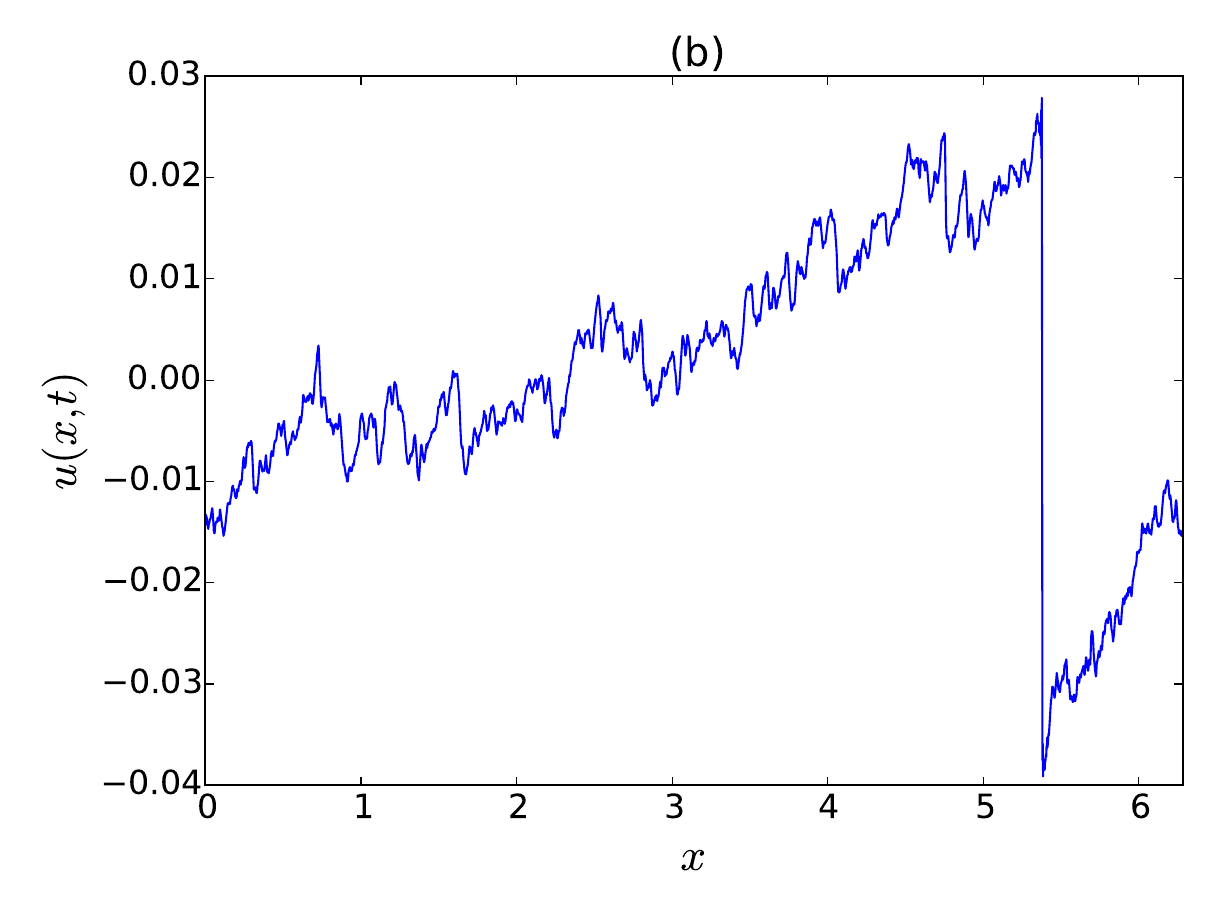}
\caption{(a) Typical realization of the velocity field $u(x,t)$ in
DNS of Burgers turbulence ($\alpha=1$ in Equation (\ref{eq:gen_burgers})).
The velocity field exhibits a sawtooth profile that
is due to the formation of large negative velocity gradients.
(b) Velocity field realization that belongs to the largest velocity
gradient attained in the DNS belonging to Tab.~\ref{tab:1}.}
\label{fig:burgers}
\end{figure}
\subsection{Strong Intermittency $\alpha=1$: Burgers Equation}
For the case $\alpha=0$, the additional nonlocality in Equation (\ref{eq:gen_burgers}) vanishes and we recover the ordinary Burgers equation with its shock-type velocity profiles.
A typical realization of the velocity field of Burgers turbulence (run \#1)
is shown in Figure~\ref{fig:burgers} (a).
The velocity field exhibits a sawtooth-like structure and forms shock fronts consisting of large negative velocity gradients. In this particular snapshot, we can distinguish three or four shocks that are connected via ramps of positive inclination and are superimposed by small-scale structures that also consist of small shocks.
Figure~\ref{fig:burgers} (b) shows the velocity field belonging to the
most extreme shock event that occurred during the simulations.
It consists of one large negative gradient event that is barely resolved
by the corresponding resolution of $N=3072$ grid points.
Events as the one depicted in Figure~\ref{fig:burgers} (b)
are extremely rare, but they do bear a particular
statistical significance.

\begin{figure}
 \centering
\includegraphics[width=0.49 \textwidth]{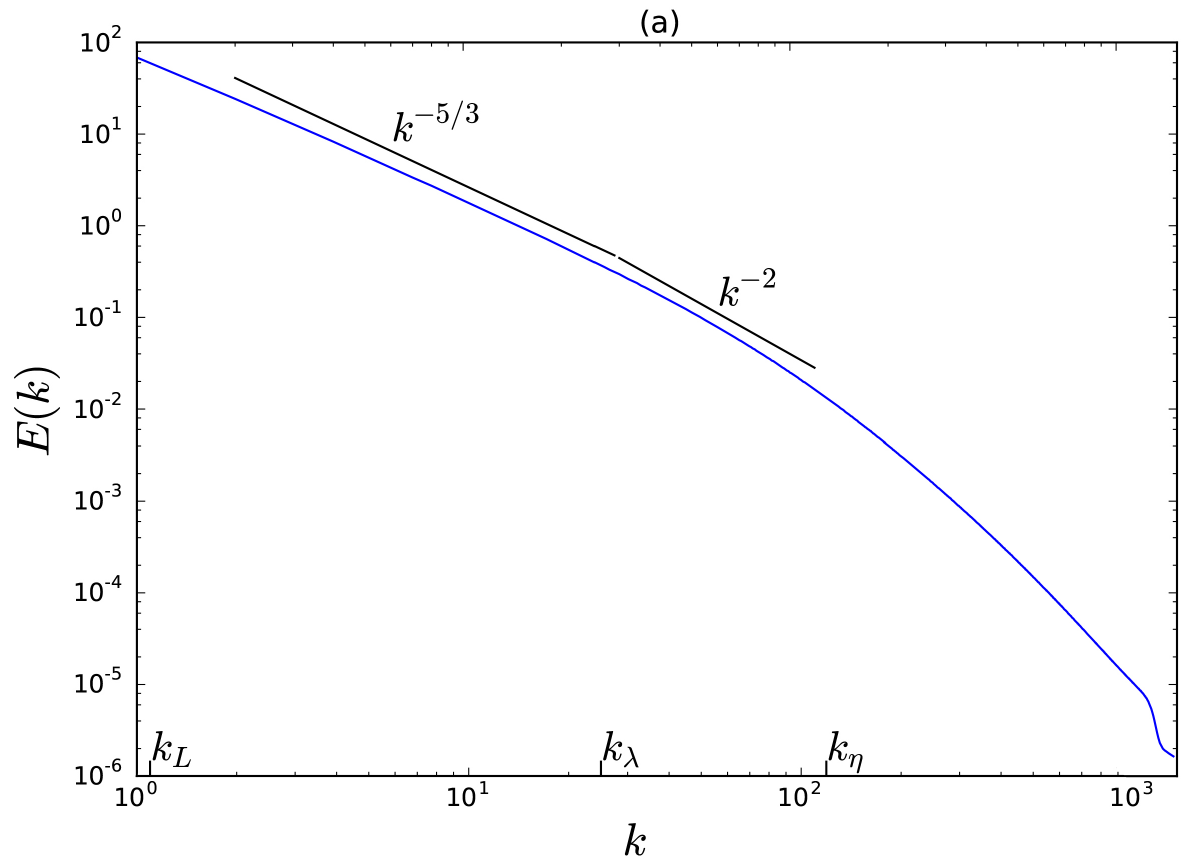}
\includegraphics[width=0.49 \textwidth]{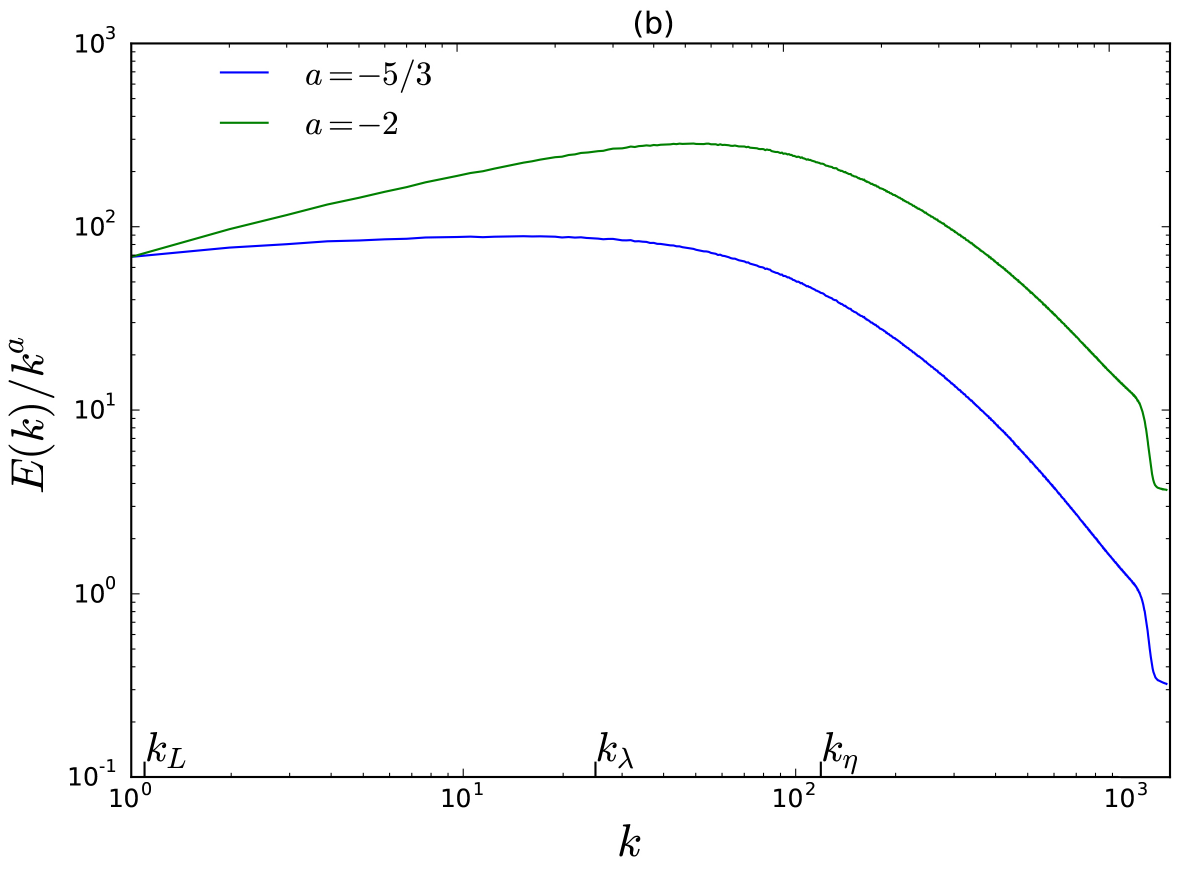}
\caption{(a) Energy spectrum $E(k,t)$ of the velocity field of
DNS of Burgers turbulence. The small wavenumbers (largest scales)
show nearly Kolmogorov-like behavior $\sim k^{-5/3}$ whereas
a shock-like behavior $k^{-2}$ can be perceived for a
small band of larger wavenumbers at the edge of the inertial range.
The dissipation range exhibits an exponential decay of the energy
spectrum. Inertial range limits have been indicated by $k_L$ and $k_{\eta}$
and are explained in detail in the plain text.
(b) Compensated energy spectra $E(k)/k^a$. The blue lines corresponds
to the Kolmogorov case $a=-5/3$ and the green line to the shock case $a=-2$.
Constant lines in the plot indicate scaling behavior of the spectrum.
The spectrum for smaller wavenumbers is quite close to the Kolmogorov spectrum.
Larger wavenumbers in the inertial range at around $60< k<80$
show more of a shock-like spectrum, also the compensated spectrum
is not as flat as in the Kolmogorov case.}
\label{fig:spec_burgers}
\end{figure}
Figure~\ref{fig:spec_burgers} (a) shows the energy spectrum $E(k)$ of simulation \#1.
Here, the inertial range is limited by the wave numbers $k_L$ and $k_{\eta}$ which are associated
to the integral length scale $L$ and the Kolmogorov dissipation length $\eta$. The small wavenumber regime can be described quite accurately by a Kolmogorov-like
spectrum $E(k) \sim k^{-5/3}$. However, as the wavenumbers increase, the spectrum drops faster than the Kolmogorov spectrum. It is tempting to propose a shock-like spectrum $E(k) \sim k^{-2}$ for intermediate wavenumbers $60<k<80$ that piles up in front of the dissipation range. Obviously, the latter does not manifest itself as clear as the $k^{-5/3}$-part, but since the spectrum drops already in front of the dissipation range and shocks represent a small-scale quantity, it seems to be a convenient explanation. We must emphasize, however, that in the original work of Cheklov and Yakhot~\cite{Chekhlov1995}, solely
the Kolmogorov-spectrum has been observed.
The latter finding might be a result of hyperviscosity and the attained high Reynolds numbers. In order to further quantify these tendencies, Figure~\ref{fig:spec_burgers} (b)
shows compensated spectra $E(k)/k^a$ with $a=-5/3$ and $a=-2$. Plateaus
in the plot correspond to scaling behavior of the spectrum and confirm
the Kolmogorov part at small wavenumbers and a small range of the shock-like part as well.

\begin{figure}[h]
  \vspace{-2ex}
  \centering
\includegraphics[width=0.54 \textwidth]{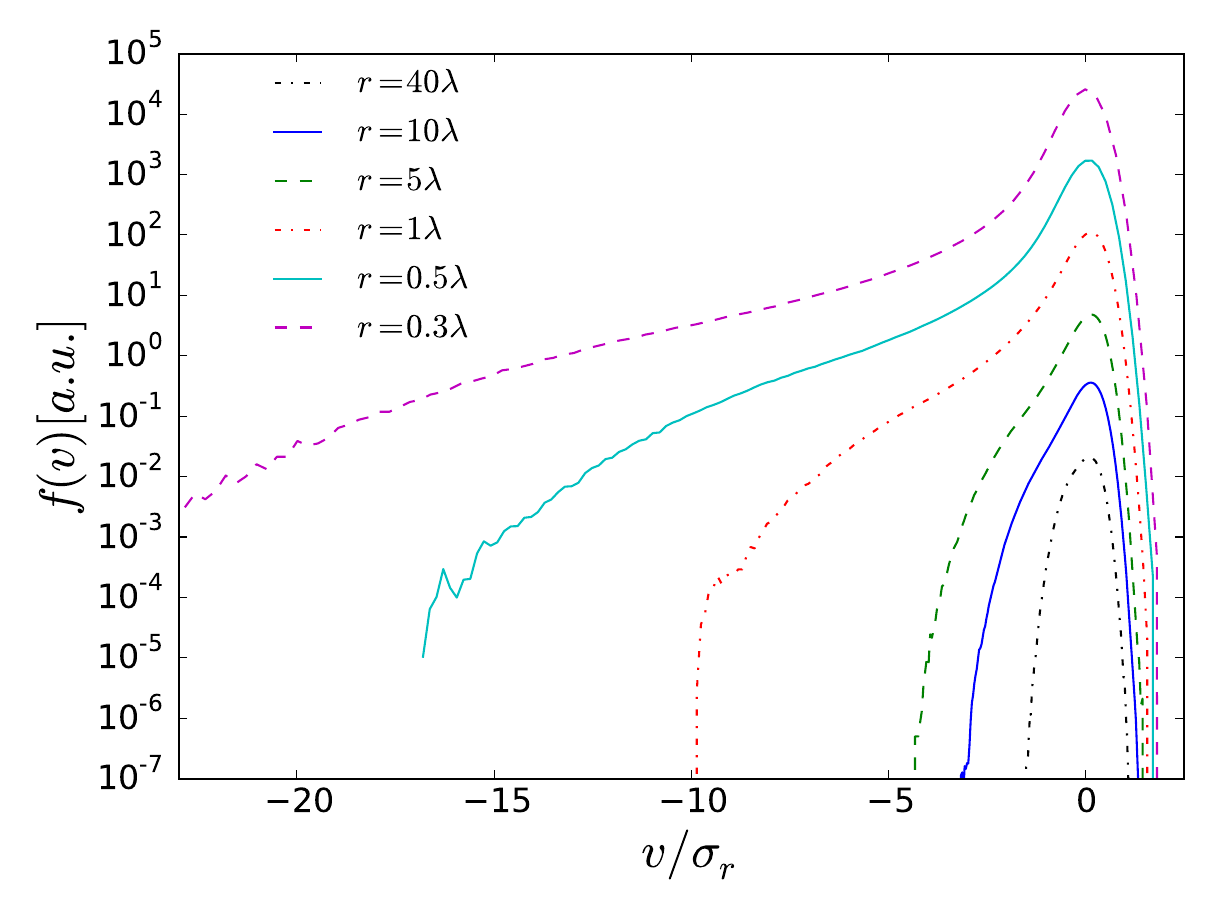}
\caption{Evolution of the velocity increment PDF in scale (multiples of the Taylor length $\lambda$) for the Burgers case $\alpha=1$. The PDFs are shifted vertically  and normed with their corresponding
standard deviation $\sigma_r$ for improved visualization. The pronounced left
part of the PDFs is dominated by small-scale shock events whereas the right part exhibits nearly self-similar behavior.}
\label{fig:scale_burgers}
\end{figure}

Figure~\ref{fig:scale_burgers} shows the evolution of the one-increment PDF $f_1(v,r)$ in scale, where $r$ is expressed in multiples of the Taylor length $\lambda$. At small scales, the PDF shows a pronounced left tail due to shock events whereas it is close to Gaussian on large scales. Here, the PDF for $r=0.3 \lambda$ seems to exhibit an algebraic part for negative increments and drops exponentially for larger negative increments. However, it is not obvious whether the algebraic part corresponds to $(v/r)^{-7/2}$ as it has been predicted for the gradient
PDF~\cite{E1999}. Due to the moderate Reynolds numbers, it is also not clear whether the exponential decay for large negative velocity increments follows the exponential decay $\sim e^{-\left(v/(\textrm{Re}\;r))\right)^{3/2}}$ predicted from instanton calculations~\cite{Balkovsky1997}.
\subsubsection{Examination of the Markov Property}
\label{sec:ex_burgers}
In this section, we seek to examine the Markov property (\ref{eq:markov_prop}) from DNS of Burgers turbulence. To this end, we briefly mention two possibilities: First, the Markov property can be
examined directly in comparing the conditioned PDF with the transition PDF
\begin{equation}
 p(v_3, L/2-\Delta r| v_2, L/2; v_1, L/2+ \Delta r) = p(v_3, L/2-\Delta r|v_2, L/2)\;,
 \label{eq:markov_direct}
\end{equation}
where the intermediate scale $L/2$ was chosen to lie well within the inertial range and $\Delta r$ can be considered as the variable step width of the process. In general, the intermediate scale can also be chosen at a different inertial range scale, but in the following we will only consider this particular case.
At this point, since we are comparing two objects of different dimensionality,
we perform cuts of the two-times conditioned PDF for fixed $v_1$. This can be done by the simple choice $v_1=0$, which offers the best statistics.
However, for a more critical examination, it is appropriate to put $v_1$ at least to $\sigma_{\infty}$,
where
\begin{equation}
 \sigma_{\infty}= \lim_{r\rightarrow \infty} \sqrt{(\delta_r v)^2 \rangle}=
 \lim_{r\rightarrow \infty} \sqrt{\langle (u(R+r)-u(R))^2 \rangle}=\sqrt{2} u_{rms}\;.
\end{equation}
\begin{figure}[h!]
 \centering
\includegraphics[width=0.49 \textwidth]{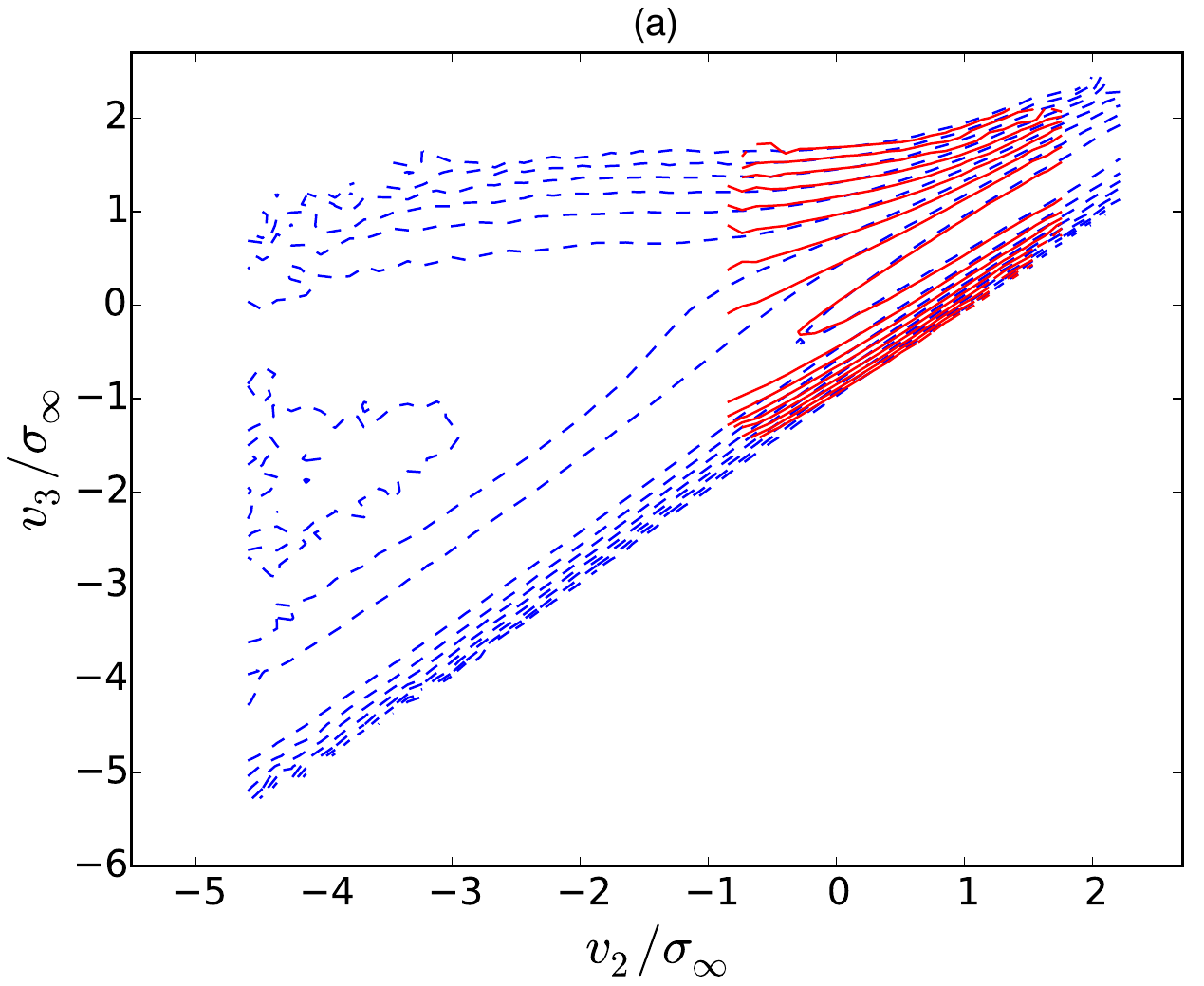}
\includegraphics[width=0.495 \textwidth]{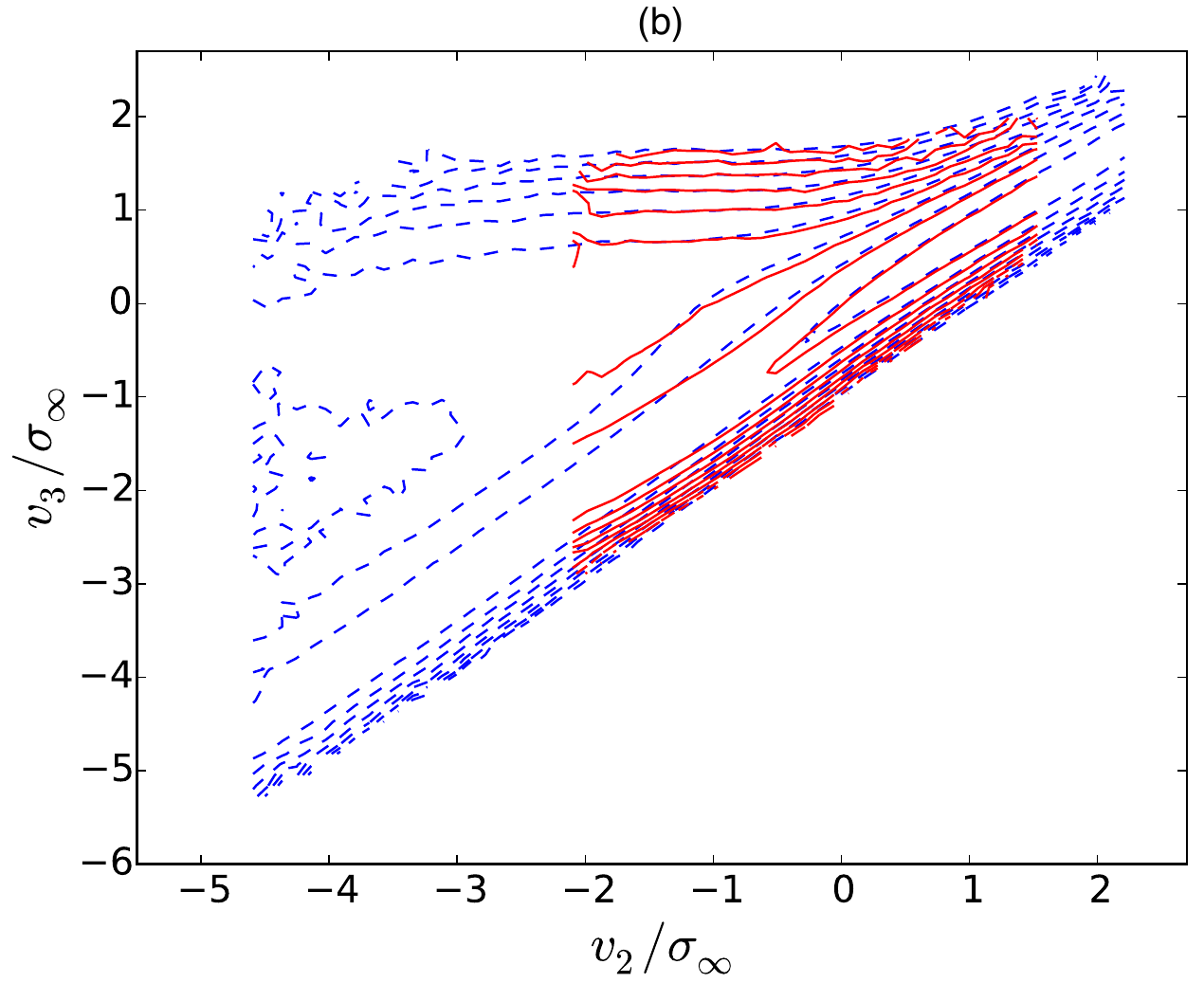}
\caption{(a) Examination of the Markov property (\ref{eq:markov_direct}) from DNS of Burgers turbulence
for $r= 2\lambda$ and $v_1=0$ via a logarithmic contour plot.
The dashed blue contour lines correspond to $p(v_3,L/2-\Delta r|v_2,L/2)$ whereas the red lines correspond to $p(v_3,L/2-\Delta r|v_2, L/2; v_1, L/2+ \Delta r)$. The Markov property is fulfilled only approximately. Although the contours exhibit an almost identical shape, they seem to be slightly displaced.
The shape of the transition PDF (blue) can be reproduced for the most part by the theoretical predictions
(\ref{eq:trans}).
(b) Same as in (a), but for $v_1 = - \sigma_{\infty}$. A broader
part of the blue dashed lines that belong to $p(v_3,L/2-\Delta r|v_2,L/2)$
can be covered in comparison to (a). Event though the red contour lines indicate insufficient statistics in the boundary regions, the Markov
property has not deteriorated in comparison to (a).}
\label{fig:markov_burgers_2}
\end{figure}
\begin{figure}[h!]
 \centering
\includegraphics[width=0.49 \textwidth]{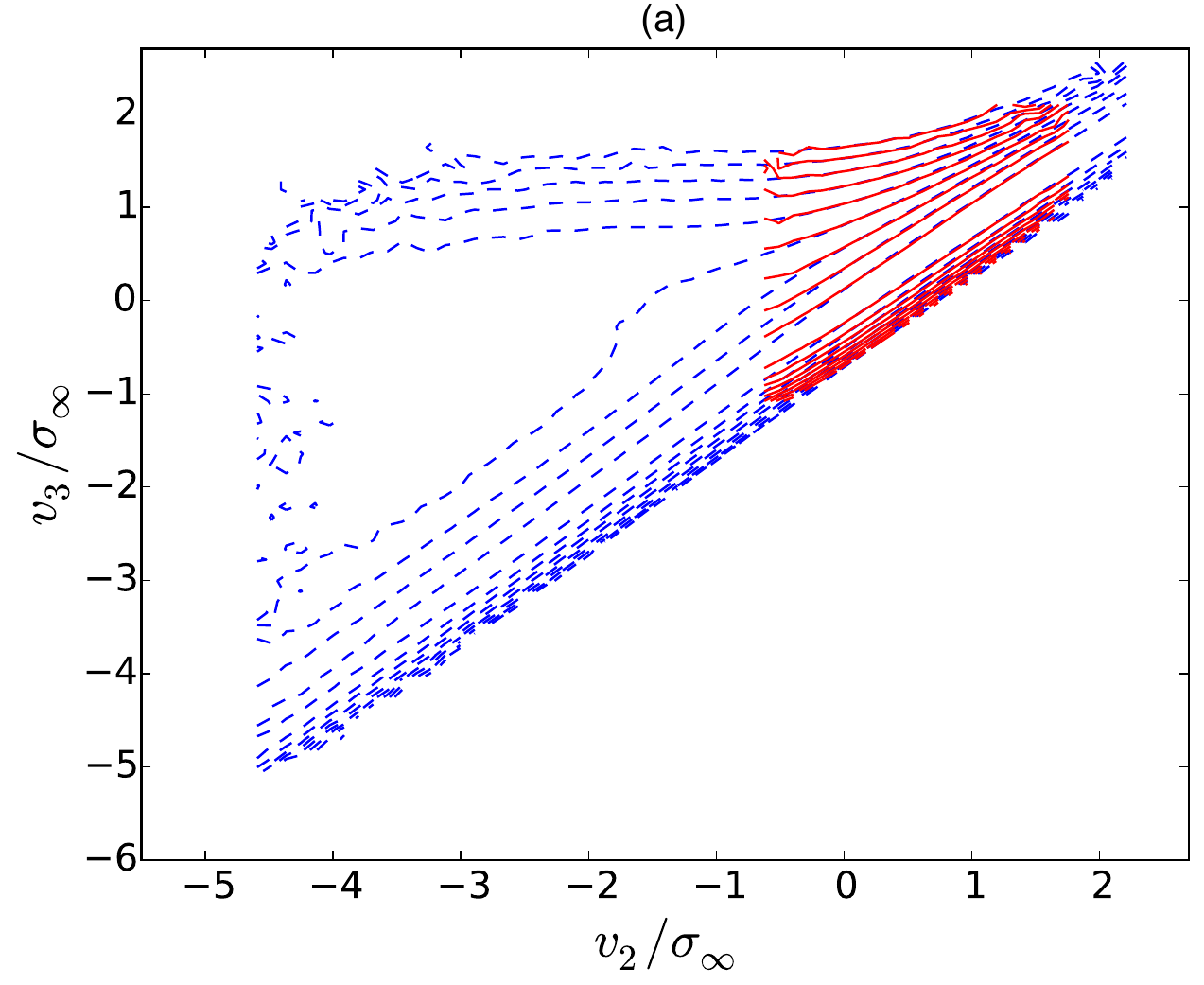}
\includegraphics[width=0.49 \textwidth]{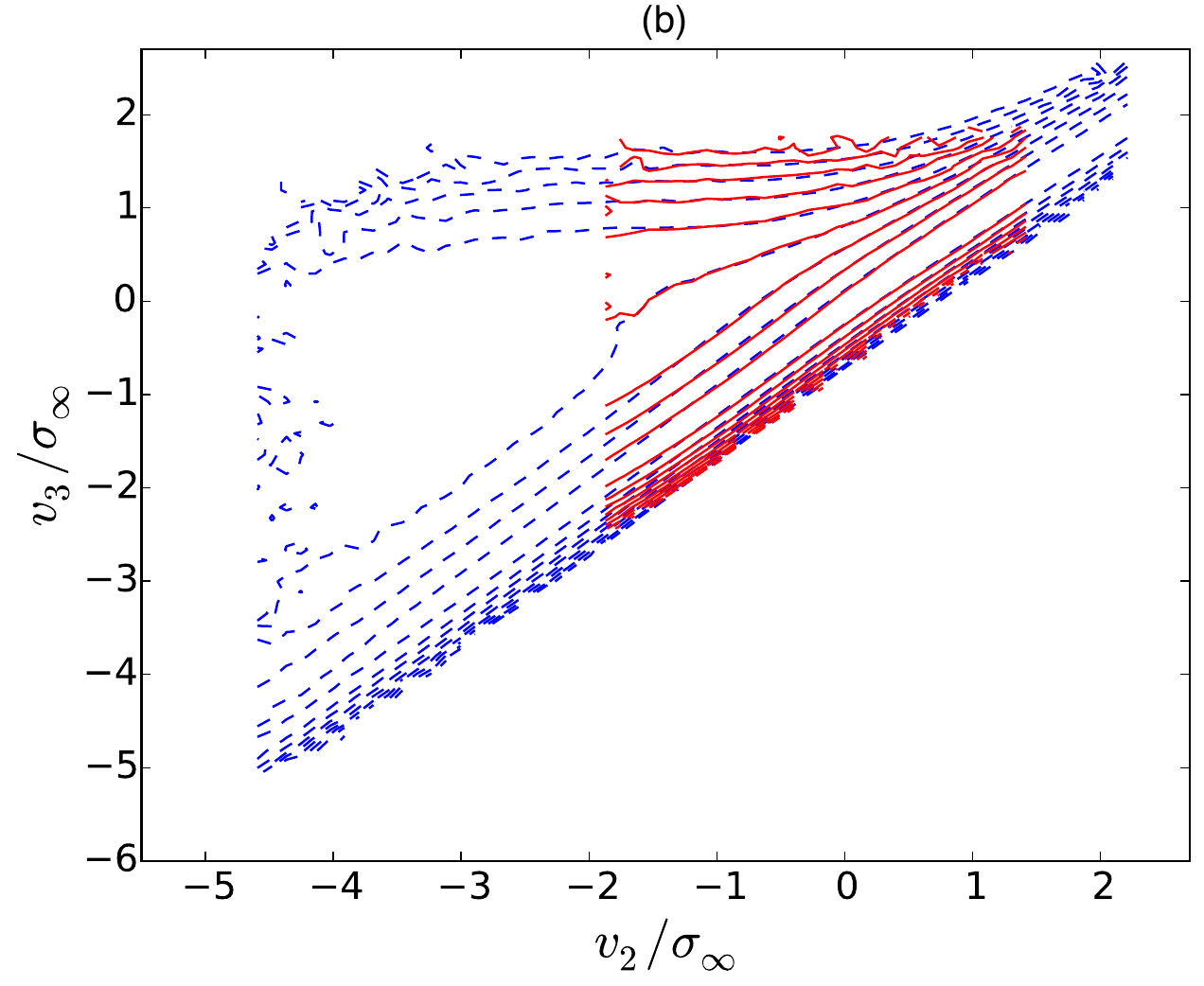}
\caption{(a) Examination of the Markov property (\ref{eq:markov_direct}) from DNS of Burgers turbulence
for $\Delta r= \lambda$ and $v_1=0$.
The Markov property seems to hold quite well for this set of parameters.
(b) Same as in (a), but for $v_1 = - \sigma_{\infty}$. A broader
part of the blue dashed lines that belong to $p(v_3,L/2-\Delta r|v_2,L/2)$
The Markov property has not deteriorated in comparison to (a).}
\label{fig:markov_burgers_1}
\end{figure}
\begin{figure}[h!]
\centering
\includegraphics[width=0.49 \textwidth]{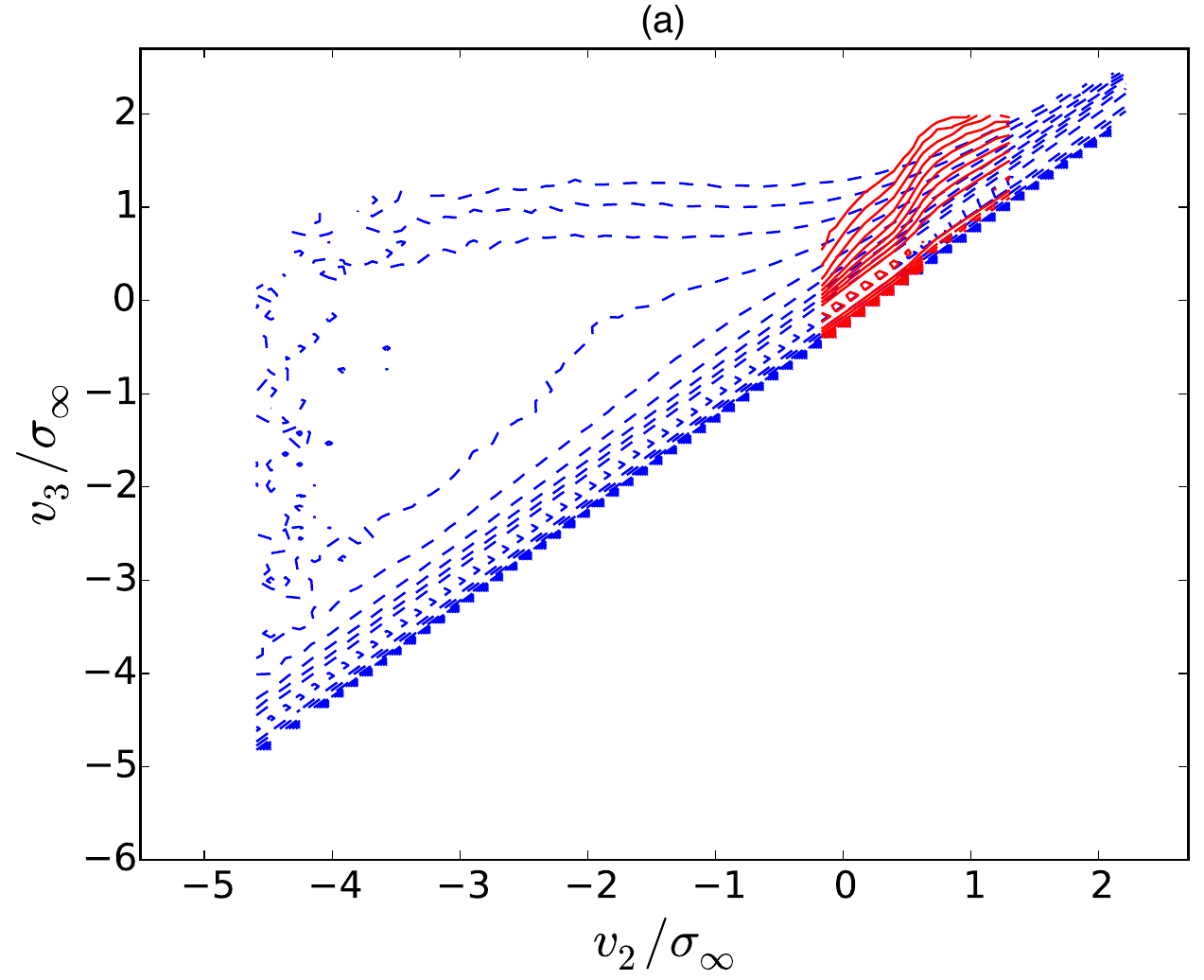}
\includegraphics[width=0.49 \textwidth]{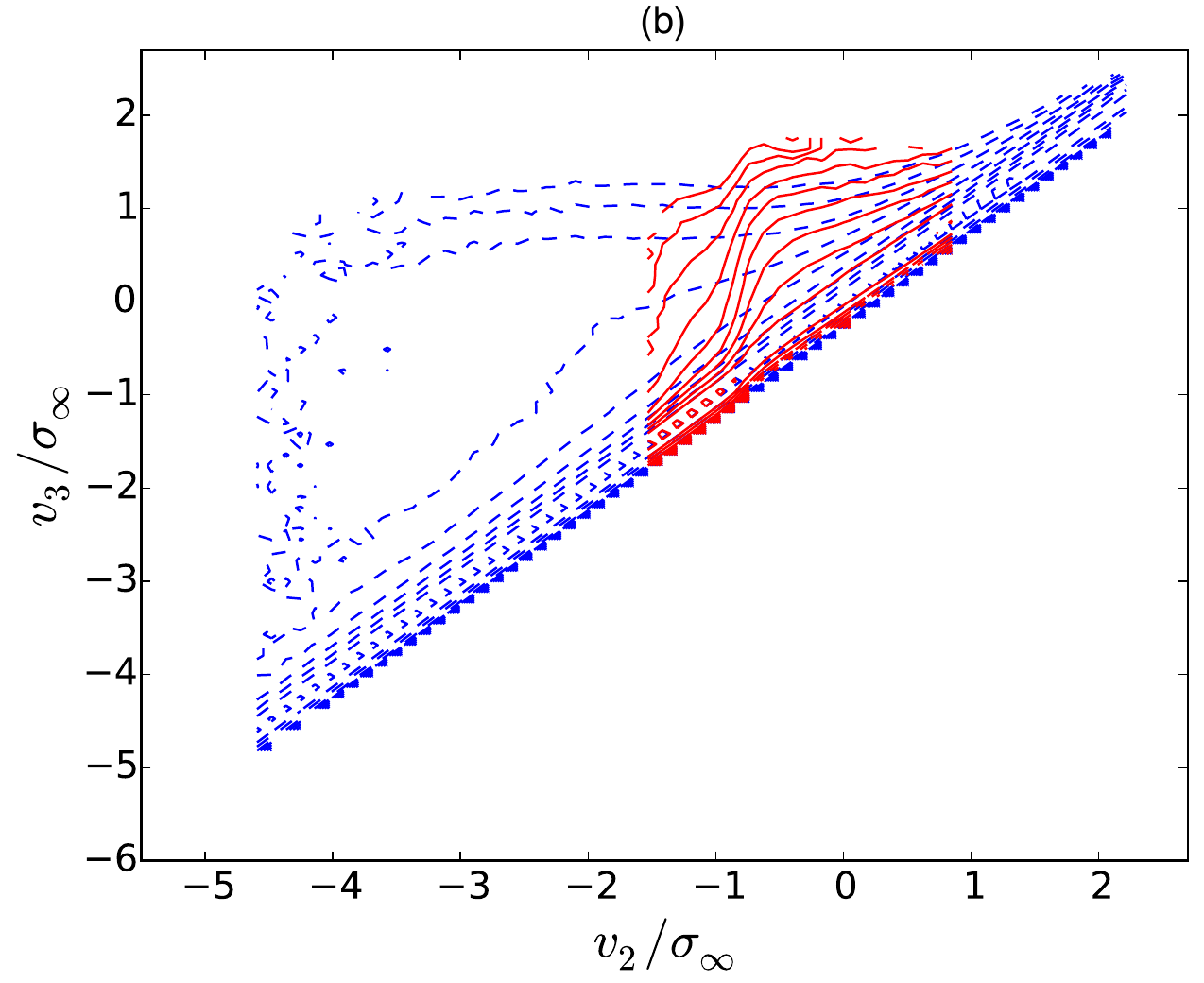}
\caption{(a) Examination of the Markov property (\ref{eq:markov_direct}) from DNS of Burgers turbulence for $\Delta r= 0.2 \lambda$ and $v_1=0$.
The Markov property is violated as the transition PDF (red) overestimates the correlations that is exhibited by the conditional PDF (blue).
(b) Same as in (a), but for $v_1 = - \sigma_{\infty}$. The violation of the Markov property becomes even more pronounced.}
\label{fig:markov_burgers_0.2}
\end{figure}
The other method is to examine the Chapman-Kolmogorov equation~\cite{risken}
\begin{equation}
  \int \textrm{d}v_2~ p(v_3,L/2-\Delta r|v_2,L/2) ~p(v_2,L/2|v_1,L/2+\Delta r )= p(v_3,L/2-\Delta r|v_1,L/2+\Delta r)\;,
  \label{eq:chap}
\end{equation}
which has the advantage that it involves solely transition PDFs and, hence, reduces itself to the comparison of two objects of equal dimensionality.

In the following, we use the first method, i.e., Equation (\ref{eq:markov_direct}).
In order to obtain a sufficient overlap between
the two PDFs, we choose $v_1=0$ and  $v_1=-1\;\sigma_{\infty}$,
since the positive increment parts of the PDFs are suppressed by shocks.
The resulting contour plots for three different $\Delta r$ are depicted in Figs.~\ref{fig:markov_burgers_2}-\ref{fig:markov_burgers_0.2}.
Here, the dashed blue lines correspond to the transition PDF $p(v_3, L/2-\Delta r|v_2, L/2)$ whereas the red lines correspond to the conditional PDF $p(v_3,L/2-\Delta r|v_2, L/2; v_1, L/2+\Delta r)$. At first glance, we can see that the shape of the transition PDF (blue) can be reproduced for the most part by the solution of (\ref{eq:km_exp_trans})
\begin{equation}
 p(v_3,r_3|v_2,r_2) = \left \{\begin{array}{ll}
		\delta\left(v_3-\frac{r_3}{r_2} v_2\right) & \quad \textrm{ for } v_2 \geq 0\;, \\
		~\\
		\frac{r_3}{r_2} \delta(v_3-v_2) +
		\left(1-\frac{r_3}{r_2}\right) \delta(v_3)  & \quad\textrm{ for } v_2 \leq 0\;.
	   \end{array} \right. \;,
\label{eq:trans}
\end{equation}
which is derived in Appendix~\ref{app:shock}.
For $v_2<0$, the additional branch $\delta(v_3)$ appears. Obviously, the contours of the transition PDFs are not pure delta functions, but are rather broad. We want to emphasize that this can be considered as an artifact of the Kramers-Moyal expansion (\ref{eq:km_exp_trans}): The solution (\ref{eq:trans}) has been constructed from the initial condition
\begin{equation}
  \lim_{r_3\rightarrow r_2} p(v_3,r_3|v_2,r_2) =\delta(v_3-v_2)\;.
\end{equation}
A second initial condition, which would be a Gaussian transition probability at large scales, cannot be imposed since the Kramers-Moyal expansion is only a first-order differential equation.
The overall structure in Figs.~\ref{fig:markov_burgers_2}-\ref{fig:markov_burgers_0.2}, however, agrees fairly well with the theoretical predictions (\ref{eq:trans}).

Concerning the Markov property itself, it seems to be best fulfilled
around the Taylor length, i.e., for $\Delta r=1\lambda$ in Figure~\ref{fig:markov_burgers_1} (a). We can also establish this finding for $v_1 = -\sigma_{\infty}$. However, at larger scale separations $\Delta r=2\lambda$ in Figure~\ref{fig:markov_burgers_2}, the Markov property slightly deteriorates. Although the shape of the transition PDF and the conditional PDF are basically the same, there is a small shift of the contour lines in the $v_3$-direction. The effect becomes even stronger for $v_1=-\sigma_{\infty}$
which can be seen from Figure~\ref{fig:markov_burgers_2} (b). In comparison to the true violation of the Markov property
in Figure~\ref{fig:markov_burgers_0.2}, however, this effect is rather small. At those smaller scale separations, here for $\Delta r=0.1 \lambda$, both PDFs possess a different shape.
Figure~\ref{fig:markov_burgers_0.2} (a) shows that the transition PDF (red) overestimates the $v_3-v_2$-correlations of the conditional PDF (blue) manifesting itself by a strong steepening of the contour lines of the transition PDF in comparison to the conditional PDF.
The effect becomes even more pronounced for $v_1 = -\sigma_{\infty}$. Nevertheless, for large negative values $v_2$, the contour lines seem to overlap again. It is therefore tempting to speculate about whether the Markov property might again be fulfilled inside the shocks, and that only the regions of extreme curvature in the vicinity of the shocks lead to the break-down of the Markov property. In this context, it is important to notice that the Taylor length $\lambda$ is located just in front of the $k^{-2}$-part of the spectrum in Figure~\ref{fig:spec_burgers}.
In the following section, we aim to quantify the break-down of the Markov property by the introduction of a distance measure between the two distributions in Equation (\ref{eq:markov_direct}).

\subsubsection{Determination of the Markov-Einstein Length}
Here, we seek to quantify the break-down of the Markov property (\ref{eq:markov_direct}). There are a variety of methods for comparing
the transition PDF and the conditional PDF~\cite{Renner2002a}. In the present study, we
will restrict ourselves to the so-called Hellinger distance~\cite{Hellinger1909},
although other methods such as the correlation distance or the Kullback-Leibler divergence have also been experimented with. The advantage of the Hellinger distance $H$ is that it forms a true metric in the space of the PDFs and can therefore be used to decide at which scale separation $\Delta r$ the Markov property significantly deteriorates.
%explicit computable
The Hellinger distance $H$ for continuous distributions is defined according to
\begin{align}\nonumber
H^2(v_2,v_1;\Delta r) =& \frac{1}{2}
 \int \textrm{d} v_3 \left( \sqrt{p(v_3,L/2-\Delta r|v_2,L/2;v_1,L/2+\Delta r)}
 -\sqrt{p(v_3,L/2-\Delta r|v_2,L/2)} \right)^2\\
 =& 1- \int \textrm{d} v_3  \sqrt{p(v_3,L/2-\Delta r|v_2,L/2;v_1,L/2+\Delta r)
 p(v_3,L/2-\Delta r|v_2,L/2)} \;.
\end{align}
Here, we made use of the identities
\begin{equation}
 \int \textrm{d} v_3 p(v_3,L/2-\Delta r|v_2,L/2;v_1,L/2 +\Delta r)
 =1 \quad \textrm{and} \quad \int \textrm{d} v_3 p(v_3,L/2-\Delta r|v_2,L/2)
 =1\;,
\end{equation}
in the last step. Hence, the Hellinger distance is symmetric in both probabilities
and is restricted to
\begin{equation}
 0 \le H(v_2,v_1;\Delta r) \le 1\;,
\end{equation}
which is a direct consequence of the Cauchy-Schwarz inequality. Another useful property of the Hellinger distance is that it can be explicitly calculated for certain types of PDFs (normal distribution, beta distribution, exponential distribution etc.).

\begin{figure}
 \centering
\includegraphics[width=0.49 \textwidth]{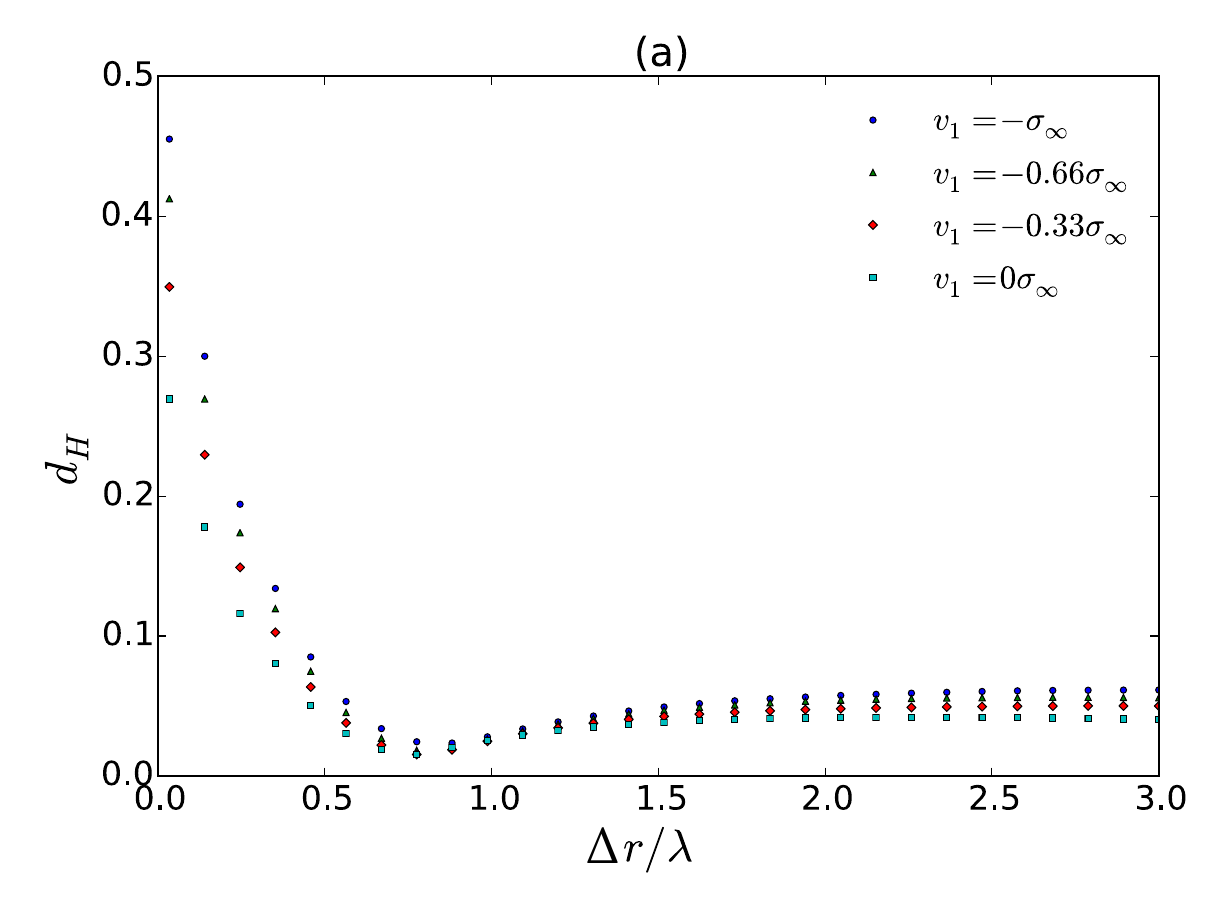}
\includegraphics[width=0.49 \textwidth]{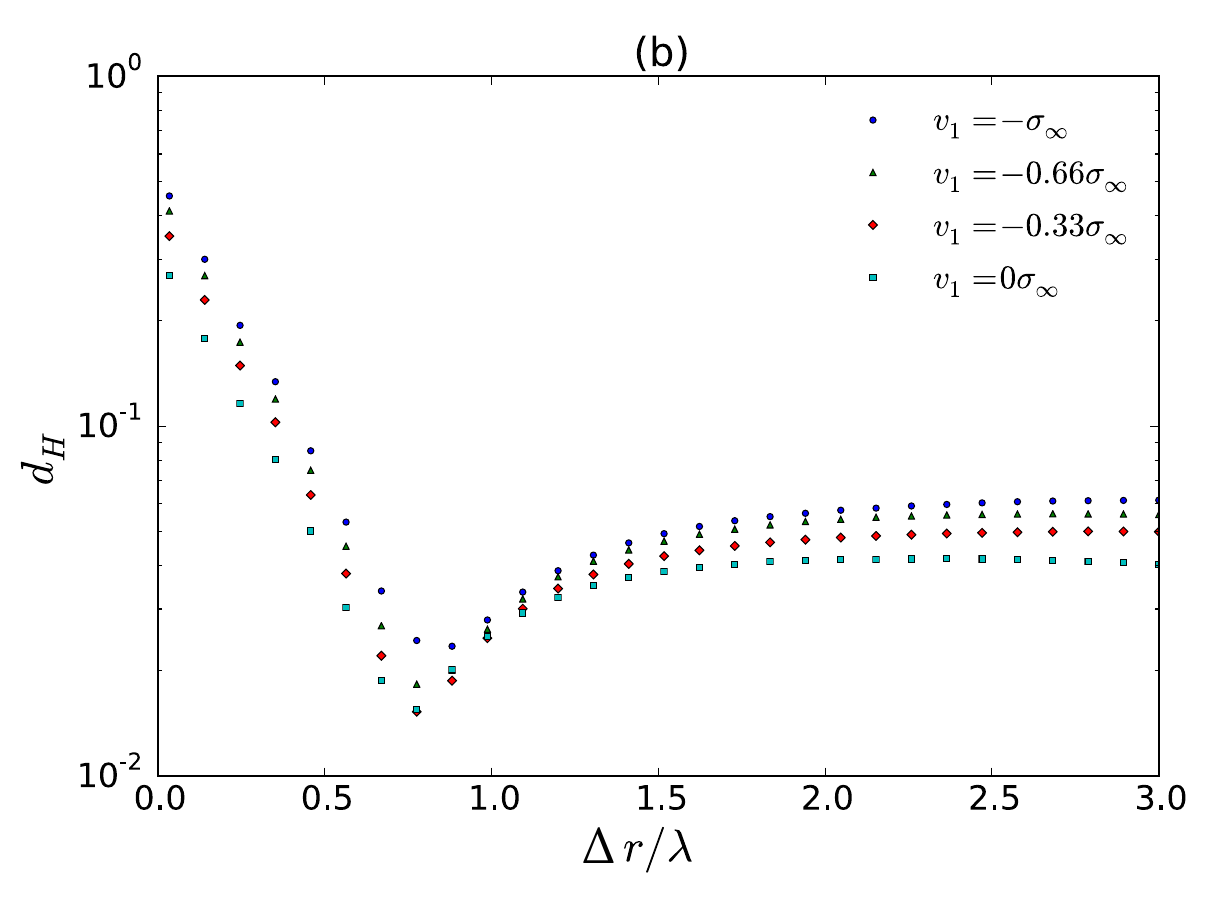}
\caption{(a) Hellinger distance $d_H(\Delta r)$ for different $v_1$ and variable step width $\Delta r$. Apparently, the Markov property only is a good approximation around $\Delta r=\lambda$. For larger $\Delta r$, the Hellinger distance slightly
increases and approaches a small constant value.
However, for smaller $\Delta r$, the Hellinger distance exhibits a clearer increase.
These tendencies become even more pronounced for $v_1=-0.33 \sigma_{\infty}, ~-0.66 \sigma_{\infty}, ,~-\sigma_{\infty}$.
(b) Semi-logarithmic plot of the Hellinger distance $d_H(\Delta r)$. For $\Delta r < 0.8 \lambda$, the Hellinger distance
seems to increase nearly exponentially.}
\label{fig:d_H_burgers}
\end{figure}

In our case, the Hellinger distance still is a function of $v_2$ if we assume that $v_1$ is fixed. Therefore, an average of the corresponding $v_2$-values
is performed in order to obtain a pure correlation measure
\begin{equation}
d_H(\Delta r,v_1)= \langle H(v_2,v_1;\Delta r)\rangle_{v_2} \;.
\label{eq:hell}
\end{equation}
%
%\begin{equation}
%  d_c(v_2,r_3-r_2) = \frac{\int \textrm{d} v_3 p(v_3, r_3,| v_2, r_2; v_1,r_1)
%  p(v_3,r_3|v_2,r_2)}{\int \textrm{d} v_3 p(v_3, r_3 | v_2,r_2)^2
%  \int \textrm{d} v_3 p(v_3, r_3 | v_2,r_2;v_1,r_1)^2} \quad  \textrm{where} \quad r_3 \ll r_2 \ll r_1.
%\end{equation}
The corresponding $\Delta r$ dependency in Figure~\ref{fig:d_H_burgers} (a) is expressed in terms of the Taylor length $\lambda$. Remarkably, the Hellinger distance $d_H(\Delta r)$ is smallest at around $\Delta r=\lambda$ and approaches a small constant that is different from zero for larger $\Delta r$. This corresponds to the observation from Figure~\ref{fig:markov_burgers_2} where the contour plots were slightly shifted in the $v_3$-direction for larger $\Delta r$.
In other words, the Markov property gets slightly better before it gets worse in the process of letting $\Delta r \rightarrow 0$. The latter observation differs from usual investigations of the Markov property in turbulence~\cite{Renner2002a}.
For smaller $\Delta r < 0.8 \lambda$, however, the Hellinger distance shows a more pronounced increase and therefore, the Markov property is clearly violated since it seems to approach $1$ in the limit $r\rightarrow 0$.
These effects become even more unambiguous for $v_1=-0.33 \sigma_{\infty}, ~-0.66 \sigma_{\infty},~-\sigma_{\infty}$. Figure~\ref{fig:d_H_burgers} (b) shows a semi-logarithmic plot of the Hellinger distance. For $\Delta r < 0.8 \lambda$, a quantitatively new behavior emerges. The latter is characterized by a close to
exponential increase of the Hellinger distance. Such behavior
implies an exponential decay of the correlations of the Markov property
Equation (\ref{eq:markov_direct}). Here, it suffices to estimate the Einstein-Markov length at around the point, where a pronounced increase sets in, i.e., $\lambda_{ME}=0.8 \lambda$.
Hence, the task of an accurate determination of the Markov-Einstein length is far from obvious. However, it can be inferred from Figure~\ref{fig:d_H_burgers} that it lies near the Taylor length as the correlation measure clearly drops at $\Delta r/\lambda \approx 1$.
\subsubsection{Determination of the Kramers-Moyal Coefficients}
In this section, we will outline a procedure which determines the Kramers-Moyal coefficients (\ref{eq:km_coeff}) from the numerically evaluated transition probabilities similar to Figure~\ref{fig:markov_burgers_1}.
\begin{figure}[h!]
  \includegraphics[width=0.333 \textwidth]{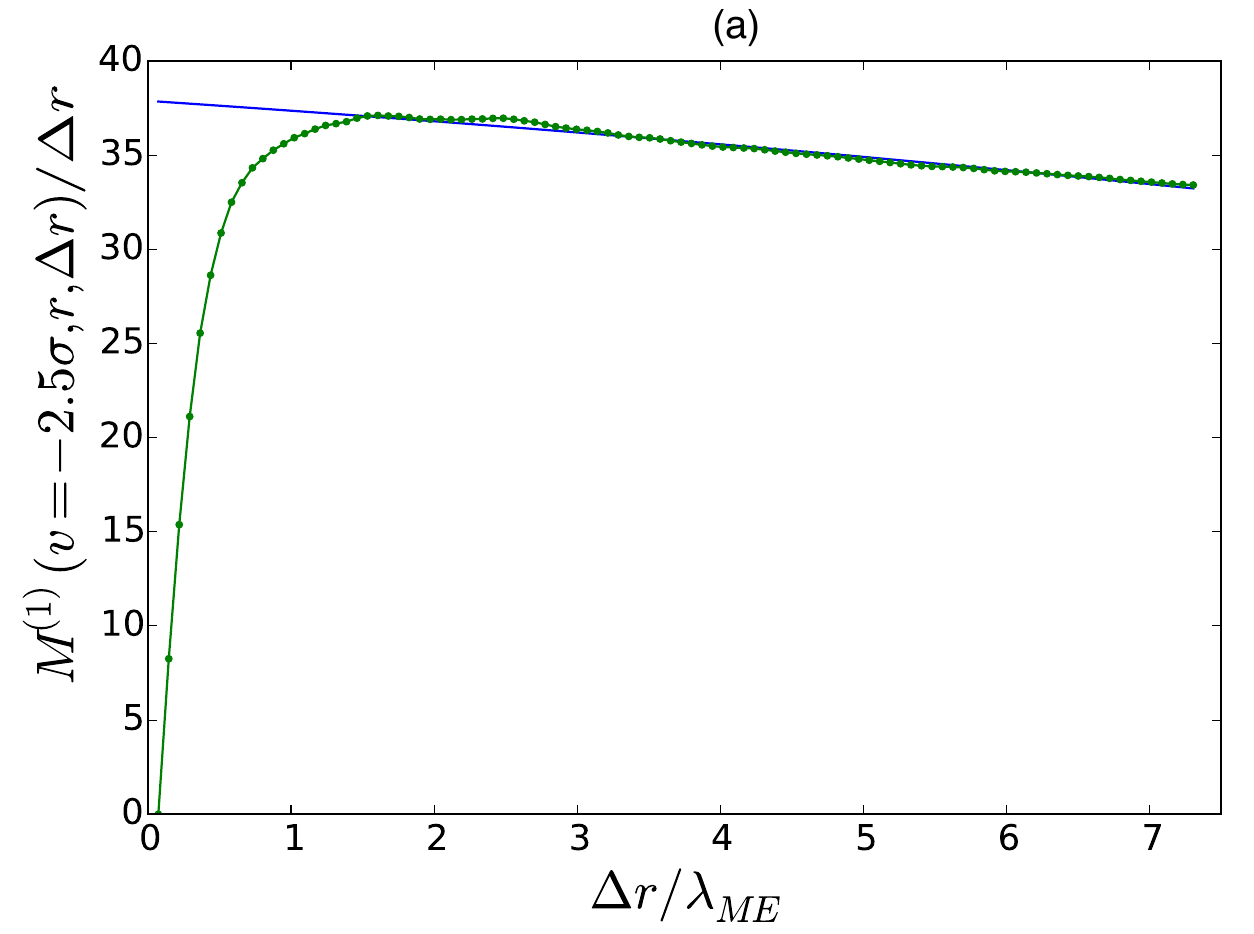}
   \includegraphics[width=0.333 \textwidth]{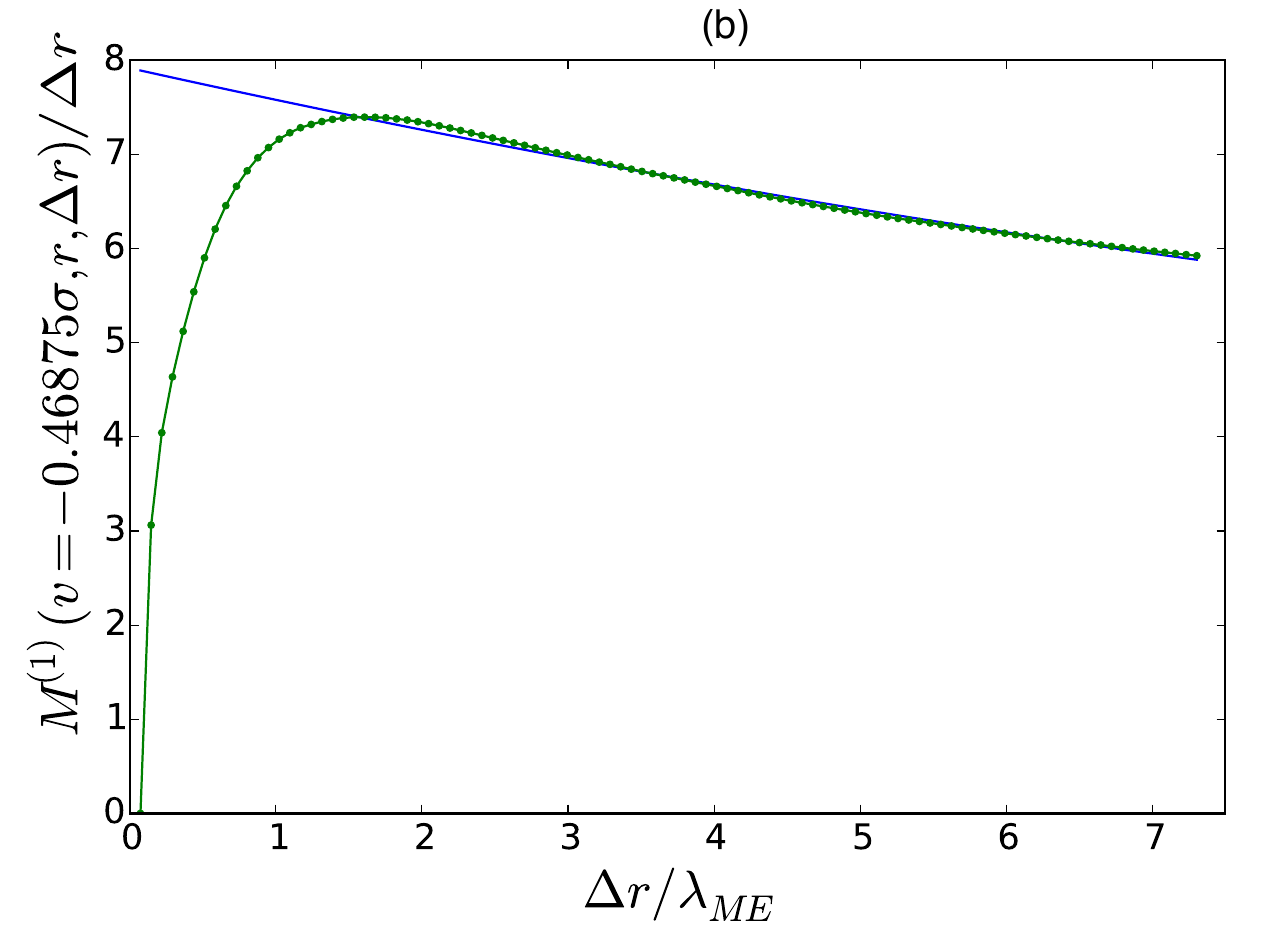}
    \includegraphics[width=0.333 \textwidth]{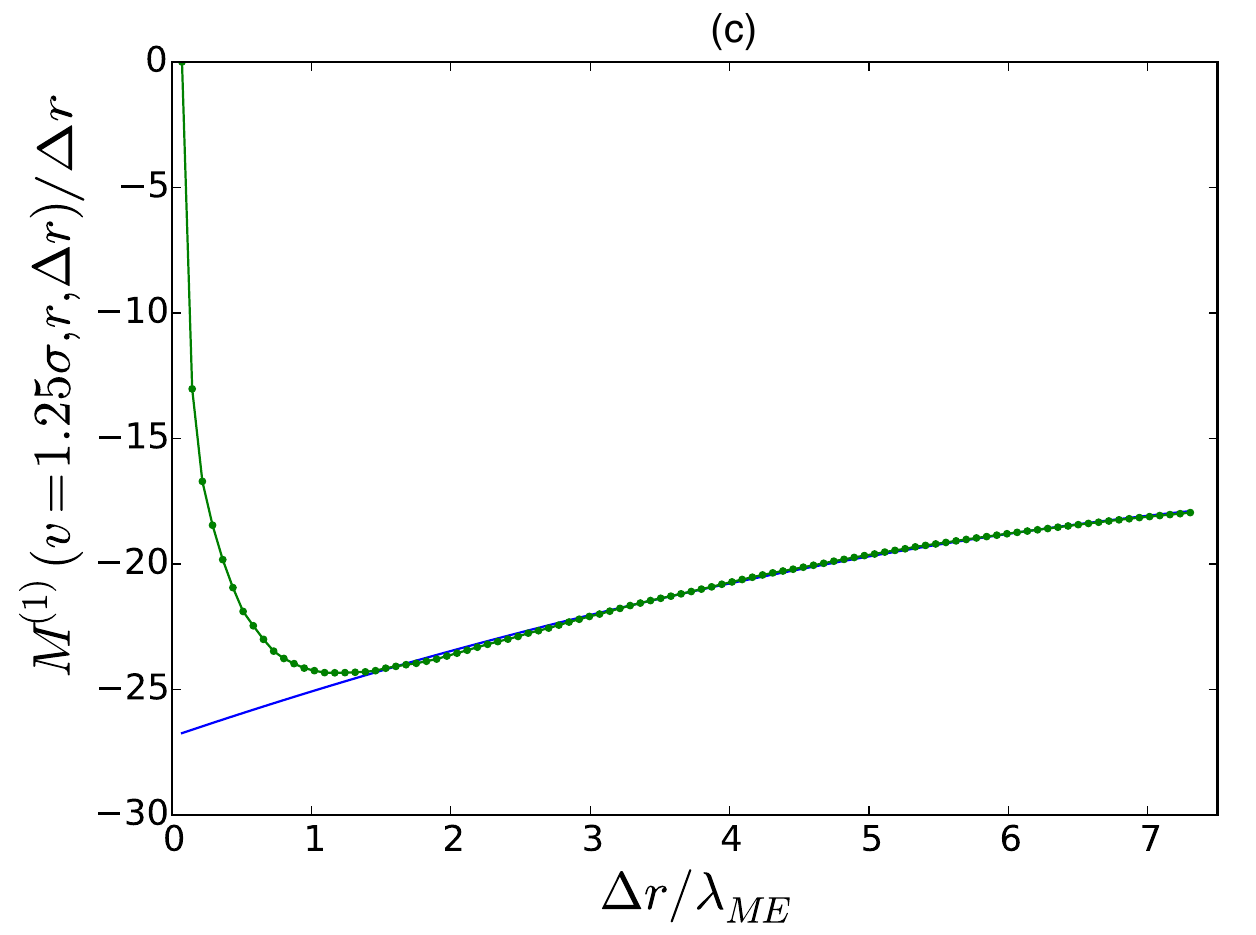}
    \caption{Conditional moments of first order divided by scale separation $\Delta r$, $M^{(1)}(v,r;\Delta r)/\Delta r$ for
    $r=L/2$, (a) $v_1=-2.5 \sigma_{\infty}$, (b) $v_1=-0.46875 \sigma_{\infty}$,
    (c) $v_1=1.25 \sigma_{\infty}$
    and variable $\Delta r$. The fits correspond to
    polynomials of second order in $\Delta r$ for $\Delta r> \lambda_{ME}$. Note that the $\Delta r$-axis has been rescaled
    by the Markov-Einstein-length $\lambda_{ME}$ and the conditional moment drops to zero for $\Delta r < \lambda_{ME}$.}
\label{fig:cond_moment}
\end{figure}
\begin{figure}[h!]
\centering
\includegraphics[width=0.5 \textwidth]{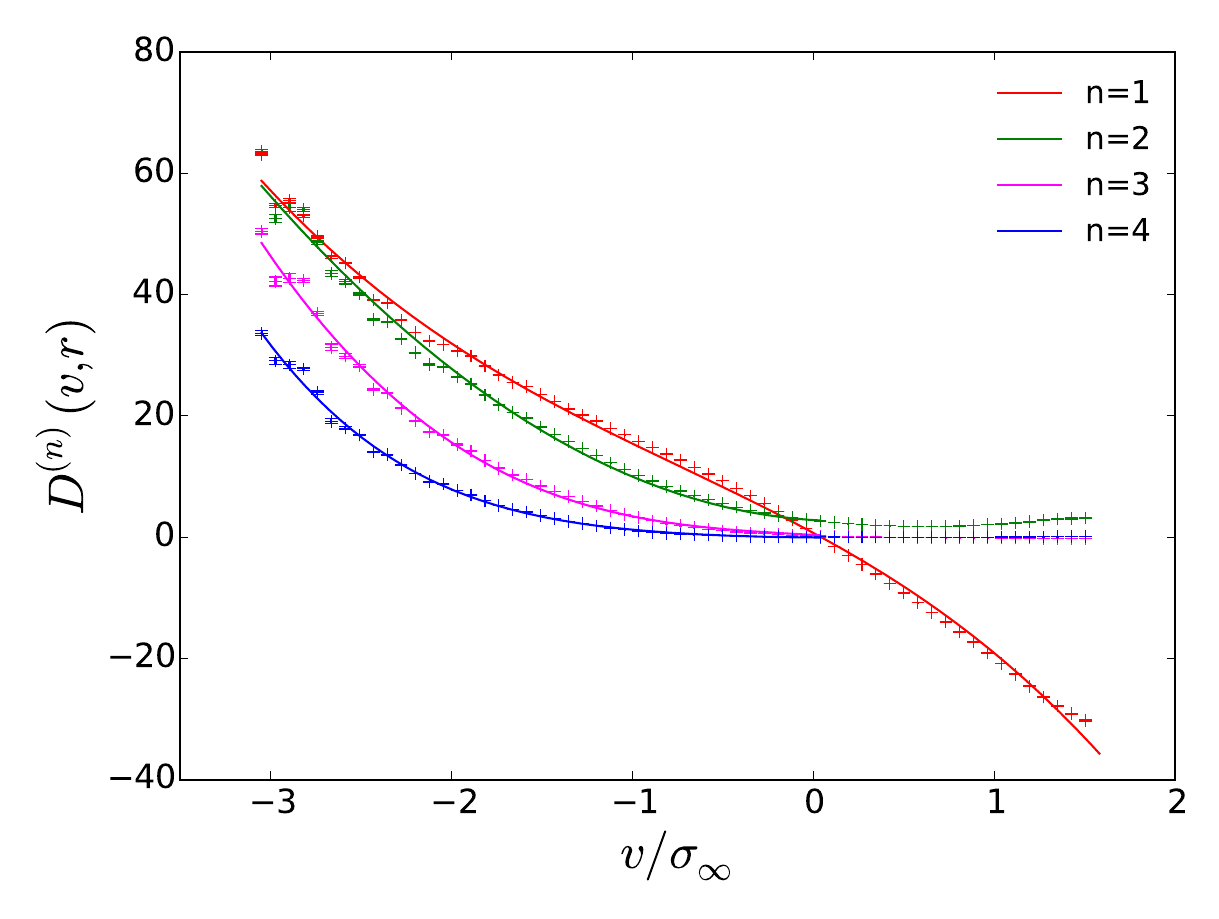}
\caption{Estimation of the Kramers-Moyal coefficients from DNS
of Burgers turbulence for $r=L/2$. The fits correspond to
polynomials of the order $n$ of the coefficient except for $n=1$ where a polynomial of order three has been used.
The reduced Kramers-Moyal coefficients have been determined according to
$K_1=1.1689 \pm 0.08$, $K_2 =0.7880 \pm 0.14$, $K_3=0.6956 \pm 0.17$ and $K_4=0.7137 \pm 0.12$}
\label{fig:km_burgers}
\end{figure}
To this end, we devise an appropriate extrapolation method for the conditional moments
\begin{equation}
 M^{(n)}(v,r;\Delta r) = \int \textrm{d} v' (v'-v)^n p(v',r-\Delta r|v, r)\;,
\end{equation}
as the Markov property becomes violated in the proximity of the Einstein-Markov length. Subsequently, the limit
\begin{equation}
 \lim_{\Delta r \rightarrow 0} \frac{M^{(n)}(v,r;\Delta r)}{\Delta r}= D^{(n)}(v, r)\;,
\end{equation}
has to be determined from the extrapolation of the conditional moments in
order to obtain the corresponding Kramers-Moyal coefficient~\cite{Renner2002a}.
As an example, we plotted the conditional moments of first order
$\frac{M^{(1)}(v,r;\Delta r)}{\Delta r}$ for three different $v_1$ in Figure~\ref{fig:cond_moment}.
$M^{(1)}(v,r;\Delta r)/\Delta r$ drops against zero for $\Delta r < \lambda_{ME}$ as the Markov property is violated. In order to extrapolate the moments, polynomial
fits of second order in $\Delta r$ were performed for $\Delta r> \lambda_{ME}$ (blue lines in the plots). The Kramers-Moyal coefficient for a particular $v$, in this case the drift coefficient, can be read off from the y-intercepts of the fits. Consequently, this procedure has to be repeated for several $v$ (and in general also different $r$) for the sake of obtaining the full functional form of the Kramers-Moyal coefficients.

The method was used for the Kramers-Moyal coefficients up to order four in Figure~\ref{fig:km_burgers}. The corresponding coefficients were fitted with
polynomials of order $n$.
The obtained reduced Kramers-Moyal coefficients $K_n$ correspond well with the theoretical predictions from Equation (\ref{eq:km_shock})
even at such small Reynolds numbers. Kramers-Moyal coefficients of higher order are detectable for negative velocity increments and Pawula's theorem~\cite{Pawula1967} is violated. The drift coefficient
$D^{(n)}(v,r)$ possesses an additional cubic $v$-dependence that has already been reported in the experiment~\cite{Renner2002a}.
In the Burgers case, only the drift coefficient is different from zero for positive increments, which underlines the self-similarity of the right tail of the PDF in Figure~\ref{fig:scale_burgers}.
The diffusion coefficient, however, possesses an additional positive intercept which turns out to be a consequence of the non-conservative forcing procedure. In fact, it can be shown that a conservative force $F(x,t)$ in Equation (\ref{eq:gen_burgers}) leads to the vanishing of the intercept.
\subsection{No intermittency $\alpha=0$: Purely Nonlocal Case}
\begin{figure}[h!]
 \centering
\includegraphics[width=0.49 \textwidth]{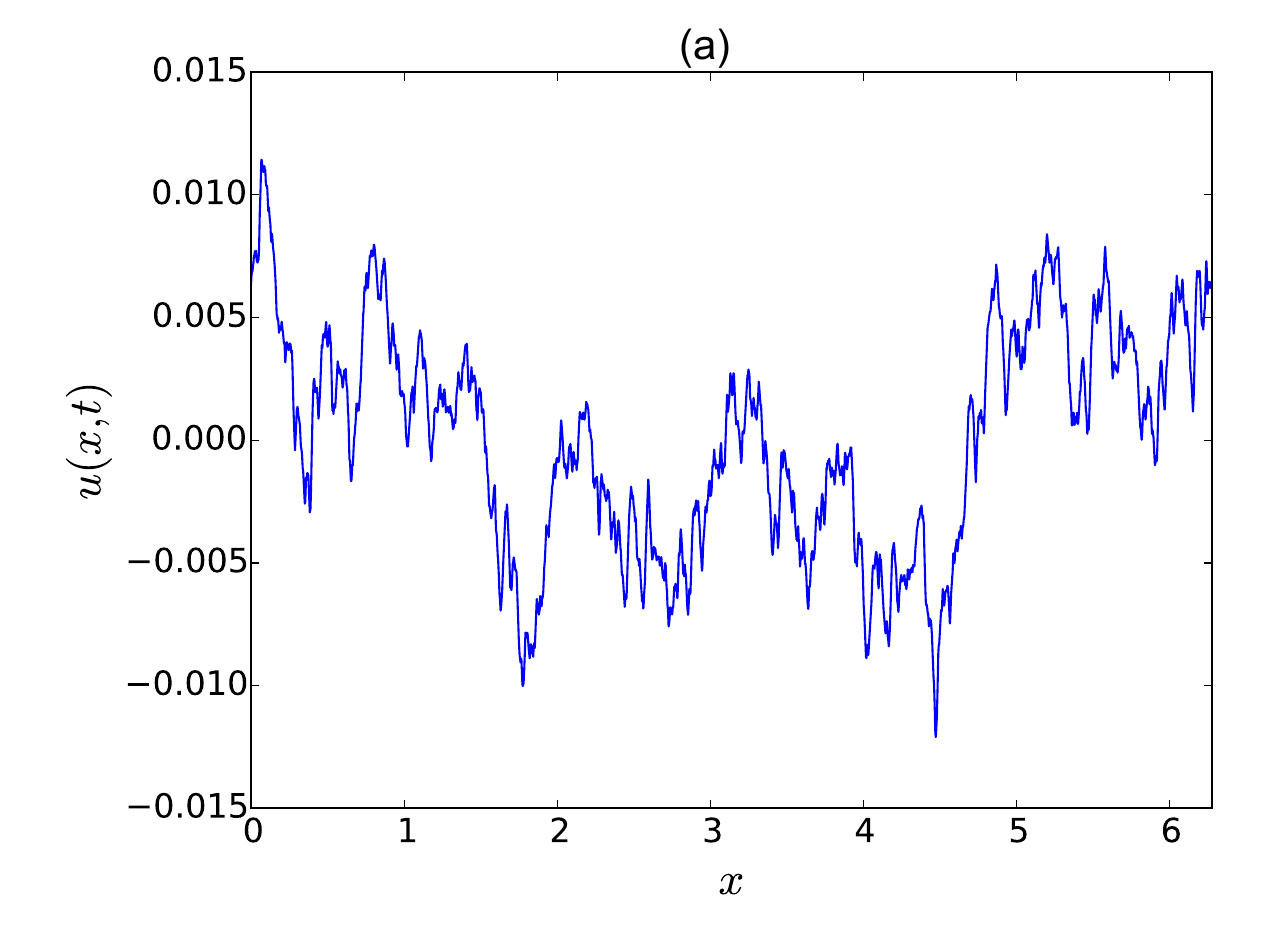}
\includegraphics[width=0.49 \textwidth]{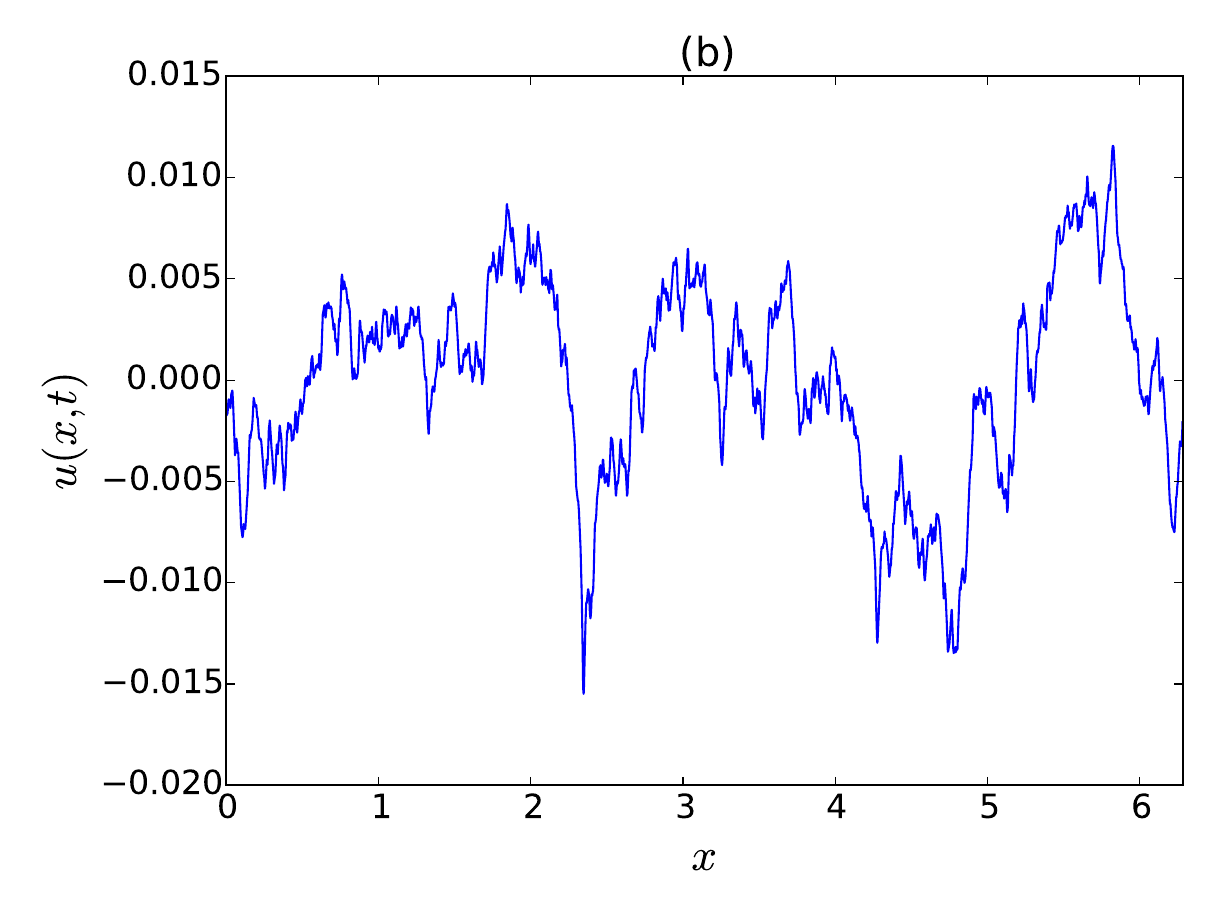}
\caption{(a) Typical realization of the velocity field $u(x,t)$ in
DNS of the purely nonlocal case ($\alpha=0$ in Equation (\ref{eq:gen_burgers})).
The velocity field is not composed of shocks as in Burgers turbulence,
but rather shows cusp-like structures.
(b) Velocity field realization that belongs to the largest velocity
field gradient that was attained in the DNS belonging to run \#2 in Tab.~\ref{tab:1}.}
\label{fig:nonlocal}
\end{figure}

In the following, we will discuss the purely nonlocal case ($\alpha=0$) of the generalized Burgers equation (\ref{eq:gen_burgers}). Figure~\ref{fig:nonlocal} shows a typical velocity field realization of the DNS of run \#2 in Tab.~\ref{tab:1}.
In contrast to the case of Burgers turbulence, no clear shock fronts can be detected and the velocity field is organized in cusp-like structures.
\begin{figure}[h!]
 \centering
\includegraphics[width=0.48 \textwidth]{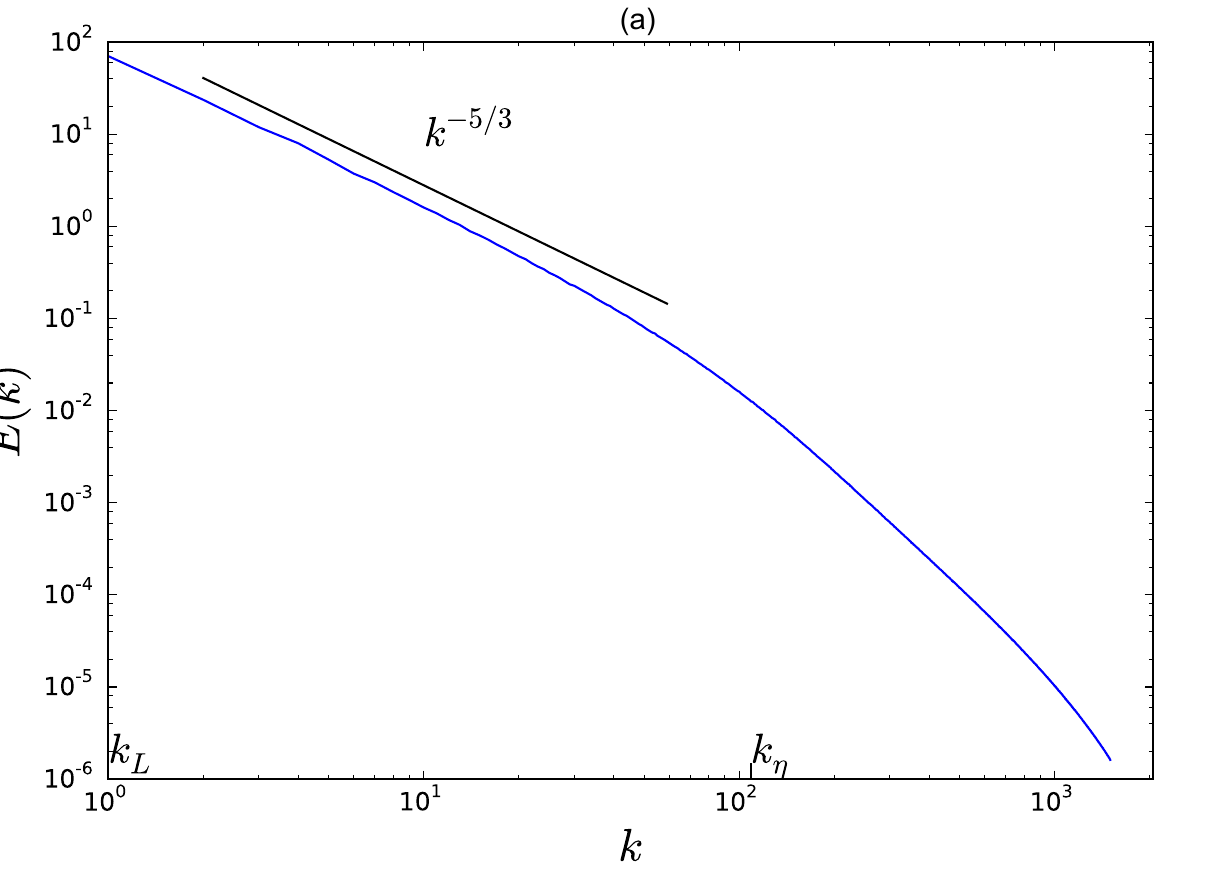}
\includegraphics[width=0.48 \textwidth]{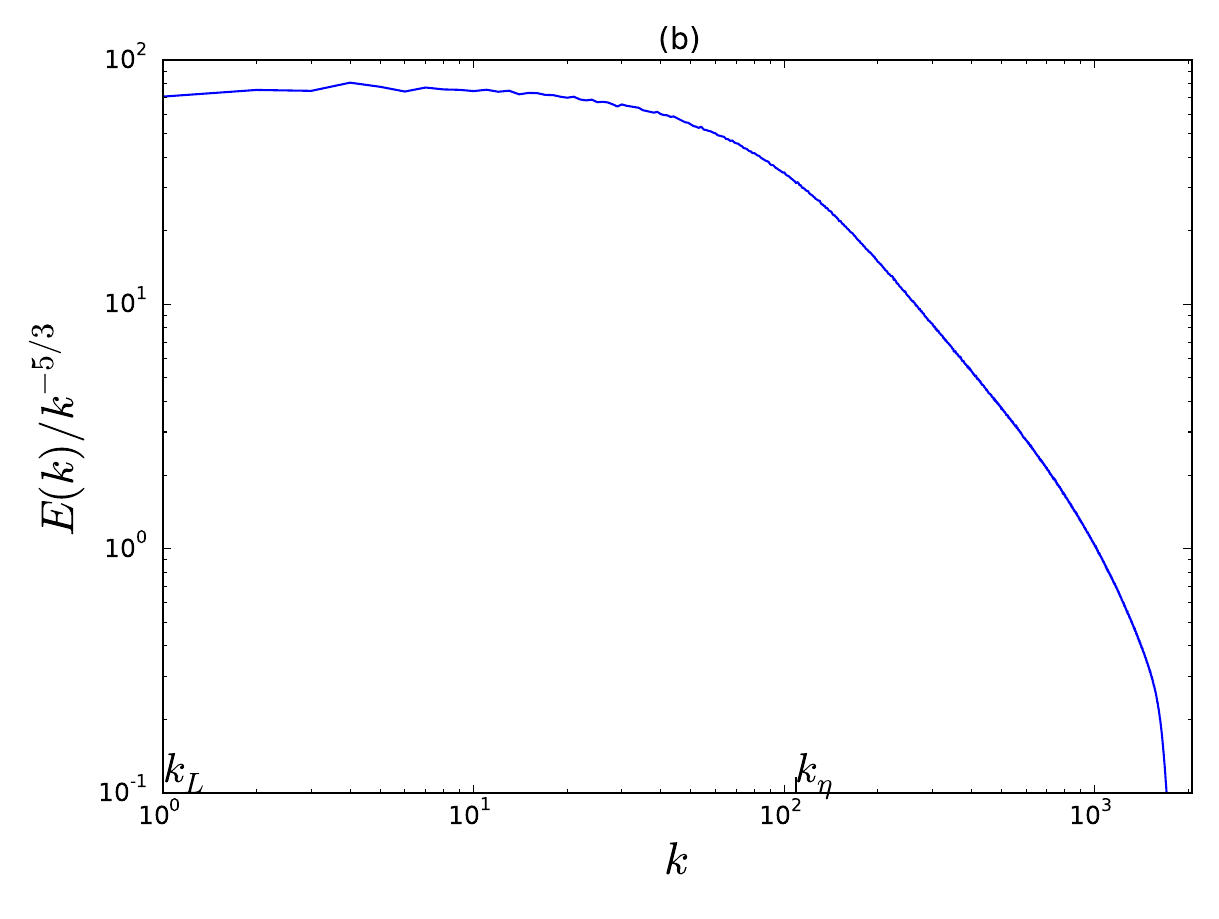}
\caption{(a) Energy spectrum $E(k,t)$ of the velocity field of
DNS of the nonlocal case. The inertial range obeys a Kolmogorov-like behavior $\sim k^{-5/3}$.
(b) Compensated energy spectrum $E(k)/k^{-5/3}$. }
\label{fig:spec_nonlocal}
\end{figure}
Apparently, the nonlocality in the generalized Burgers equation (\ref{eq:gen_burgers})
leads to entirely different singular structures. The latter manifest themselves also in the spectrum $E(k)$ in Figure~\ref{fig:spec_nonlocal} which exhibits a close to Kolmogorov-type inertial range. Moreover, the PDFs $f_1(v,r)$ at different scales $r$ in Figure~\ref{fig:km_nonlocal} (a) are purely self-similar functions that are close to Gaussian. Hence, the purely nonlocal case is characterized by the absence of intermittency. The self-similarity of the one-increment PDF is in agreement with the heuristic arguments established
in Section~\ref{sec:markov} \emph{i.}.

\subsubsection{Examination of the Markov Property}
\begin{figure}[h!]
 \centering
\includegraphics[width=0.49 \textwidth]{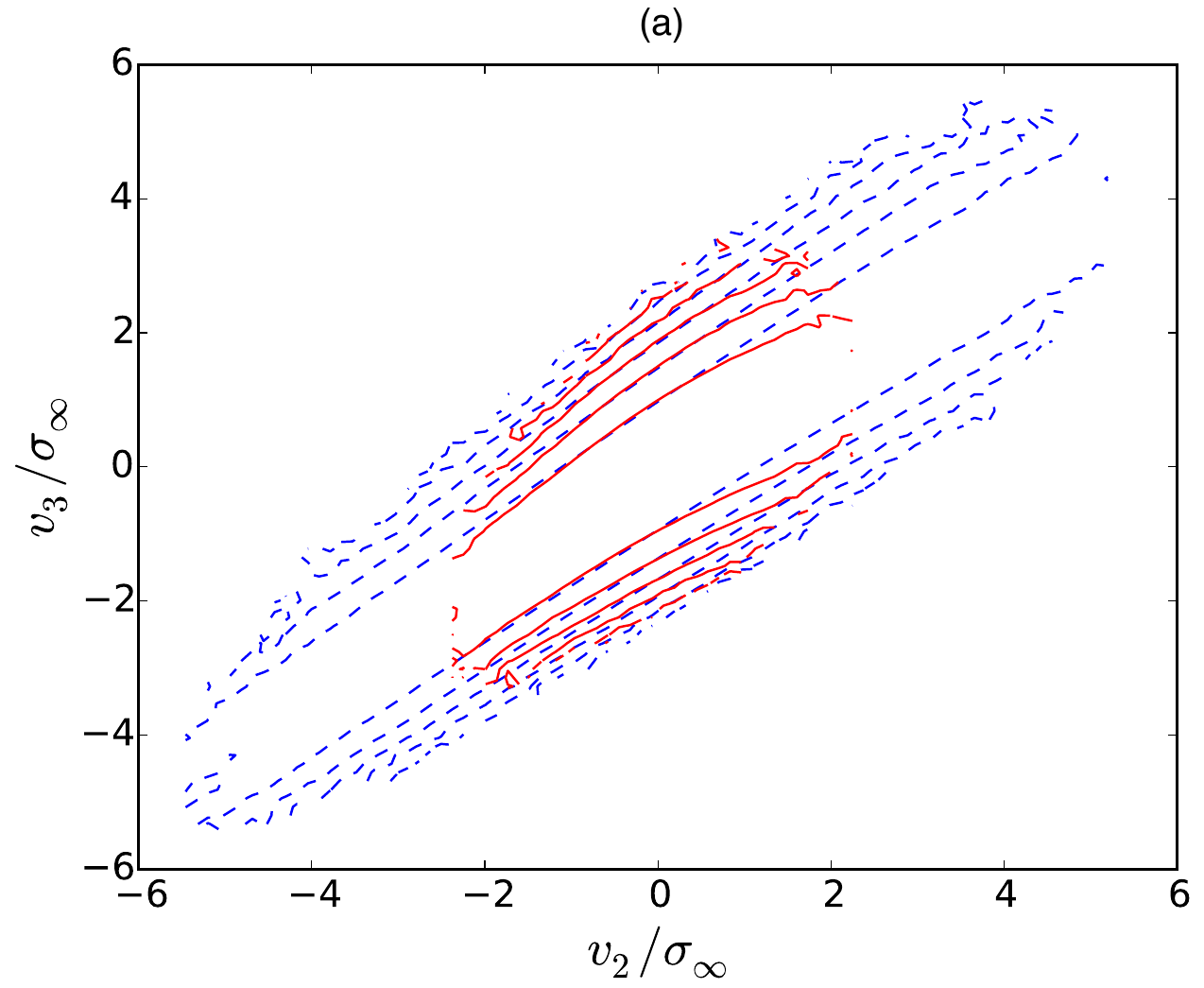}
\includegraphics[width=0.49 \textwidth]{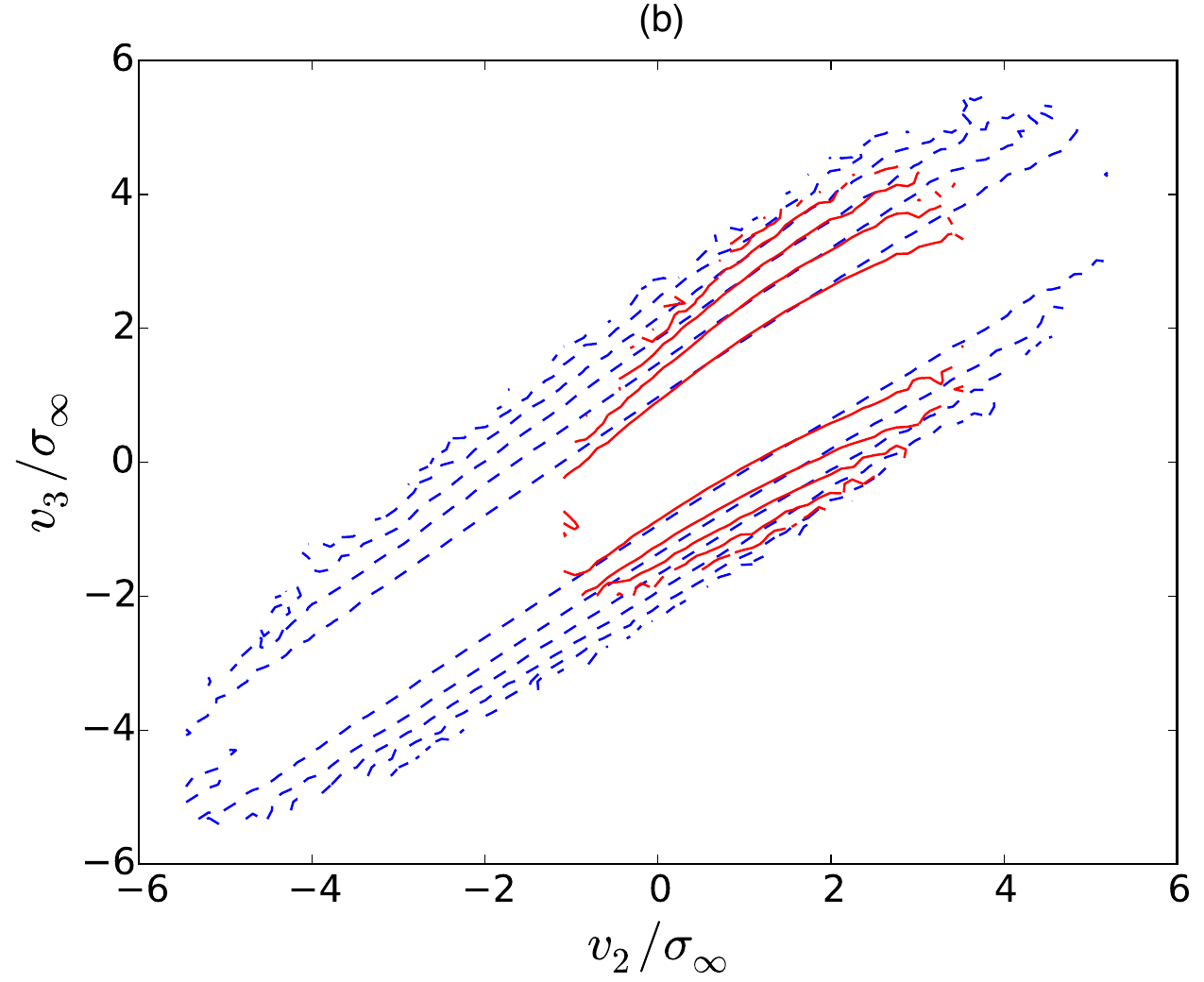}
\caption{(a) Examination of the Markov property (\ref{eq:markov_direct}) from DNS of the nonlocal case
for $\Delta r= 2.2\lambda$ and $v_1=0$ via a logarithmic contour plot.
(b) Same as in (a), but for $v_1 = \sigma_{\infty}$. The shape of the conditional PDF (red) does not change significantly in comparison to (a).}
\label{fig:markov_non_2.2}
\end{figure}

\begin{figure}[h!]
 \centering
\includegraphics[width=0.49 \textwidth]{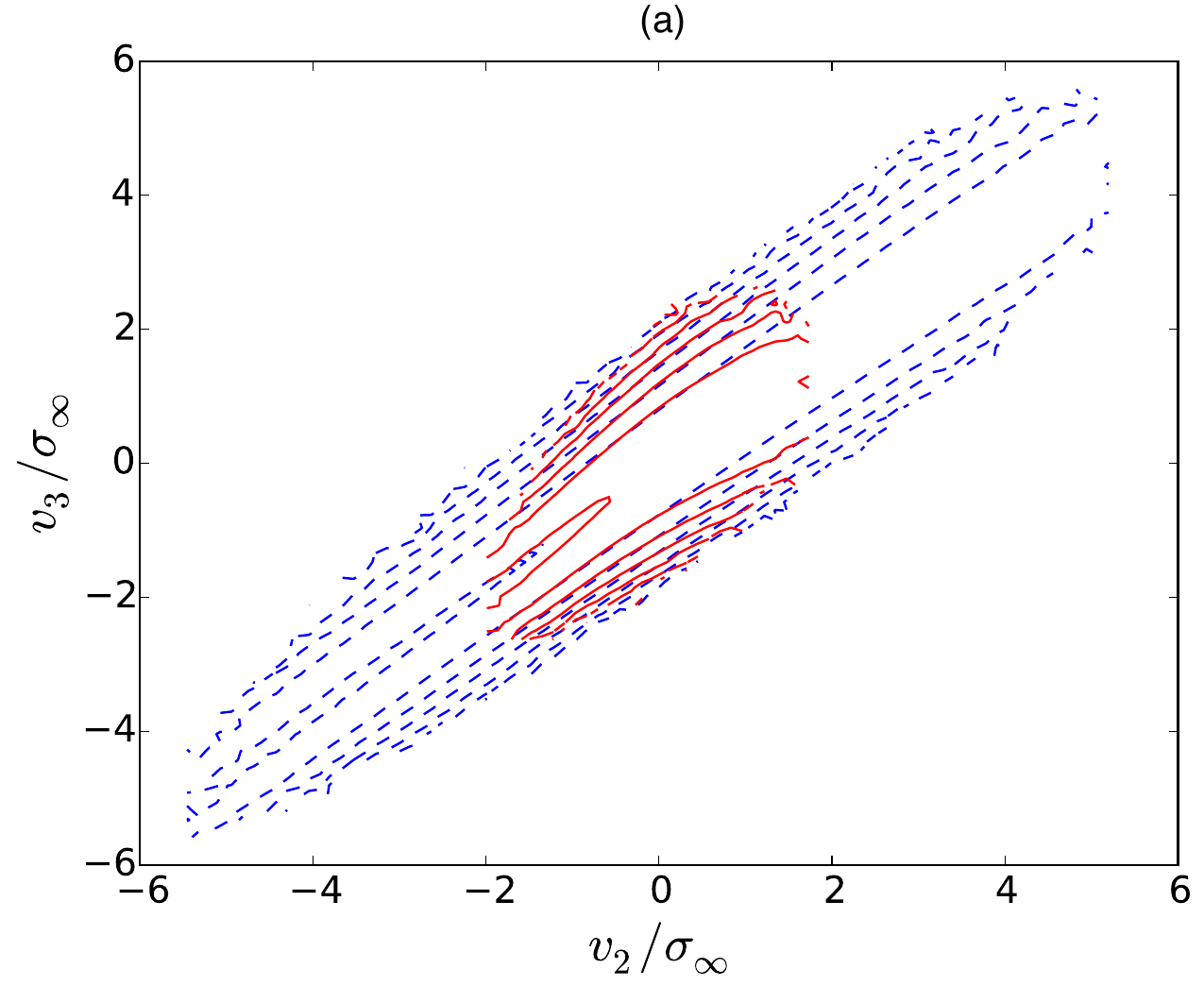}
\includegraphics[width=0.49 \textwidth]{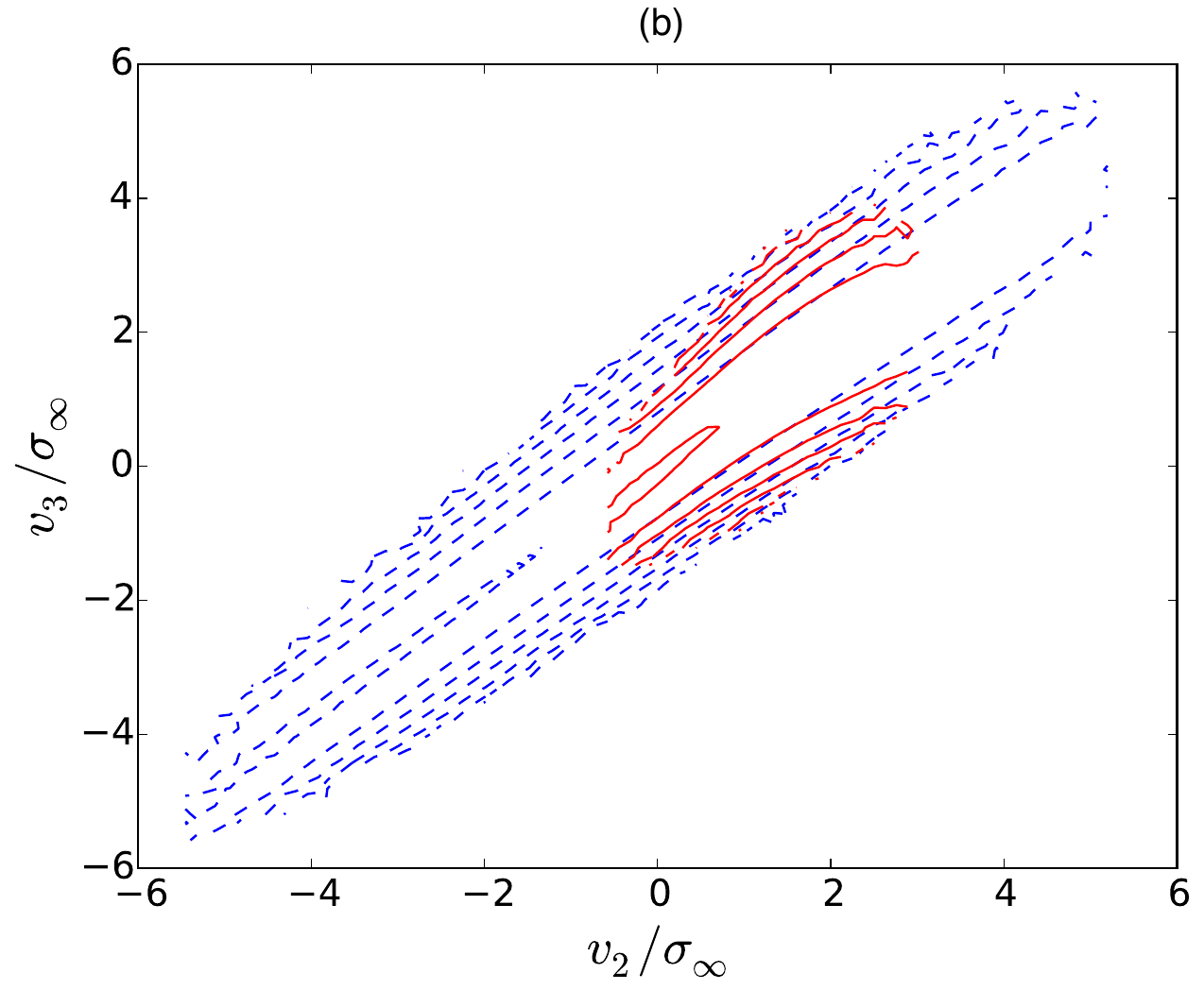}
\caption{(a) Examination of the Markov property (\ref{eq:markov_direct}) from DNS of the nonlocal case
for $\Delta r= 1\lambda$ and $v_1=0$ via a logarithmic contour plot.
(b) Same as in (a), but for $v_1 = \sigma_{\infty}$.}
\label{fig:markov_non_1}
\end{figure}

\begin{figure}[h!]
\includegraphics[width=0.49 \textwidth]{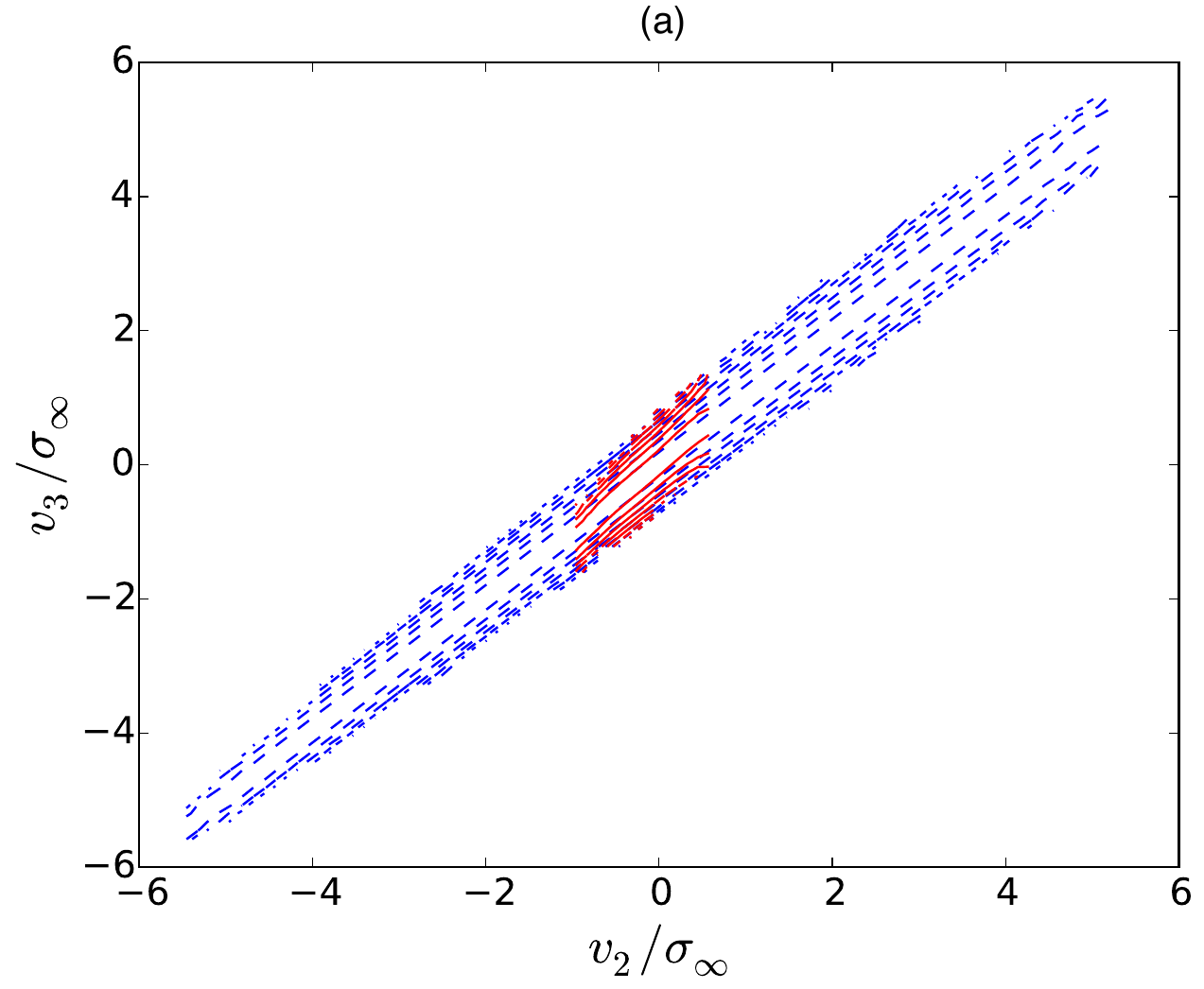}
\includegraphics[width=0.49 \textwidth]{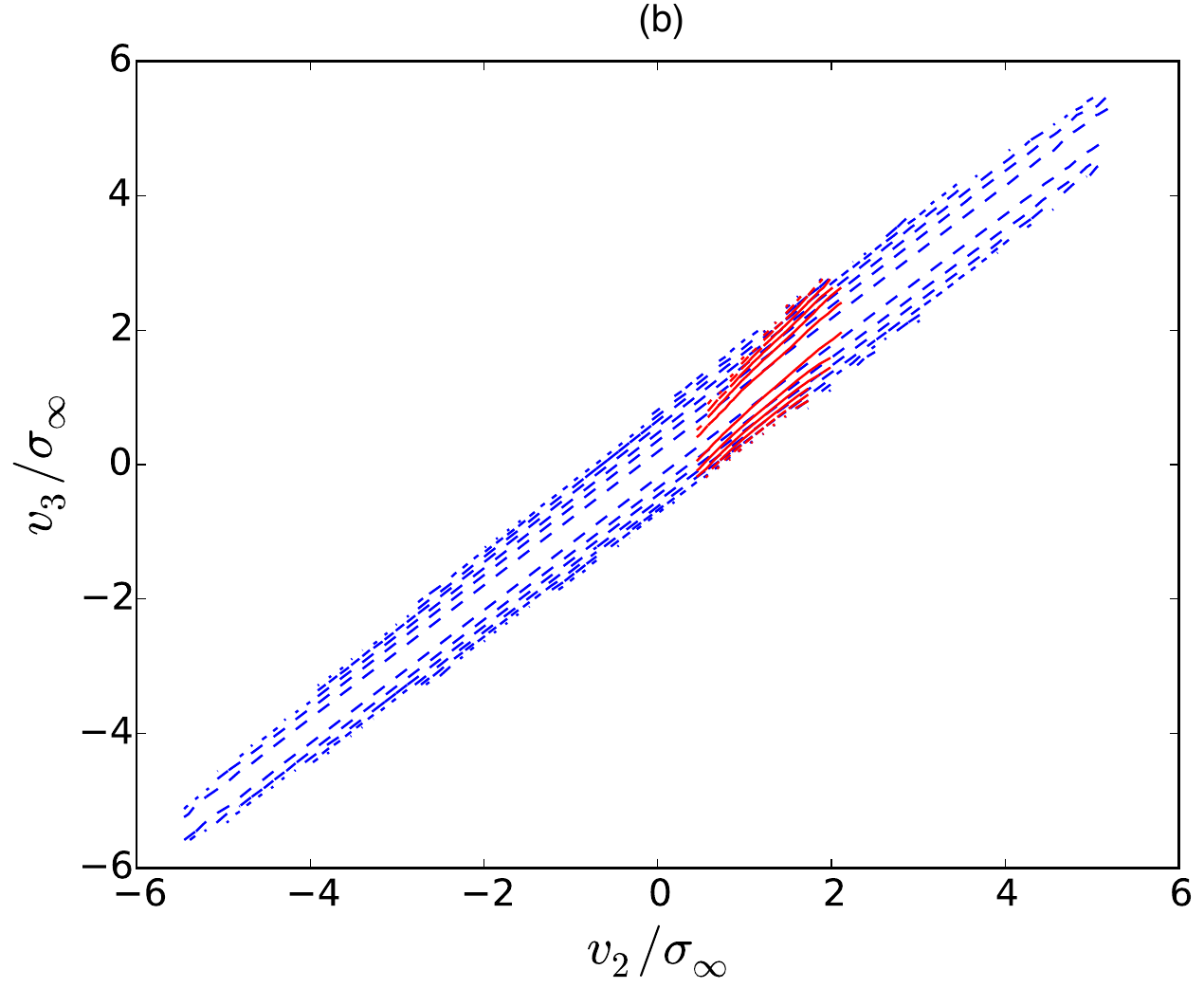}
\caption{(a) Examination of the Markov property (\ref{eq:markov_direct}) from DNS of the nonlocal case
for $\Delta r= 0.2\lambda$ and $v_1=0$ via a logarithmic contour plot.
(b) Same as in (a), but for $v_1 = \sigma_{\infty}$. The Markov property is violated.}
\label{fig:markov_non_0.2}
\end{figure}

Figs.~\ref{fig:markov_non_2.2}-\ref{fig:markov_non_0.2} show the contour plots of the one-time conditioned PDF (blue) and the two-time conditioned PDF (red).
For this particular case, the transition PDF (blue) possesses a solely diagonal shape in contrast to the pure Burgers case in Section~\ref{sec:ex_burgers}, which possessed a $\delta(v_3)$-part for negative $v_2$.
The Markov property is fulfilled to a great extend for $\Delta r=2.2\lambda$ and $\Delta r=1\lambda$ in Figs.~\ref{fig:markov_non_2.2} and~\ref{fig:markov_non_1}. For smaller scale separations, i.e., $\Delta r=0.2 \lambda$ in
Figure~\ref{fig:markov_non_0.2}, the Markov property is broken:
the two-times conditional PDF (red) appears steeper than the transition PDF (blue), which underestimates the correlations between $v_3$ and $v_2$.
The contours are shown for two different slices of $p(v_3,L/2-\Delta r|v_2,L/2;v_1,L/2-\Delta r)$, namely $v_1=0$ in (a) and $v_1=1 \sigma$. Contrary to the Burgers case in Figs.~\ref{fig:markov_burgers_2}-\ref{fig:markov_burgers_0.2}, the exact $v_1$-position of the slice is quite unimportant and only alters the significant statistics, not the shape of the PDFs.

\subsubsection{Determination of the Markov-Einstein Length}
\begin{figure}[h!]
 \centering
\includegraphics[width=0.49 \textwidth]{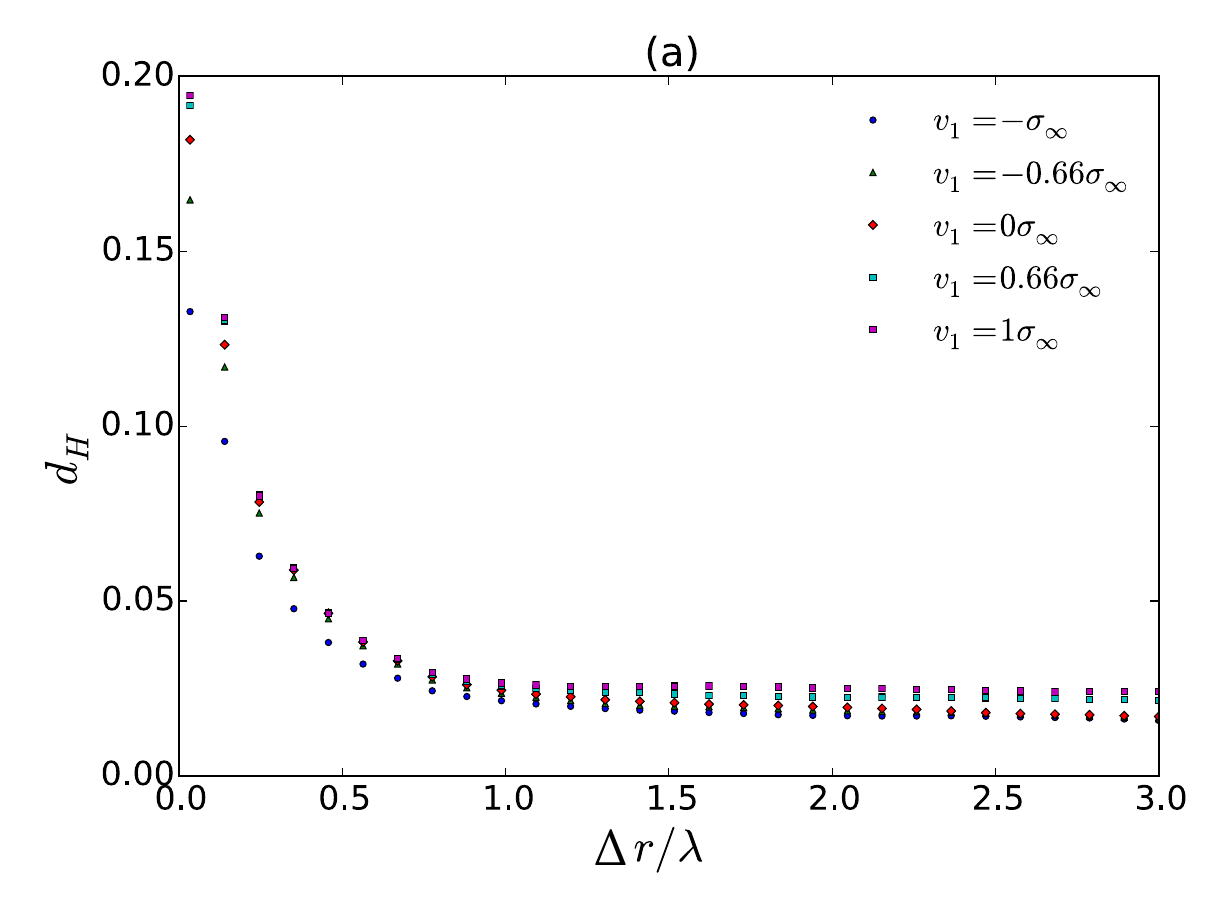}
\includegraphics[width=0.49 \textwidth]{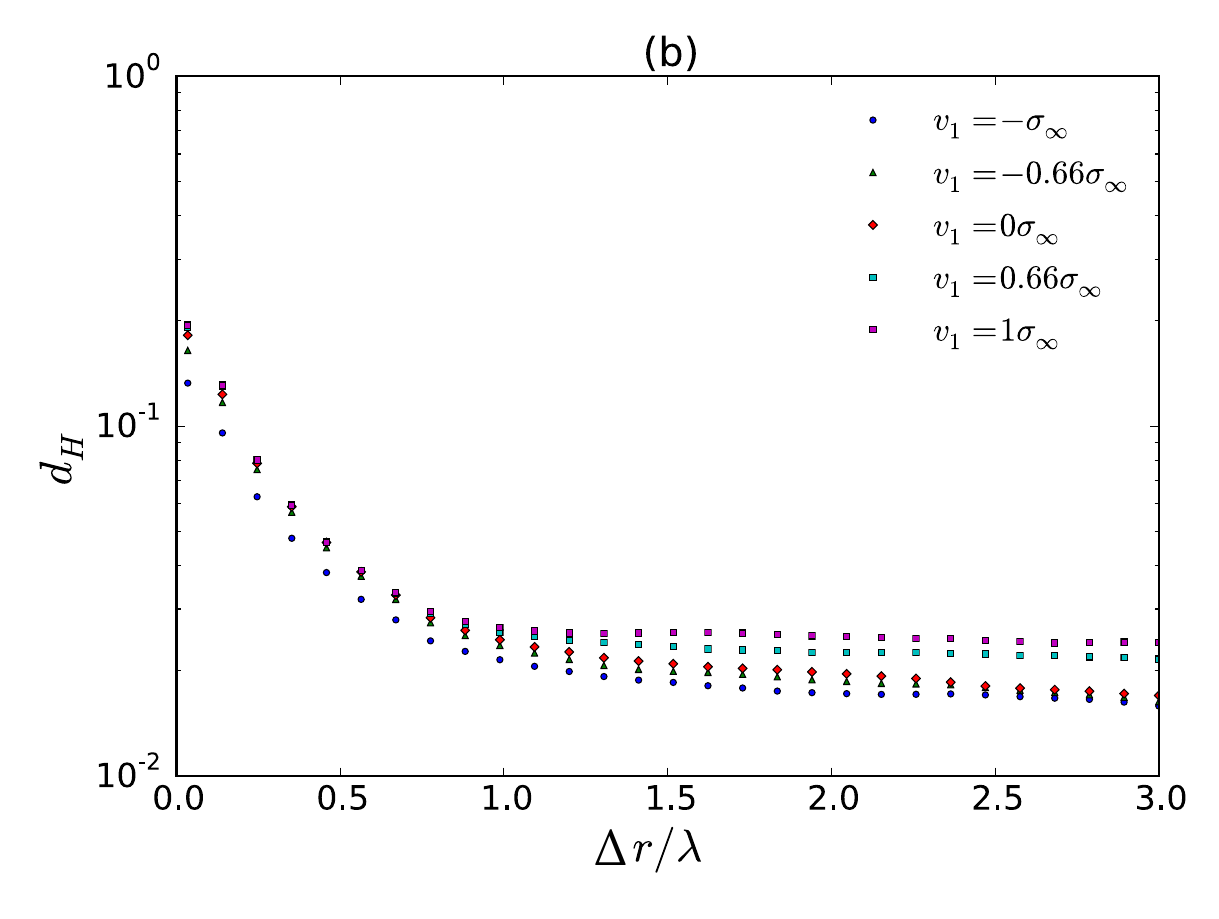}
\caption{(a) Hellinger distance $d_H(\Delta r)$ for different $v_1$ and variable step width $\Delta r$ for the purely nonlocal case.
The Hellinger is close to zero and only increases at around $\Delta r \approx \lambda$
(b) Semi-logarithmic plot of the Hellinger distance $d_H(\Delta r)$.}
\label{fig:d_H_non}
\end{figure}

Figure~\ref{fig:d_H_non} presents the Hellinger distance $d_H(\Delta r)$ from Equation~(\ref{eq:hell}) for the purely nonlocal case. The Hellinger distance is close to zero for large scale separations $\Delta r$. In contrast to the Burgers case in Figure~\ref{fig:d_H_burgers}, which exhibited a clear drop of the Hellinger distance at around $\Delta r=\lambda$, the Markov property is a good approximation for all larger scale separations.
Here, the Hellinger distances increase at around $\Delta r=\lambda$. It must be stressed that this behavior differs significantly from the Burgers case: whereas the Hellinger distances in Figure~\ref{fig:d_H_burgers} decrease at around $\Delta r \approx \lambda$ and then increase, i.e., they exhibit a clear minimum, the nonlocal case exhibits a steady increase at these scales.
Figure~\ref{fig:d_H_burgers} (b) shows a semi-logarithmic plot of the Hellinger distance $d_H(\Delta r)$. Whether the increase of the Hellinger distance is exponential is somewhat hard to anticipate. However, the increase is not as violent as in the Burgers case in Figure~\ref{fig:d_H_burgers} (b). Accordingly, the Markov property in the nonlocal case is not deteriorating as fast as in the Burgers case. The determination of the Markov-Einstein length $\lambda_{ME}$ for the nonlocal case proves to be quite challenging since the Hellinger distance is steadily increasing for smaller $\Delta r$. In the following, we estimate $\lambda_{ME}=\lambda$, in order to determine the Kramers-Moyal coefficients of the nonlocal case.
\subsubsection{Determination of the Kramers-Moyal Coefficients}
\begin{figure}[h!]
\centering
 \vspace{-19ex}
\includegraphics[width=0.499 \textwidth]{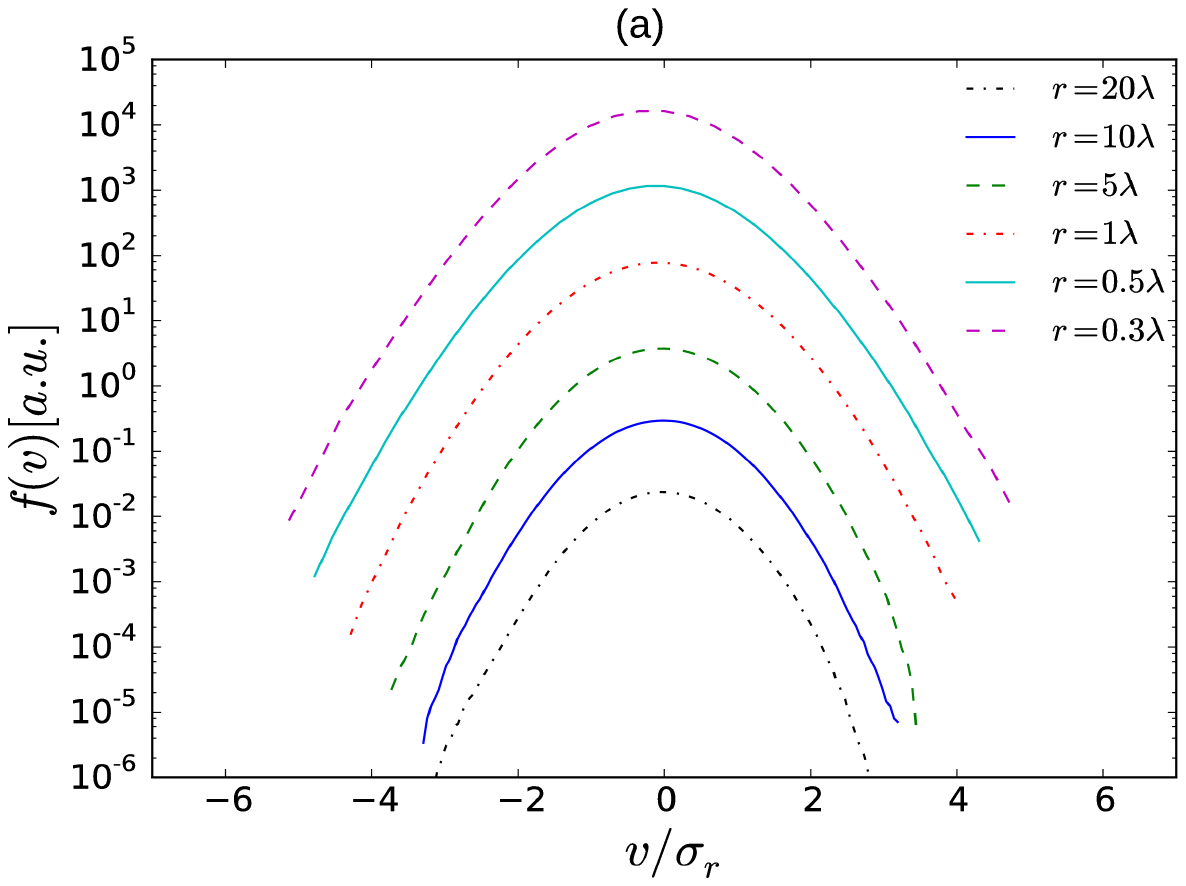}
\raisebox{+1.5pt}{\includegraphics[width=0.489 \textwidth]{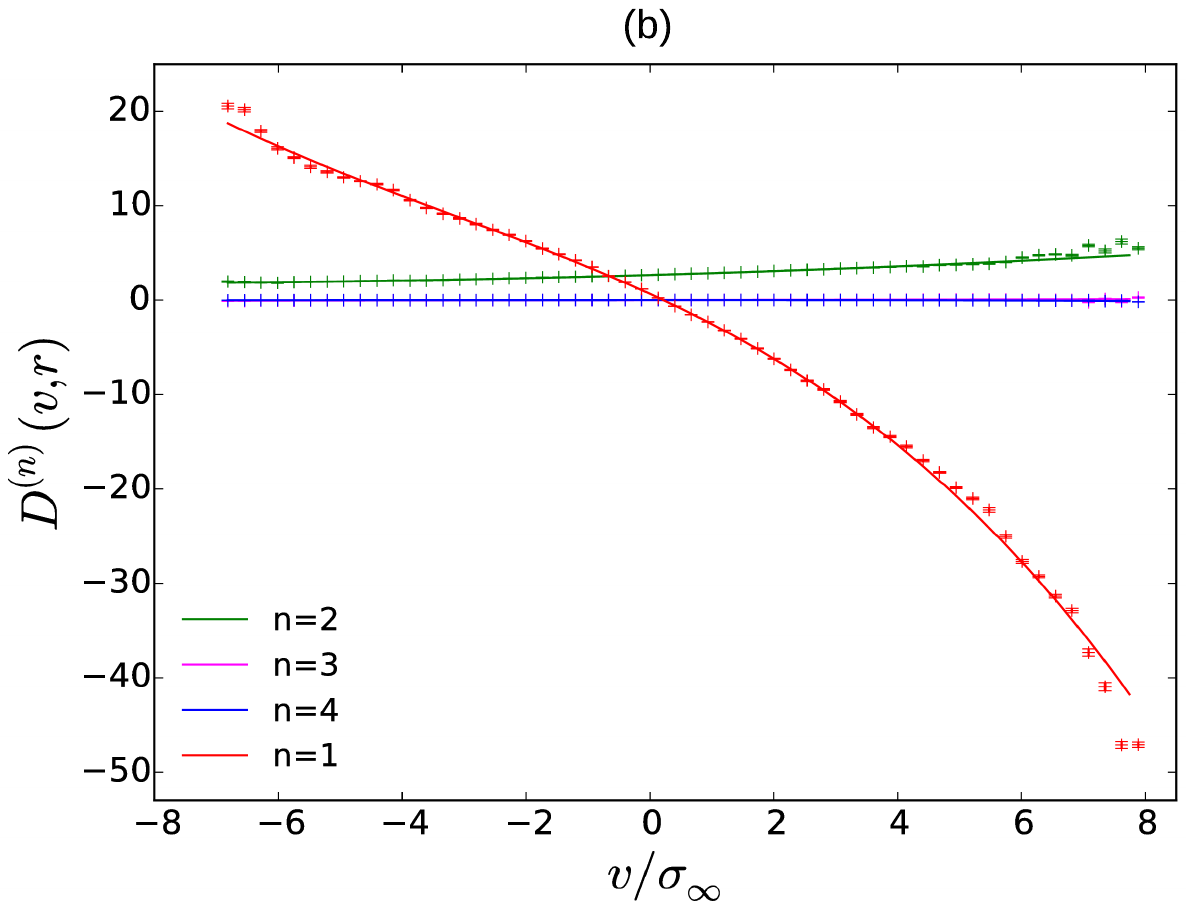}}
\vspace*{-25mm}
\caption{(a) Evolution of the velocity increment PDF in scale for the purely nonlocal case $\alpha=0$.
The PDFs exhibit self-similarity in the inertial range.
(b) Estimation of the Kramers-Moyal coefficients $D^{(n)}(v,r=L/2)$ from DNS of the
purely nonlocal case $\alpha=0$. The fits correspond to
polynomials of the order $n$ of the coefficient except for $n=1$ where a polynomial of order three has been used.
The reduced Kramers-Moyal coefficients have been determined according to
$K_1=0.3108 \pm 0.0002$, $K_2 =0.0021 \pm 0.0001$, $K_3=(2.64\pm 0.01 )\times 10^{-5}$ and
$K_4=(2.28 \pm 2.56) \times 10^{-5}$.}
\label{fig:km_nonlocal}
\end{figure}
The Kramers-Moyal coefficients for the nonlocal case are depicted in Figure~\ref{fig:km_nonlocal} (b). The drift coefficient $D^{(1)}(v,r)$ possesses
a slightly cubic dependence. The corresponding reduced Kramers-Moyal coefficient $K_1=0.3108 \pm 0.0002$ is close to the K41 prediction $K_1=1/3$. The diffusion coefficient points further into the direction of the K41 phenomenology: it shows only a slight quadratic dependence on $v$ and has a rather linear shape. Accordingly, the reduced Kramers-Moyal coefficient of order two is rather small,
$K_2 =0.0021 \pm 0.0001$. Higher-order coefficients, $n=3,4$ are even smaller
($K_3=(2.64\pm 0.01 )\times 10^{-5}$,
$K_4=(2.28 \pm 2.56) \times 10^{-5}$). Furthermore, for $n>4$, the coefficients strongly deviate from the scaling $D^{(n)}(v,r)\sim v^n/r$, e.g., the reduced Kramers-Moyal coefficients become negative.
Hence, the obtained Kramers-Moyal coefficients are in agreement with the self-similarity of the one-increment PDF in Figure~\ref{fig:km_nonlocal} (a).
Here, the PDFs are depicted for different $r$ in the inertial range. They can be reproduced by a self-similar function
\begin{equation}
  f_1(v,r) = \frac{1}{(\langle \varepsilon \rangle r)^{\alpha}}
  g \left(\frac{v}{(\langle \varepsilon \rangle r)^{\alpha}}\right)\;,
\end{equation}
with $\alpha \approx 0.31$ and $g$ being very close to a Gaussian (not depicted in the figure). Hence,
the nonlocal case is in very close correspondence to the K41 phenomenology.
\begin{figure}[h!]
  \includegraphics[width=0.49 \textwidth]{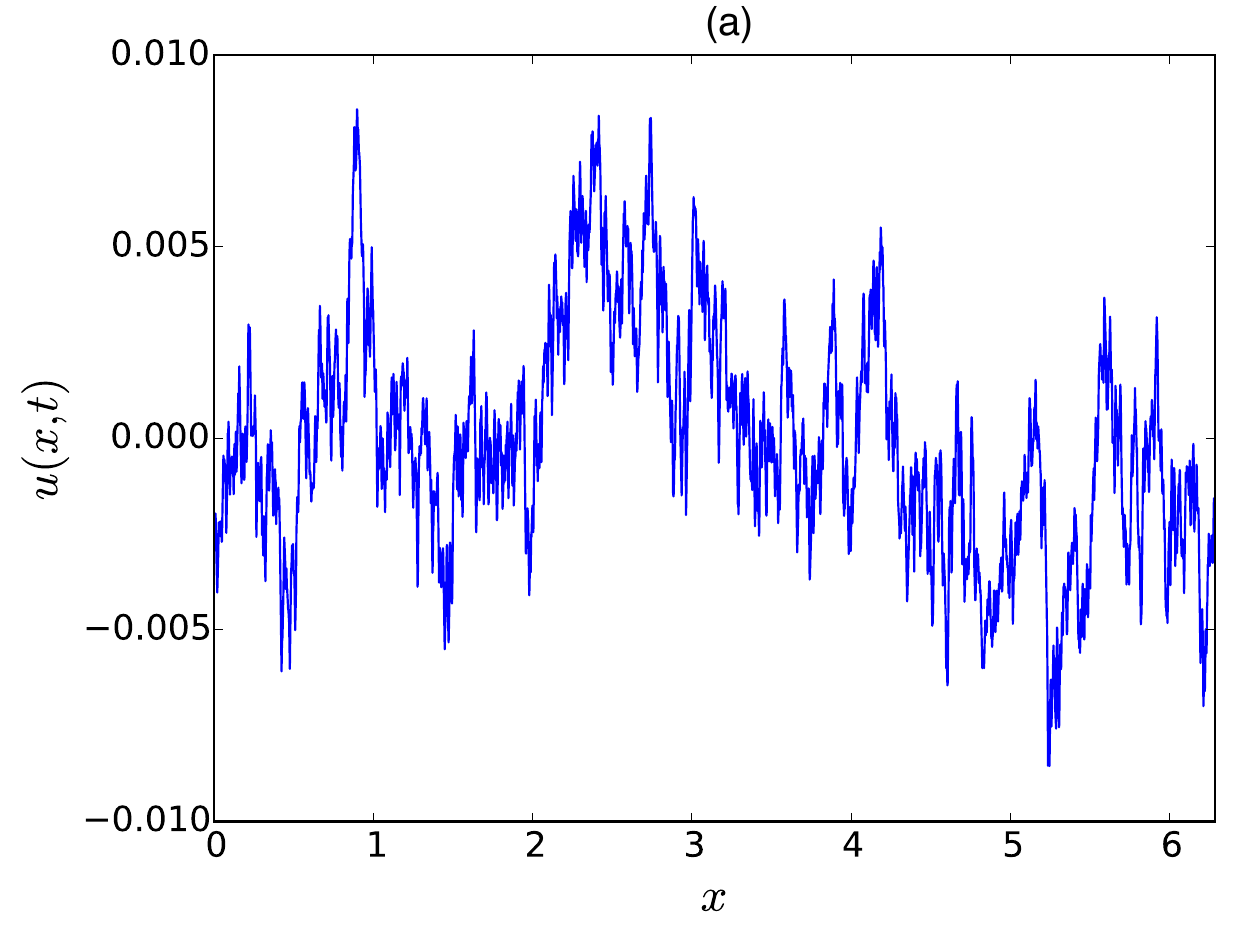}
  \includegraphics[width=0.49 \textwidth]{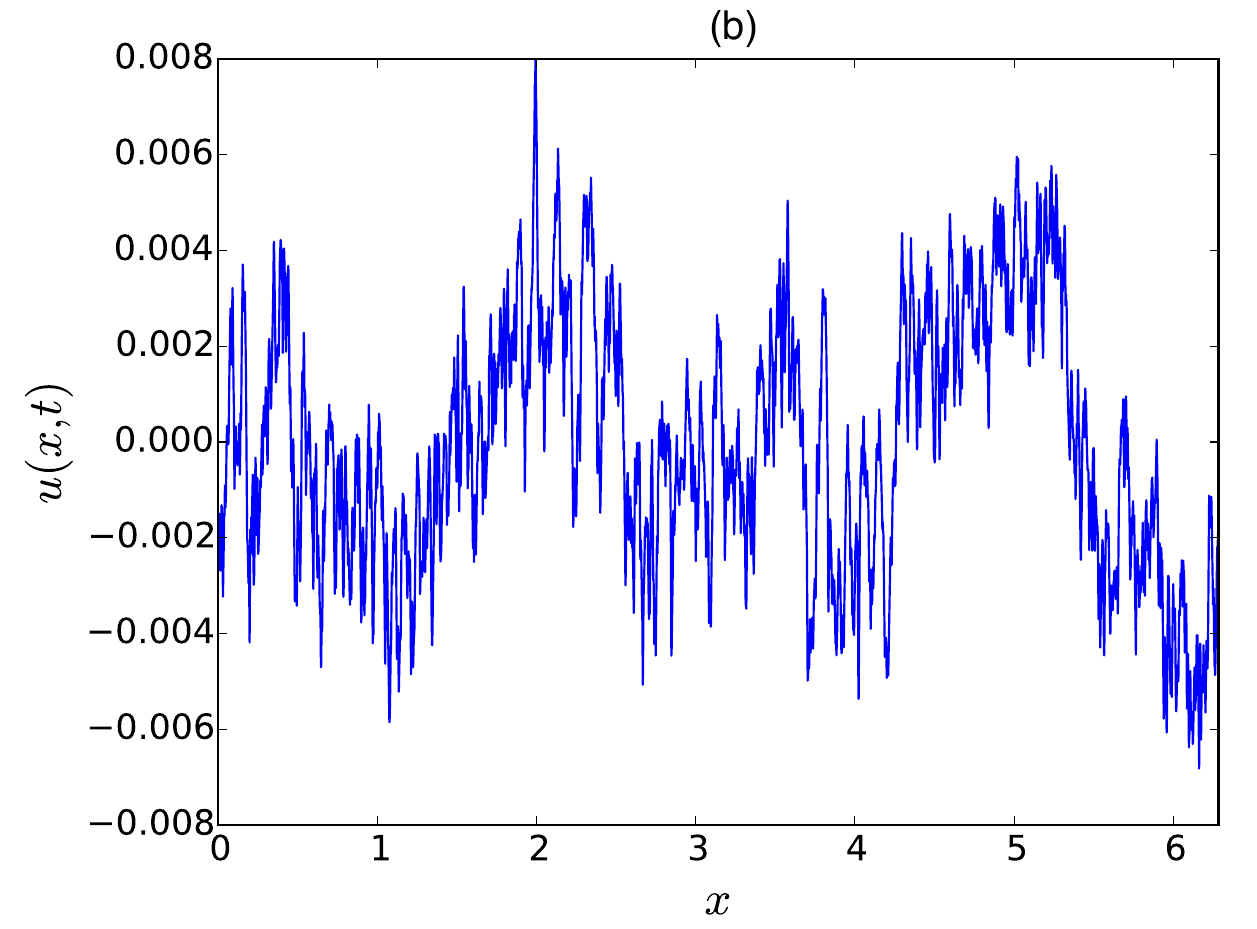}
  \caption{(a) Typical realization of the velocity field $u(x,t)$ in
  DNS of the intermediate case ($\alpha=0.15$ in Equation (\ref{eq:gen_burgers})).
  (b) Velocity field realization that belongs to the largest velocity
  field gradient that was attained in the DNS.}
\label{fig:intermediate}
\end{figure}
\subsection{Intermediate Case $\alpha=0.15$}

Seminal numerical investigations~\cite{Zikanov1997} already highlighted the fact that the intermediate case exhibits statistical behavior that is close to the intermittency behavior encountered in ordinary hydrodynamic turbulence.
The latter behavior manifested itself by a skewed velocity gradient PDF as well as by structure function exponents that deviated considerably from the predictions suggested by the K41 theory.

Figure~\ref{fig:intermediate} (a) shows a snapshot of the velocity field obtained from DNS run \#3. In contrast to run \#2 in Figure~\ref{fig:nonlocal}, the intermediate case reveals a rather saw-tooth like velocity field profile. Nonetheless, no clear shock-like structures can be distinguished.
\begin{figure}[h]
  \vspace{-2ex}
  \centering
\includegraphics[width=0.54 \textwidth]{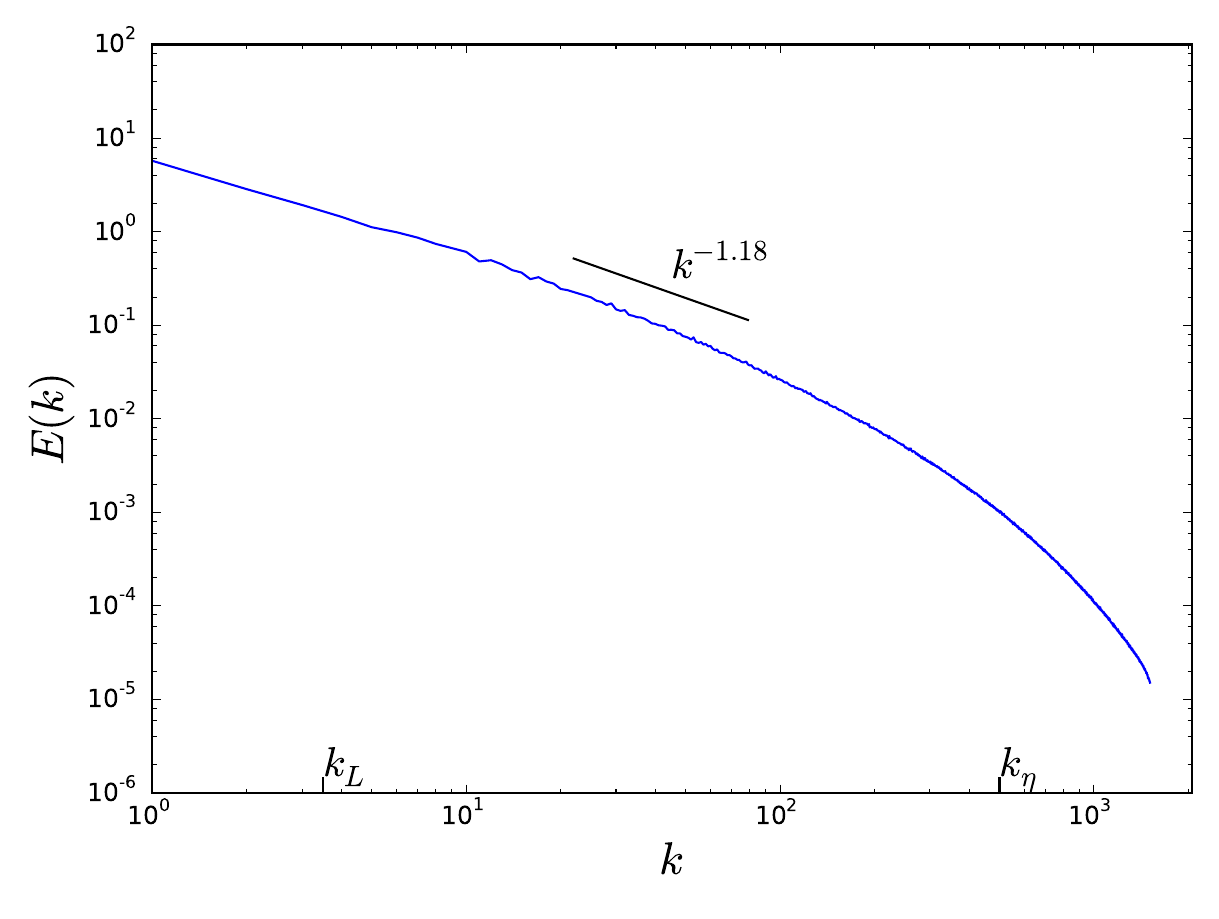}
\caption{Energy spectrum $E(k)$ from DNS of the intermediate regime ($\alpha=0.15$). The line indicates the spectral slope $k^{-1.18}$. No clear power-law behavior of the energy spectrum could be detected. The Kolmogorov spectrum might as well appear for larger $k$ behind the indicated spectral slope.}
\label{fig:spec_int}
  \vspace{-2ex}
\end{figure}
Figure~\ref{fig:spec_int} shows the energy spectrum of run \#2. It deviates considerably from the Kolmogorov prediction, i.e., $\sim k^{-5/3}$. Firstly, no clear power-law behavior could be detected. The fitted line corresponds to
$\sim k^{-1.18}$, however, a Kolmogorov spectrum might as well be attained at higher $k$-values. Zikanov et al.~\cite{Zikanov1997} propose an energy spectrum of the form $\sim k^{-1.5861}$ which is closer to $k^{-5/3}$ than the simulations performed here. At this point, it is not clear, whether the small Reynolds numbers attained in run \#2 are responsible for the underestimation of the spectral energy decay.

\subsubsection{Examination of the Markov Property}
\begin{figure}[h!]
 \centering
\includegraphics[width=0.49 \textwidth]{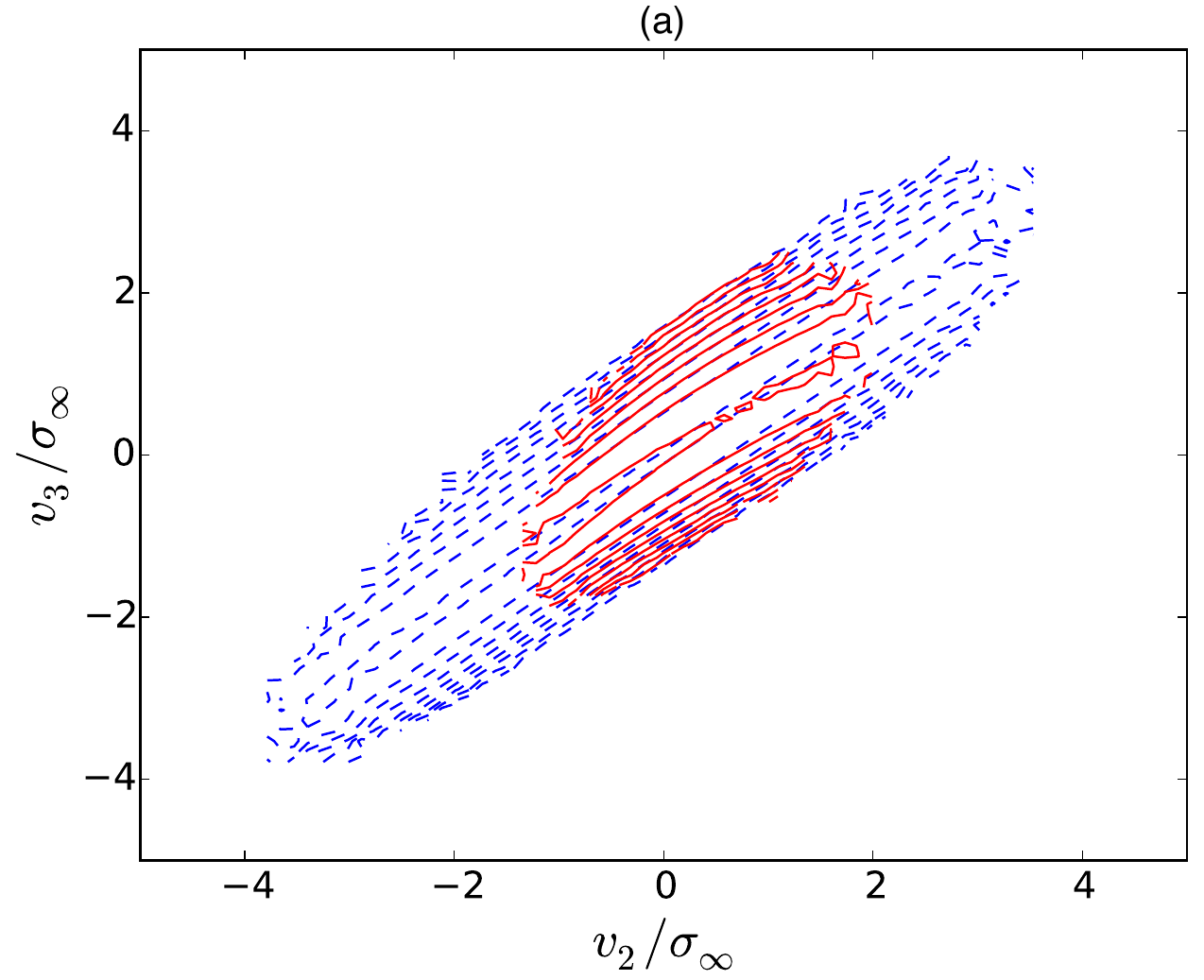}
\includegraphics[width=0.49 \textwidth]{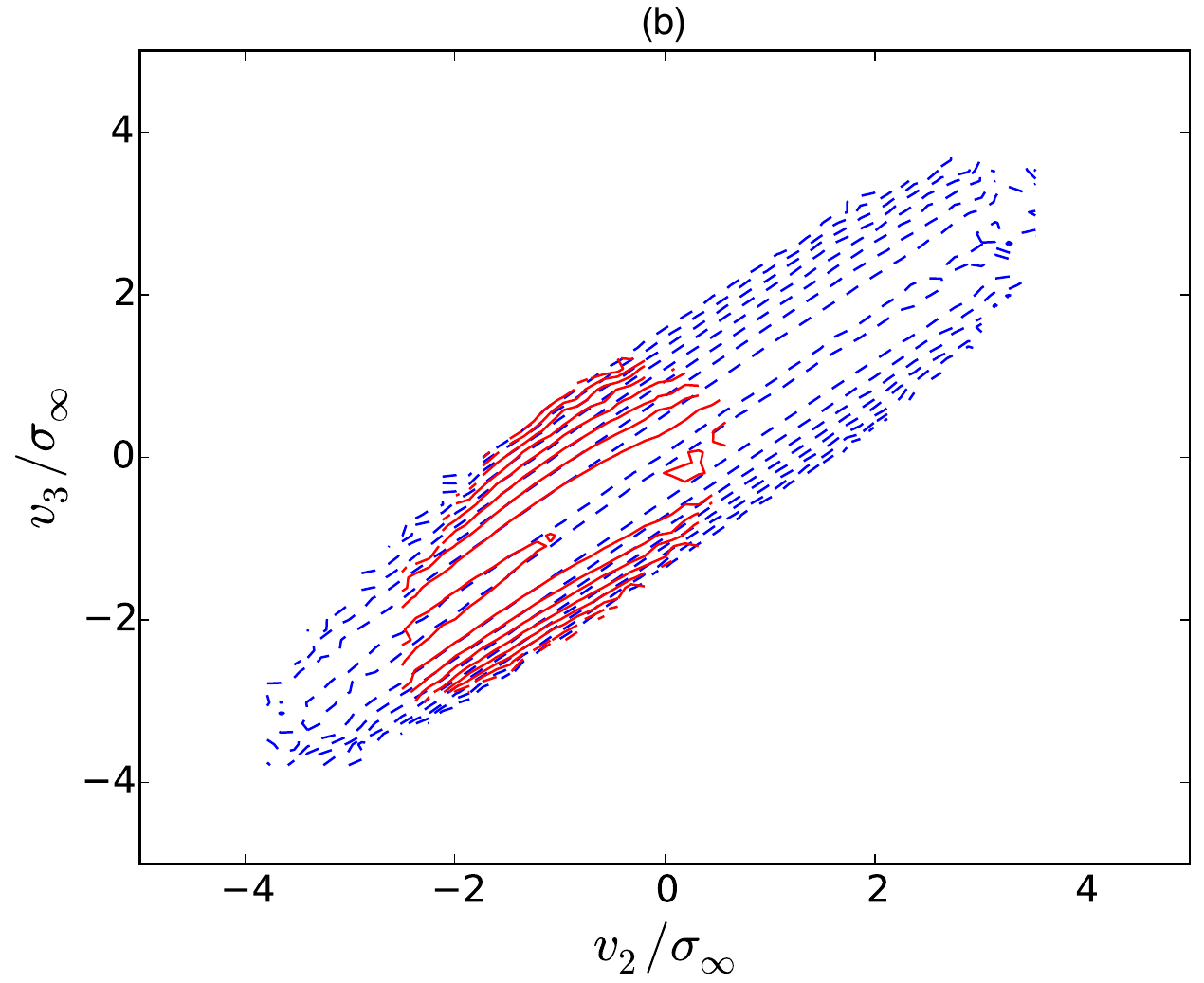}
\caption{(a) Examination of the Markov property (\ref{eq:markov_direct}) from DNS of the intermediate case
for $r= 1\lambda$ and $v_1=0$ via a logarithmic contour plot.
(b) Same as in (a), but for $v_1 = \sigma_{\infty}$.}
\label{fig:markov_int_1}
\end{figure}

\begin{figure}[h!]
\includegraphics[width=0.49 \textwidth]{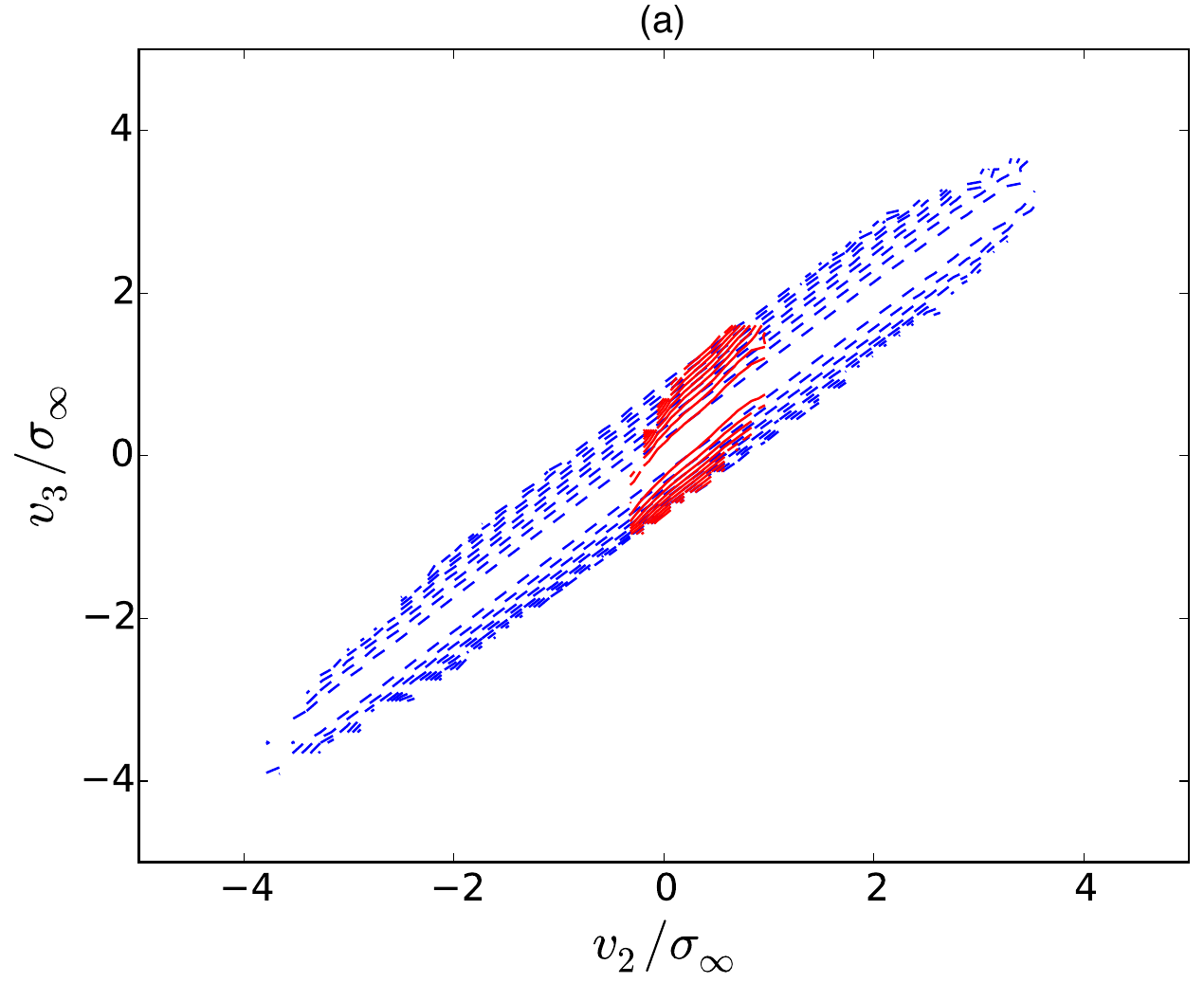}
\includegraphics[width=0.49 \textwidth]{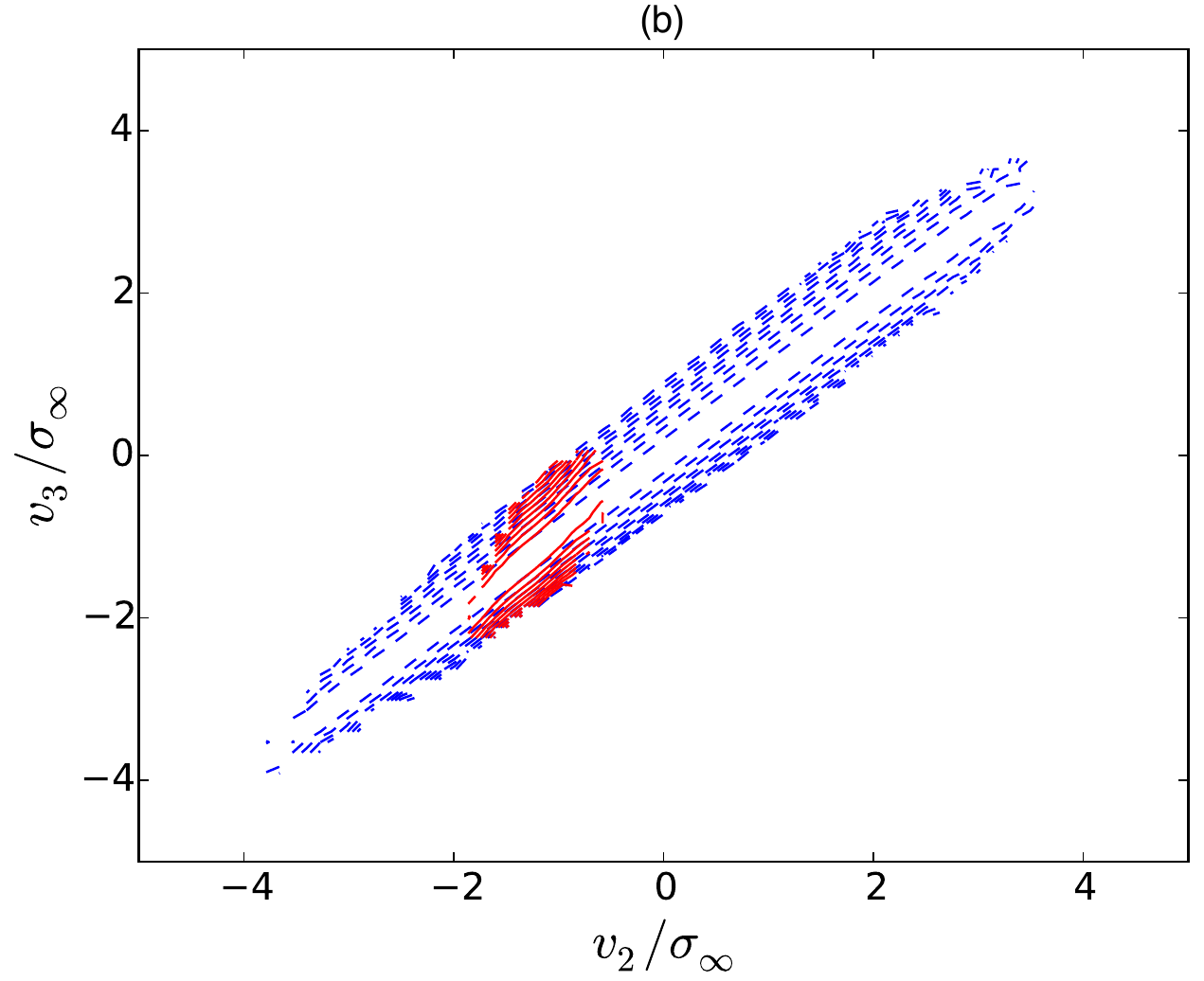}
\caption{(a) Examination of the Markov property (\ref{eq:markov_direct}) from DNS of the intermediate case
for $\Delta r= 0.2\lambda$ and $v_1=0$ via a logarithmic contour plot.
(b) Same as in (a), but for $v_1 =  \sigma_{\infty}$. The Markov property is violated.}
\label{fig:markov_int_0.2}
\end{figure}

The Markov property is fulfilled for the intermediate case as well, provided that the scale separation $\Delta r$ is not too small.
\begin{figure}[h!]
\includegraphics[width=0.49 \textwidth]{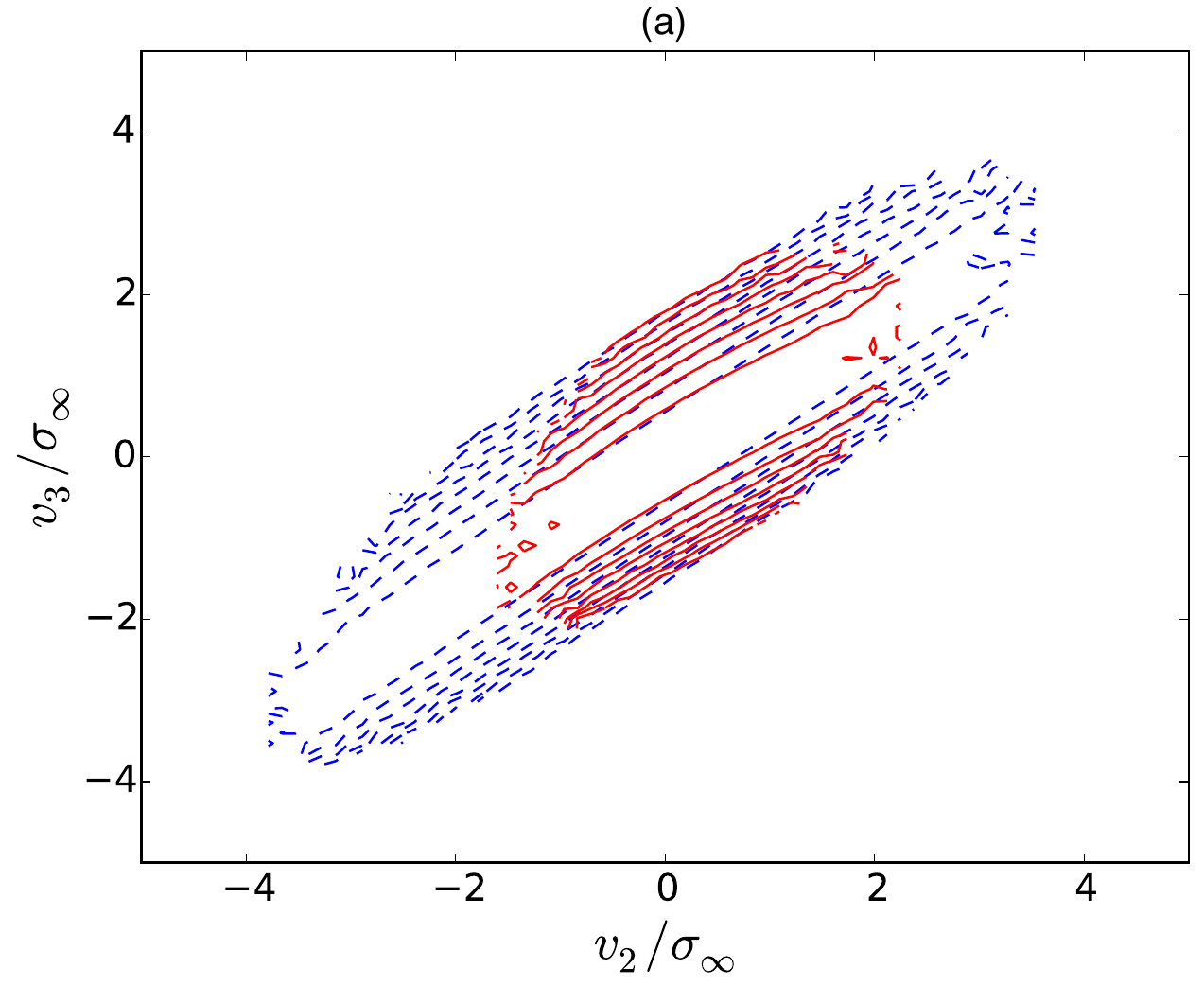}
\includegraphics[width=0.49 \textwidth]{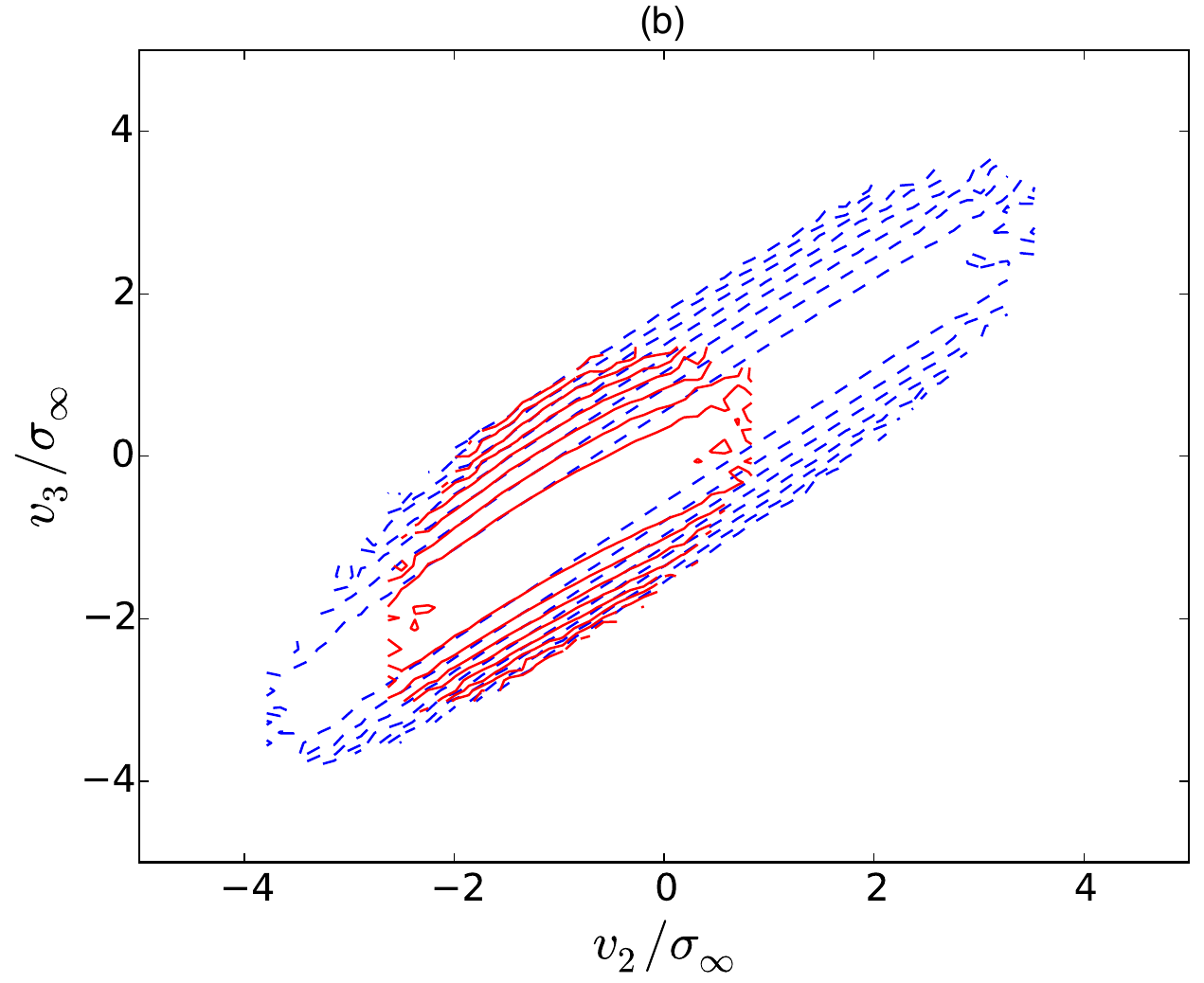}
\caption{(a) Examination of the Markov property (\ref{eq:markov_direct}) from DNS of the intermediate case
for $\Delta r= 2.2\lambda$ and $v_1=0$ via a logarithmic contour plot.
(b) Same as in (a), but for $v_1 = - \sigma_{\infty}$. The shape of the conditional PDF (red) does not change significantly in comparison to (a)}
\label{fig:markov_int_2.2}
\end{figure}
Figs.~\ref{fig:markov_int_2.2}-\ref{fig:markov_int_1} show the corresponding contour plots for $\Delta r=1.5 \lambda$ and $\Delta r=\lambda$. The shape of the PDFs qualitatively agree with those of the purely nonlocal case, however for smaller scale separations one might guess certain differences.
The Markov property is not fulfilled for $\Delta r=0.2 \lambda$, which can be deduced from Figure~\ref{fig:markov_int_0.2}. The transition PDF (blue) underestimates the $v_3$-$v_2$-correlations of the two-times conditional PDF (red).
\subsubsection{Determination of the Markov-Einstein Length}
\begin{figure}[h!]
 \centering
\includegraphics[width=0.49 \textwidth]{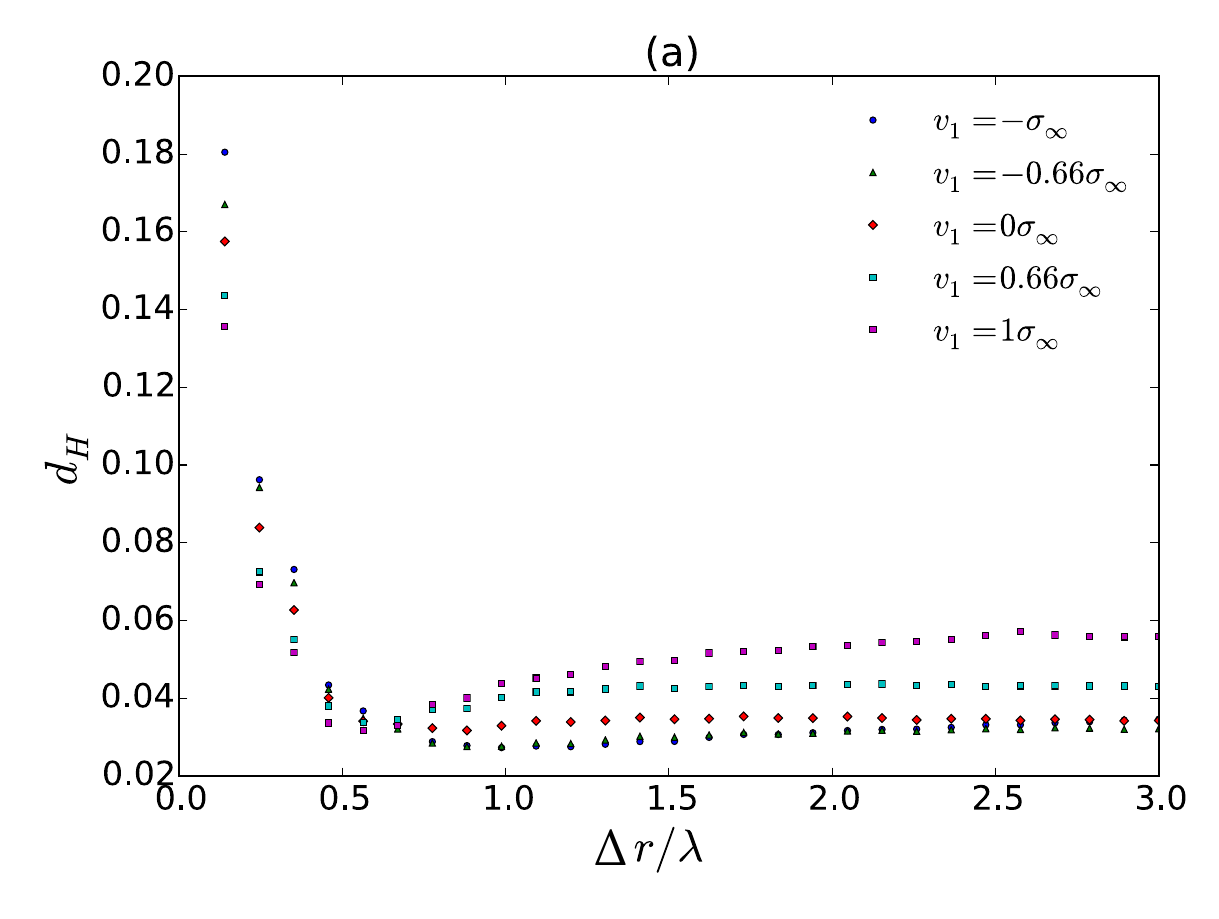}
\includegraphics[width=0.49 \textwidth]{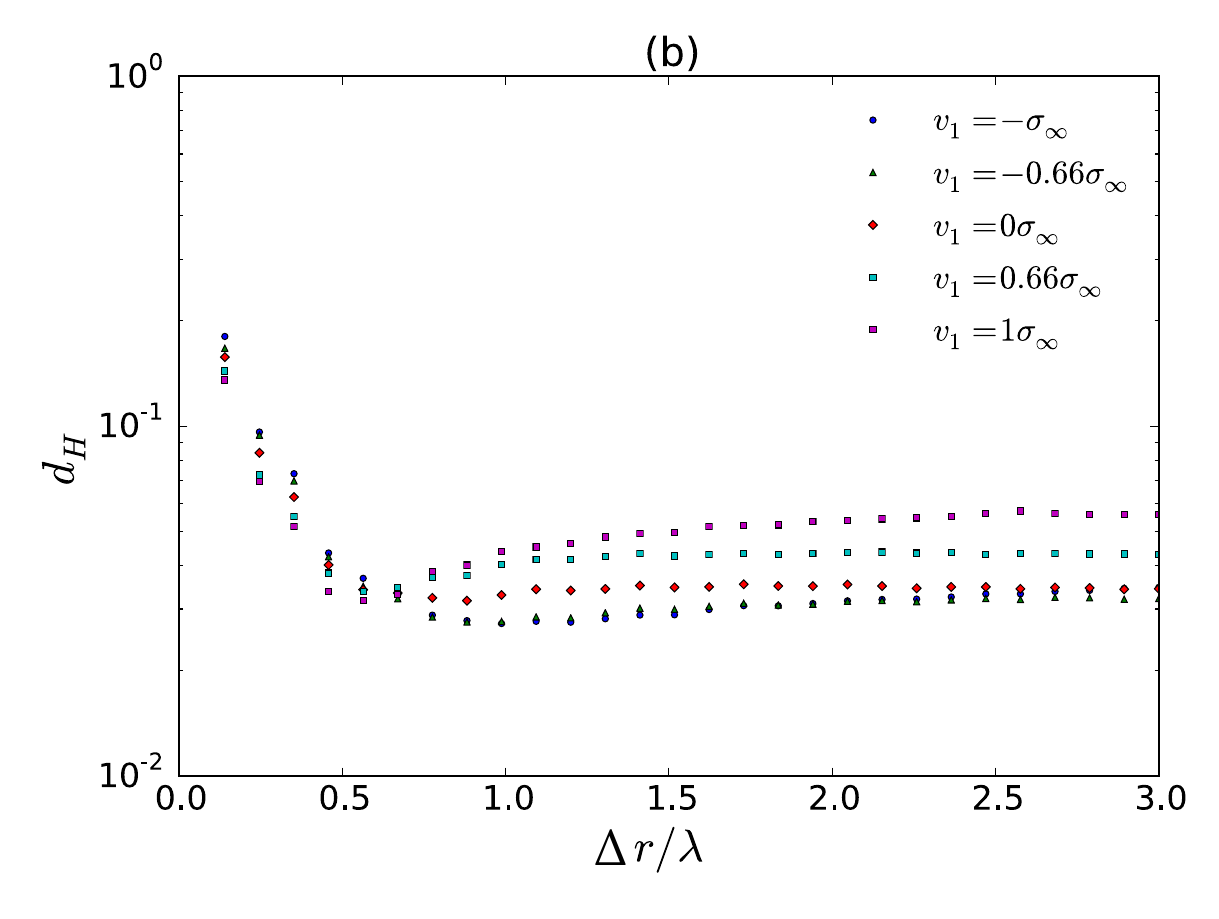}
\caption{(a) Hellinger distance $d_H(\Delta r)$ for different $v_1$ and variable step width $\Delta r$.
Apparently, the Markov property
is a good approximation only around $\Delta r=\lambda$. For larger $\Delta r$, the Hellinger distance slightly increases and approaches a small constant value.
However, for smaller $\Delta r$, the Hellinger distance exhibits a more pronounced increase. Therefore, the Markov
property is clearly violated
These effects become even more pronounced for $v_1=-0.33 \sigma_{\infty}, ~-0.66 \sigma_{\infty},
,~-\sigma_{\infty}$.
(b) Semi-logarithmic plot of the Hellinger distance $d_H(\Delta r)$. For $r < 0.8 \lambda$, the Hellinger distance
seems to increase nearly exponentially.}
\label{fig:d_H_int}
\end{figure}

\begin{figure}[h!]
  \vspace{-19ex}
\centering
\includegraphics[width=0.499 \textwidth]{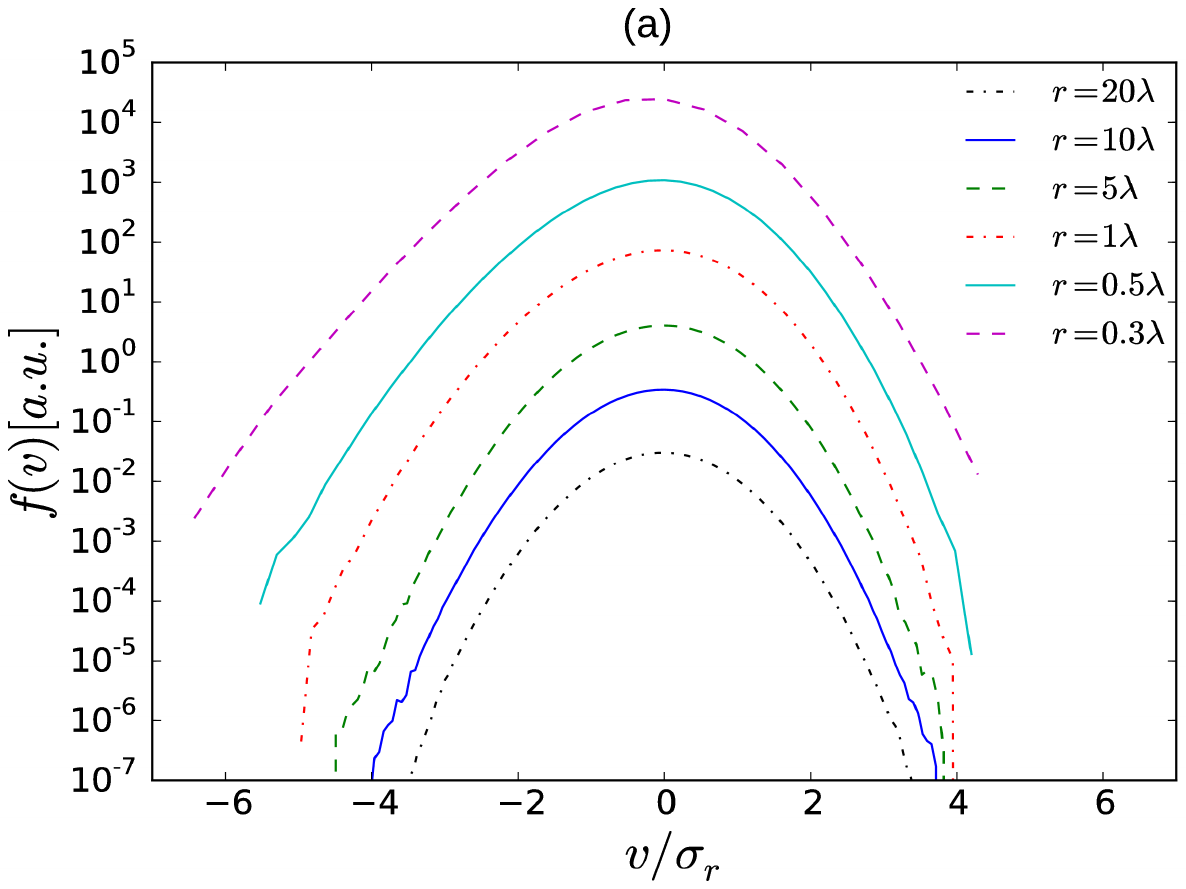}
\raisebox{+1.5pt}{\includegraphics[width=0.489 \textwidth]{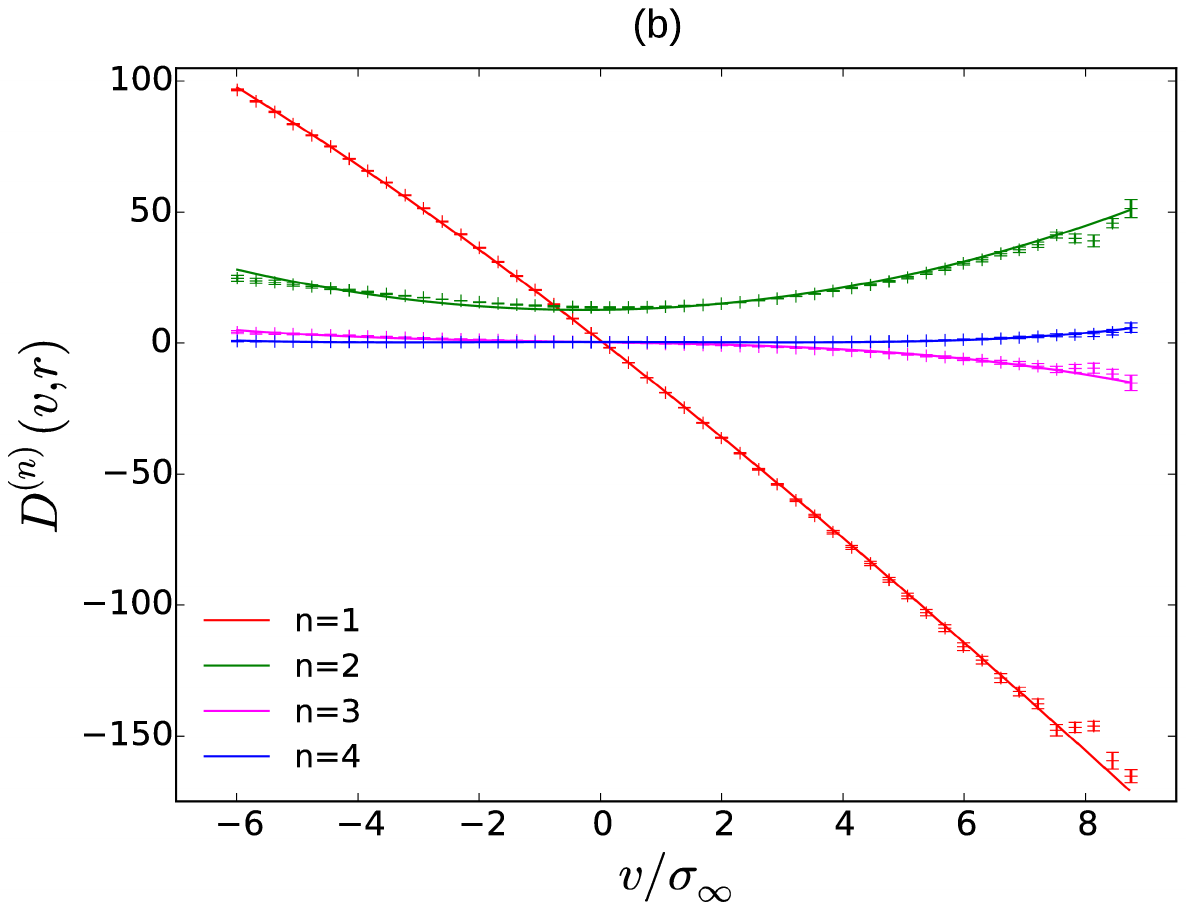}}
\vspace*{-25mm}
\caption{(a) Evolution of the velocity increment PDF in scale for the intermediate case $\alpha=0.15$.
The PDFs show a slight asymmetry at small scales.
(b) Estimation of the Kramers-Moyal coefficients from DNS of the
intermediate case $\alpha=0.15$. The fits correspond to
polynomials of the order $n$ of the coefficient except for $n=1$ where a polynomial of order three has been used.
The reduced Kramers-Moyal coefficients have been determined according to
$K_1=0.4356 \pm 0.0007$, $K_2 =0.0208 \pm 0.0004$, $K_3=0.0014\pm 0.0001 $ and
$K_4=0.00041 \pm 0.00001$.}
\label{fig:km_intermediate}
\end{figure}

Figure~\ref{fig:d_H_int} shows the Hellinger distance (\ref{eq:hell}) for
five different $v_1$. It is remarkable that for $v_1 < 0$, the Helligner distance is smaller than for $v_1 \ge 0$, which only changes for small $r$. This is also a first hint for the asymmetry of velocity increments in the intermediate case. The latter can be further quantified in considering the evolution in scale of the one -increment PDF in Figure~\ref{fig:km_intermediate} (a) which shows that extreme large negative velocity increments at small scales are more common than positive ones.
%
%\begin{figure}[h!]
%\centering
%\includegraphics[angle=270, width=0.68 \textwidth]{figures/km_dns.pdf}
%  \includegraphics[angle=270, width=0.73 \textwidth]{figures/logkm_dns.pdf}~~~~~~
%\caption{(a) Reduced Kramers-Moyal coefficients for DNS of the generalized Burgers equation
%(\ref{eq:gen_burgers}). The magnitude of the coefficients for Burgers turbulence ($\alpha=1$)
%are in the range of 1 which agrees with the phenomenological predictions. The intermediate ($\alpha=0.15$) and the purely nonlocal case ($\alpha=0$) can hardly be distinguished from one another in this representation. The semi-logarithmic plot (b), however, reveals that the reduced Kramers-Moyal
%coefficients for the intermediate case follow Yakhot's mean field theory (see Section~\ref{sec:relation}\emph{v.)}) for higher order of $n$. Note that higher-order ($n>5$) coefficients for the Burgers and the purely nonlocal case
%obtained accurately due to poor polynomial fits (Burgers case) or considerable deviations from Equation (\ref{eq:km_final}) (nonlocal case).}
%\label{fig:km_dns}
%\end{figure}
%
Concerning the Hellinger distance itself, we can see that, in contrast to the nonlocal case, the Hellinger distance exhibits a rather pronounced minimum
at around $\Delta r \approx 0.6 \lambda$. This behavior thus resembles somewhat to the behavior encountered in the pure Burgers case and, hence, can be attributed to nonlinear shock generation. Henceforward, the determination of the Markov-Einstein length is rather simple and it can be estimated according to $\lambda_{ME}=0.6 \lambda$. The latter finding is supported by the semi-logarithmic plot in Figure~\ref{fig:d_H_int} (b) as well.
\begin{figure}[h!]
\centering
  \includegraphics[width=0.73 \textwidth]{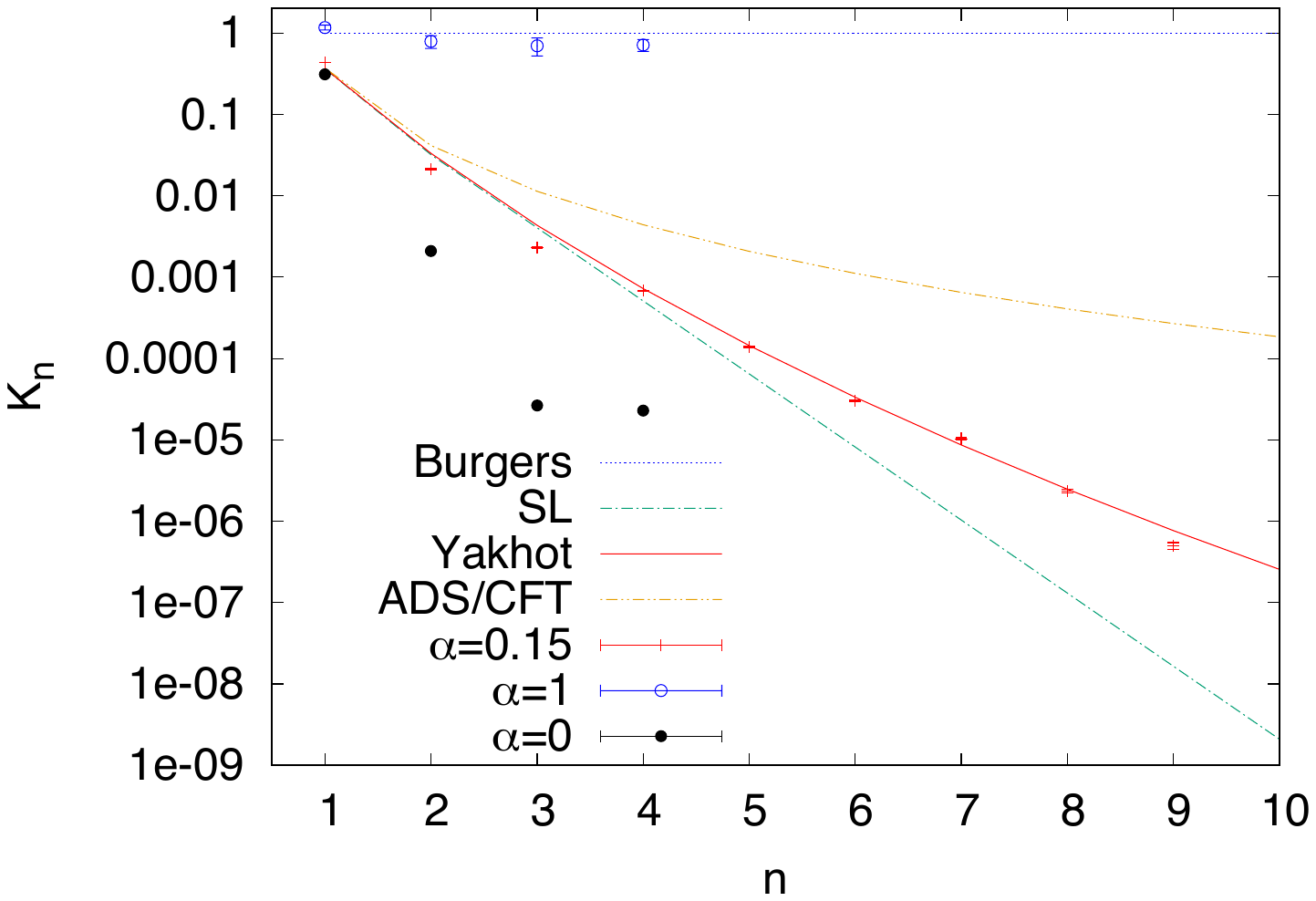}
\caption{Semi-logarithmic plot of reduced Kramers-Moyal coefficients from DNS of the generalized Burgers equation
(\ref{eq:gen_burgers}). The coefficients for Burgers turbulence ($\alpha=1$)
are in the range of 1 which agrees with the phenomenological predictions. The semi-logarithmic plot reveals that the reduced Kramers-Moyal
coefficients for the intermediate case ($\alpha=0.15$) follow Yakhot's mean field theory \emph{v.)}.
Note that higher-order $n>5$ coefficients for the Burgers and the purely nonlocal case ($\alpha=0$)
could not be accurately obtained due to poor polynomial fits (Burgers case) or considerable
deviations from Equation \ref{eq:km_final} (nonlocal case).}
\label{fig:km_dns}
\end{figure}
\subsubsection{Determination of the Kramers-Moyal Coefficients}
The Kramers-Moyal coefficients from run \#3 are shown in Figure~\ref{fig:km_intermediate} (a). The second Kramers-Moyal coefficient exhibits a clear parabolic dependence on $v$ in contrast to the purely nonlocal case, which revealed a linear dependence. The corresponding reduced Kramers-Moyal coefficient $K_2 =0.0208 \pm 0.0004$, hence, is roughly ten times bigger than the one determined from the nonlocal case $K_2 =0.0021 \pm 0.0001$. The tendency for these intermittency effects also reflects itself in higher-order Kramers-Moyal coefficients $n=3,4$. The latter coefficients, even though they appear rather flat, follow a clear $v$-polynomial of order $3$ and $4$, respectively.

Moreover, in contrast to the two previous runs \#1 and \#2, Kramers-Moyal coefficients up to order $9$ could significantly be detected and their reduced Kramers-Moyal coefficients could be determined. The evolution in scale of the one-increment PDF in Figure~\ref{fig:km_intermediate} (b) shows deviations from self-similarity at small scales and is slightly skewed.
%Tab. compares all obtained reduced Kramers-Moyal coefficients to theoretical values of the corresponding phenomenologies.
\section{Quantification of Small-Scale Intermittency on the Basis of Kramers-Moyal Coefficients}
We compare the reduced Kramers-Moyal coefficients from run \#1-\#3 in Figure~\ref{fig:km_dns} to the ones predicted by the phenomenological models
from Section~\ref{sec:markov}.
Higher order $K_n$ of the intermediate case correspond well to Yakhot's intermittency model. However, it must be stressed that the first three coefficients deviate from Yakhot's predictions. Moreover, in calculating the third-order structure function $\zeta_3= 3K_1 -3K_2+K_3\approx 1.266$, we observe that it does not follow Kolmogorov's law. Hence, an accurate description via the Kramers-Moyal coefficients has to resolve the spatial dependence of $D^{(n)}(v,r)$ as well in order to extract the $1/r$ dependence.

Higher-order coefficients ($n=3,4$) can be detected significantly for negative increments due to rare large-negative gradient events. Moreover, only the drift coefficient is different from zero for positive increments whereas all higher order coefficients drop to zero for $v>0$. It should be noted that $D^{(2)}$ possesses an additional intercept
that is due to the non-conservative forcing procedure.
The reduced Kramers-Moyal coefficients have been
obtained via polynomial fits (see caption in Figure~\ref{fig:km_burgers}). Moreover, the evolution of the one-increment PDF in scale
is depicted in Figure~\ref{fig:km_burgers} (a) and shows a pronounced left tail due to shock events
\cite{Balkovsky1997,E1999}.

Concerning the purely nonlocal case ($\alpha=0$), we observe a
self-similar evolution of the one-increment PDF in scale which
can be seen from Figure~\ref{fig:km_nonlocal} (a). As self-similar behavior is characterized by a single drift coefficient $D^{(1)}$, higher order Kramers-Moyal coefficients should be close to zero.
In fact, Figure \ref{fig:km_nonlocal} (b) shows that $D^{(3)}$ and $D^{(4)}$
are rather small. Furthermore, the diffusion coefficient $D^{(2)}$ is linear in $v$ and has only a small reduced Kramers-Moyal coefficient $K_2=0.0021 \pm 0.0001$. Finally, $ K_1=0.3108 \pm 0.0002$ suggests that the purely nonlocal case can be described quite accurately by the K41 theory \emph{i.)},
which has already been reported in~\cite{Zikanov1997}. Turning to the intermediate case $\alpha=0.15$, Figure~\ref{fig:km_intermediate} (b) indicates
that $D^{(2)}$ shows a pronounced parabolic form in contrast to the aforementioned purely nonlocal case.
In addition, higher order coefficients $D^{(3)}$ and $D^{(4)}$ possess also a slight cubic and quartic $v$-dependence.
Obviously, the latter coefficients are rather small compared to the Burgers case in Figure~\ref{fig:km_burgers} (b). In order to discuss this behavior
quantitatively, we have added the numerically obtained reduced Kramers-Moyal coefficients
to the phenomenological predictions in Figure~\ref{fig:km} (a) and (b).
Figure~\ref{fig:km_dns} reveals that the reduced Kramers-Moyal coefficients up
to order 9 possess an asymptotic behavior that is consistent with Yakhot's mean field theory~\cite{yakhot:2001}. In particular, one can clearly distinguish between the green (She-Leveque model) and red curve (Yakhot's model), which is usually not possibly on the basis of a mere scaling exponent analysis.

\section{Conclusion and Outlook}
The present paper underlines the importance of
the multi-scale approach devised by Friedrich and Peinke~\cite{Friedrich1997} which is capable of capturing the general effects of anomalous scaling in turbulence embodied in Equation (\ref{eq:km_final}). An admissible description of intermittency in turbulence, however, should take into account an infinite number of Kramers-Moyal coefficients, which has been demonstrated by the semi-logarithmic plots of the reduced Kramers-Moyal coefficients in Figure~\ref{fig:km} (b) and~\ref{fig:km_dns}. Further work will be dedicated to the investigation of higher-order
Kramers-Moyal coefficients in the experiment and in DNS of 3D turbulence~\cite{Friedrich2017}. In
this context, the presented semi-logarithmic plot in Figure~\ref{fig:km_dns}
might as well be a more accurate method for the determination of possible
scaling behavior than the usual structure function plot Figure
\ref{fig:scale_exp}. This would open the possibility to decide which of the
various phenomenological models is best suited to describe 3D Navier-Stokes
turbulence. In the case of artificial generalized Burgers turbulence, the Yakhot's mean field model~\cite{yakhot:2001} could clearly be confirmed as the most accurate candidate. Moreover, the Kramers-Moyal approach should yield important insights in the ongoing discussion about different intermittency behavior between longitudinal and transverse structure functions. Here, the simple rescaling relation between longitudinal and transverse structure functions~\cite{grauer-homann-pinton:2012,Friedrich_2016} might be extended to allow for different intermittency in tuning the corresponding set of reduced Kramers-Moyal
coefficients in Equation (\ref{eq:km_final}).

%\section{Description of the intermittency models by a Kramers-Moyal expansion}
%Interpretation of Markov process, KM expansion, different KM coefficients for different
%phenomenologies, violation of Pawula's theorem, Rainer's log-plot

%\section{Evaluation of Kramers-Moyal coefficients from numerical simulations}
%Burgers, generalized Burgers equation switching on the non-locality, Navier-Stokes etc.

\section*{Acknowledgement}

%%%%%%%%%%%%%%%%%%%%%%%%%%%%%%%%%%%%%%%%%%
%% optional
%\supplementary{The following are available online at \linksupplementary{s1}, Figure S1: title, Table S1: title, Video S1: title.}

% Only for the journal Methods and Protocols:
% If you wish to submit a video article, please do so with any other supplementary material.
% \supplementary{The following are available at \linksupplementary{s1}, Figure S1: title, Table S1: title, Video S1: title. A supporting video article is available at doi: link.}

%%%%%%%%%%%%%%%%%%%%%%%%%%%%%%%%%%%%%%%%%%
\funding{J.F. acknowledges funding from the Humboldt Foundation within a Feodor-Lynen fellowship and benefited from financial support through the Project IDEXLYON of the University of Lyon in the framework of the French program ``Programme Investissements d'Avenir'' (ANR-16-IDEX-0005). Parts of this research were supported by the DFG-Research Unit FOR 1048,
project B2.}
% \url{https://search.crossref.org/funding}
%%%%%%%%%%%%%%%%%%%%%%%%%%%%%%%%%%%%%%%%%%
\acknowledgments{J.F. is grateful for discussions with J. Peinke, A. Pumir, and O. Kamps.}

%%%%%%%%%%%%%%%%%%%%%%%%%%%%%%%%%%%%%%%%%%
\conflictsofinterest{The authors declare no conflict of interest.}

%%%%%%%%%%%%%%%%%%%%%%%%%%%%%%%%%%%%%%%%%%
%% optional
\appendixtitles{no} %Leave argument "no" if all appendix headings stay EMPTY (then no dot is printed after "Appendix A"). If the appendix sections contain a heading then change the argument to "yes".
\appendix
\section{Solutions for the Transition Probabilities of Burgers Turbulence}
\label{app:shock}
In the following, we want to discuss solutions of the transition PDF
for certain phenomenological models of turbulence. The point of departure
is the Kramers-Moyal expansion for the transition PDF (\ref{eq:km_exp_trans}).
We first discuss the solutions of the Kramers-Moyal
coefficients of the Burgers phenomenology for ramp and shock solutions (\ref{eq:km_shock}).
\subsection{Shock Solution}
The Burgers Kramers-Moyal expansion for the transition PDF of the shocks reads
 \begin{equation}
\frac{\partial}{\partial r}  p(v,r|v',r') = -\sum_{n=1}^\infty \frac{1}{n!}
     \frac{\partial^n}{\partial v^n} \frac{v^n}{r}  p(v,r|v',r')\;,
\end{equation}
A solution of this equation can be obtained from its Dyson series representation
\begin{align}\nonumber
 p(v,r|v',r')
 =&\delta(v-v') +\int_{r}^{r'} \textrm{d} r_1 \hat L_{KM}(v,r_1) \delta(v-v')\\ \nonumber
 ~&+ \int_{r}^{r'}\textrm{d} r_1 \int_{r}^{r_1} \textrm{d} r_2
 \hat L_{KM}(v,r_1)\hat L_{KM}(v,r_2) \delta(v-v')+ \ldots
 \\ \nonumber
 =&\delta(v-v') +\int_{r}^{r'} \textrm{d} r_1 \frac{\hat L }{r_1} \delta(v-v')
 + \int_{r}^{r'}\textrm{d} r_1 \int_{r}^{r_1} \textrm{d} r_2 \frac{\hat L^2}{r_1 r_2}  \delta(v-v')+ \ldots
 \\ \nonumber
 =& \delta(v-v') + \ln \frac{r'}{r} \hat L
      \delta(v-v')+ \frac{1}{2!} \left(\ln \frac{r'}{r}\right)^2 \hat L^2 \delta(v-v') +\ldots \\
 =& \exp\left[ \ln \frac{r'}{r} \hat L \right] \delta(v-v')\;,
\label{eq:dyson}
\end{align}
where the differential operator $\hat L$ is defined according to
\begin{equation}
 \hat L=\sum_{n=1}^\infty \frac{K_n}{n!}
       \frac{\partial^n}{\partial v^n} v^n\;.
\end{equation}
For Burgers-shocks, we have $K_n=1$ for all $n$ and thus the operator reads
\begin{equation}
 \hat L = -\sum_{n=1}^\infty \frac{1}{n!}
     \frac{\partial^n}{\partial v^n} {v^n}\;.
\end{equation}
We can now let this operator act on the delta function and obtain
\begin{equation}
 \hat L \delta(v-v')= -\sum_{n=1}^\infty \frac{1}{n!}
     \frac{\partial^n}{\partial v^n} {v'^n} \delta(v-v')\;,
\end{equation}
where we put the sifting property of the delta function into use. Now, we can write
the delta function in its Fourier representation and obtain
\begin{align}\nonumber
 \hat L \delta(v-v')=& -\sum_{n=1}^\infty \frac{1}{n!}
     \frac{\partial^n}{\partial v^n} {v'^n} \int
\frac{\textrm{d}u}{2\pi} e^{iu(v-v')} = - \sum_{n=1}^\infty \int
\frac{\textrm{d}u}{2\pi}\frac{(iuv')^n}{n!} e^{iu(v-v')}\\
=& \int
\frac{\textrm{d}u}{2\pi} e^{iu(v-v')} - \int
\frac{\textrm{d}u}{2\pi} e^{iuv'} e^{iu(v-v')}=\delta(v-v') - \delta(v)\;,
\end{align}
Another application of the operator yields
\begin{equation}
 \hat L^2 \delta(v-v')=\hat L\delta(v-v') - \hat L\delta(v)
 =-\delta(v-v') + \delta(v)\;,
\end{equation}
Inserting this result into Equation (\ref{eq:dyson}) yields
\begin{align}\nonumber
p(v,r|v',r')=& \delta(v-v') - \ln \frac{r}{r'}(\delta(v-v') - \delta(v))
+ \frac{1}{2!} \left( \ln \frac{r}{r'} \right)^2
     (\delta(v) - \delta(v-v')) +\ldots \\
     =& e^{\ln \frac{r}{r'}} \delta(v-v') +(1-e^{\ln \frac{r}{r'}}) \delta(v)\;.
\end{align}
The transition PDF for negative increments $v$ of the Burgers phenomenology
thus reads
 \begin{equation}
   p(v,r|v',r') = \frac{r}{r'} \delta(v-v') + \left(1-\frac{r}{r'}\right) \delta(v) \;.
   \label{eq:trans_shock}
 \end{equation}
It can be verified that this solution yields the correct Kramers-Moyal coefficients
\begin{align}\nonumber
 D^{(n)}(v',r')=&\frac{1}{n!} \lim_{r \rightarrow r'} \frac{1}{r'-r} \int \textrm{d} v (v-v')^n
 p(v,r|v',r') = \frac{1}{n!} \int \textrm{d} v (v-v')^n
 \frac{\partial p(v,r|v',r')}{\partial r}\Bigg |_{r=r'} \\
 =& \frac{1}{n!}\int \textrm{d} v (v-v')^n \left(\frac{1}{r'} \delta(v-v')
 -\frac{1}{r'} \delta(v) \right)= - \frac{(-1)^n}{n!}\frac{v'^n}{r'}\;.
\end{align}
At the same time it is also provable that the transition PDF (\ref{eq:trans_shock})
satisfies the coincidence property
\begin{equation}
 \lim_{r \rightarrow r'}  p(v,r|v',r')= \delta(v-v')\;.
\end{equation}
\subsection{Ramp Solution}
The Fokker-Planck equation for the transition PDF of positive velocity increments
in Burgers turbulence reads
\begin{equation}\label{eq:fpe_ramp}
\frac{\partial}{\partial r} p(v,r|v',r')= -\frac{\partial}{\partial v} \frac{v}{r}
 p(v,r|v',r')\;,
\end{equation}
which is a first-order partial differential equation. Therefore, we can obtain a solution through the method of characteristics (see next section). The same solution can also be acquired from the Dyson series (\ref{eq:dyson}) and involves interesting commutation relations of the Fokker-Planck operator in Equation (\ref{eq:fpe_ramp}).
\subsubsection{Solution by the Method of Characteristics}
The method of characteristics~\cite{Courant1962} suggests that we write the transition PDF in Equation (\ref{eq:fpe_ramp}) in dependence of the parameter $\lambda$ as $p(v(\lambda),r(\lambda)|v',r')$ which can be derived
with respect to $\lambda$ according to
\begin{equation}
 \frac{\textrm{d}}{\textrm{d}\lambda}p(v(\lambda),r(\lambda)|v',r')
 = \frac{\partial p(v(\lambda),r(\lambda)|v',r')}{\partial r(\lambda)} \dot r(\lambda)+
 \frac{\partial p(v(\lambda),r(\lambda)|v',r')}{\partial v(\lambda)} \dot v(\lambda)\;.
\end{equation}
Comparing this to Equation (\ref{eq:fpe_ramp}) we obtain the following ordinary differential equations
\begin{equation}
 \dot r(\lambda)=1 \qquad \dot v(\lambda) = \frac{v(\lambda)}{r(\lambda)}
 \qquad   \frac{\textrm{d} p(v(\lambda),r(\lambda)|v',r')}{\textrm{d}\lambda}=
 -\frac{p(v(\lambda),r(\lambda)|v',r')}{r(\lambda)}\;.
\end{equation}
Integrating the second  and the third equation from $r$ to $r'$ with the initial condition
$\delta(v-v')$ for $p$ yields
\begin{equation}
 v(\lambda)=v \frac{r(\lambda)}{r'} \qquad p(v(\lambda),r(\lambda)|v',r')= \delta(v-v') \frac{r'}{r(\lambda)}\;,
\end{equation}
Therefore, the transition for positive velocity increments $v$ of the Burgers phenomenology reads
\begin{equation}
 p(v,r|v',r')=\delta \left(v-\frac{r}{r'}v' \right)\;.
 \label{eq:trans_ramp}
\end{equation}
Again, the transition PDF yields the correct Kramers-Moyal coefficients (\ref{eq:km_coeff}) and satisfies the coincidence property.
\subsubsection{Solution from Dyson Series}
The Dyson series (\ref{eq:dyson}) for Equation (\ref{eq:fpe_ramp}) reads
\begin{equation}
p(v,r|v',r')= \delta(v-v') - \ln \frac{r}{r'}
     \frac{\partial}{\partial v} v \delta(v-v')+ \frac{1}{2!} \left( \ln \frac{r}{r'} \right)^2
     \left(
     \frac{\partial}{\partial v} v \right)^2\delta(v-v') +\ldots
     \label{eq:dyson_shock}
\end{equation}
We thus have to evaluate the following operator products
\begin{equation}
 \underbrace{ \frac{\partial}{\partial v} v \frac{\partial}{\partial v} v\ldots  \frac{\partial}{\partial v} v}_{n\textrm{-times}}
 \delta(v-v')\;.
 \label{eq:operator_product}
\end{equation}
Using the sifting property of the delta function, we rewrite Equation (\ref{eq:operator_product}) according to
\begin{equation}
 \underbrace{ \frac{\partial}{\partial v} v \frac{\partial}{\partial v} v\ldots
 \frac{\partial}{\partial v} v}_{n\textrm{-times}}
 = \frac{\partial}{\partial v^n} v^n - (n-1)
 \frac{\partial^{n-1}}{\partial v^{n-1}} v^{n-1} + \ldots =
 \sum_{k=0}^{n-1} {n-1 \choose k} \frac{\partial^{n-k}}{\partial v^{n-k}} v^{n-k} (-1)^k\;.
\end{equation}
Writing the delta function in its Fourier-representation, we acquire
\begin{align} \nonumber
 p(v,r|v',r') =& \int \frac{\textrm{d}u}{2 \pi} \left[ 1+ \sum_{n=0}^{\infty} \sum_{k=0}^{n-1}
 \frac{\left(-\ln \frac{r}{r'} \right)^n}{n!}
 \frac{\partial^{n-k}}{\partial v^{n-k}} v^{n-k}  (-1)^k \right] e^{iu(v-v')}\\ \nonumber
 =&  \int \frac{\textrm{d}u}{2 \pi} \left[ 1+  \sum_{n=0}^{\infty} \sum_{k=0}^{n-1}
 \frac{\left(-\ln \frac{r}{r'} \right)^n}{n!} (iuv')^{n-k} \right] e^{iu(v-v')} \\ \nonumber
 =& \int \frac{\textrm{d}u}{2 \pi} \sum_{n=0}^{\infty} \frac{(-iuv')^n}{n!}
 \left(\exp\left[\ln \frac{r}{r'}\right] -1\right)^n e^{iu(v-v')} \\
 =&  \int \frac{\textrm{d}u}{2 \pi} \exp\left[-iuv'(\frac{r}{r'}-1)\right] e^{iu(v-v')}=\delta \left(v-\frac{r}{r'}v' \right)\;,
\end{align}
which is the same solution that we gained from the method of characteristics (\ref{eq:trans_ramp}).
\section{Solution for the Transition Probability of the K41 Phenomenology}
The Fokker-Planck equation for the transition PDF of the K41 phenomenology discussed in Section (\ref{sec:markov}) under \emph{i.)} reads
\begin{equation}\label{eq:fpe_K41}
\frac{\partial}{\partial r} p(v,r|v',r')= -\frac{\partial}{\partial v} \frac{v}{3r}
 p(v,r|v',r')\;.
\end{equation}
Again, this equation can be solved with the method of characteristics~\cite{Courant1962}
and we reach the system of equations
\begin{equation}
 \dot r(\lambda)=1 \qquad \dot v(\lambda) = \frac{v(\lambda)}{3r(\lambda)}
 \qquad   \frac{\textrm{d} p(v(\lambda),r(\lambda)|v',r')}{\textrm{d}\lambda}=
 -\frac{p(v(\lambda),r(\lambda)|v',r')}{3r(\lambda)}\;,
\end{equation}
which has the solution
\begin{equation}
 p(v,r|v',r')=\delta \left(v-\frac{r^{1/3}}{r'^{1/3}}v' \right)\;.
\end{equation}
\section{Solution for the Transition Probability of the K62 Phenomenology}
The K62 phenomenology was already discussed in Section~\ref{sec:markov} under
\emph{ii.)} and is equivalent to a Fokker-Planck equation
\begin{equation}
\frac{\partial}{\partial r} p(v,r|v',r')= \left[-\frac{\partial}{\partial v} \frac{3+\mu }{9 r} v
-\frac{\partial^2}{\partial v^2}\frac{\mu}{18 r}v^2\right] p(v,r|v',r')\;.
\label{eq:FPE-log}
\end{equation}
We introduce
\begin{equation}
 A= \frac{3+\mu}{9} \qquad  B=-\frac{\mu}{18 }\;,
\end{equation}
and choose our ansatz as a log-normal distribution of the form
\begin{equation}
 p(v,r|v',r')= \frac{1}{\sqrt{2\pi Q(r,r')}v}\exp
 \left [-\frac{\left(\ln \frac{v}{v'} -K(r,r')\right)^2}{2 Q(r,r')} \right] \;,
 \label{eq:ansatz}
\end{equation}
where $K(r,r')$ and $Q(r,r')$ are functions that are
yet to be determined by Equation (\ref{eq:FPE-log}).
Deriving Equation (\ref{eq:ansatz}) with respect
to $r$ yields
\begin{equation}
 \frac{\partial}{\partial r} p(v,r|v',r')=
 \left[ -\frac{\dot Q}{2Q} + \frac{\left(\ln \frac{v}{v'} -K\right)}{Q} \dot K
 +\frac{\left(\ln \frac{v}{v'} -K\right)^2}{2Q^2} \dot Q \right]p(v,r|v',r')\;,
\end{equation}
where the dot indicates a derivative with respect to $r$.\\
The right-hand side of Equation (\ref{eq:FPE-log}) is evaluated as follows
\begin{equation}
 \frac{\partial}{\partial v} v p(v,r|v',r') =
 -\frac{\left(\ln \frac{v}{v'} -K\right)}{Q}p(v,r|v',r')\;,
\end{equation}
and
\begin{equation}
  \frac{\partial^2}{\partial v^2} v^2 p(v,r|v',r') =\left[-\frac{1}{Q}
 -  \frac{\left(\ln \frac{v}{v'} -K\right)}{Q}
 + \frac{\left(\ln \frac{v}{v'} -K\right)^2}{Q^2} \right] p(v,r|v',r')\;.
\end{equation}
The determining equation for $K(r,r')$ and $Q(r,r')$ thus reads
\begin{equation}
-\frac{\dot Q}{2Q} + \frac{\left(\ln \frac{v}{v'} -K\right)}{Q} \dot K
 +\frac{\left(\ln \frac{v}{v'} -K\right)^2}{2Q^2} \dot Q =
-\frac{B}{Qr}  +\left(\frac{A}{r}-\frac{B}{r}\right) \frac{\left(\ln \frac{v}{v'} -K\right)}{Q}
  +\frac{B}{r} \frac{\left(\ln \frac{v}{v'} -K\right)^2}{Q^2}\;,
\end{equation}
and we can read off the following ordinary differential equations for $K(r,r')$ and $Q(r,r')$
\begin{align}
 \dot Q(r,r')=&\frac{2B}{r}\;,\\
 \dot K(r,r') =& \left(\frac{A}{r} -\frac{B}{r} \right)\;,
\end{align}
which can be integrated according to
\begin{align}
 Q(r,r')=&2b \ln\frac{r}{r'}\;,\\
 K(r,r') =& a\ln\frac{r}{r'}\;,
\end{align}
where
\begin{equation}
 a=A-B= \frac{1}{3}+\frac{\mu}{6} \qquad b=-B=\frac{\mu}{18}\;.
\end{equation}
The exact solution for the transition probability of this particular Fokker-Planck equation
of the K62 phenomenology then reads
\begin{equation}
 p(v,r|v',r')= \frac{1}{\sqrt{4\pi b \ln\frac{r}{r'}}v}\exp \left [ -\frac{\left(\ln \frac{v}{v'}  - a
 \ln\frac{r}{r'}\right)^2}{4b \ln \frac{r}{r'}} \right] \;.
 \label{eq:trans_K62}
\end{equation}
We find that for $r \rightarrow r'$, the transition probability approaches a delta function, according to
\begin{equation}
 \lim_{r \rightarrow r' } p(v,r|v',r') =\frac{1}{v} \delta \left(\ln \frac{v}{v'} \right)= \delta(v-v')\;,
\end{equation}
which is in accordance with the coincidence property.

In the following we investigate the relation of the transition PDF (\ref{eq:trans_K62}) to the one-increment PDF proposed by Yakhot~\cite{yakhot:2006} as well as by Castaing~\cite{Castaing1990}. In using the so-called Mellin transform, Yakhot was able to derive the one-increment PDF directly from the structure functions of the K62 phenomenology. Furthermore, he assumed that the PDF follows a Gaussian distribution at large scales, e.g., for $r=1$ he stated that $f_1(v,r=1)=\frac{e^{-v^2/2}}{\sqrt{2\pi}}$. Yakhot's formula can be obtained from the transition PDF (\ref{eq:trans_K62}) in setting $r'=1$
and making use of the reduction property of the two-increment PDF
\begin{equation}
  f_1(v,r)= \int \textrm{d}v' p(v,r|v',r'=1)f_1(v',r'=1)=
  \frac{1}{\pi v \sqrt{8 b \ln r}} \int \textrm{d}v' e^{-v'^2/2}
  \exp \left [ -\frac{\left(\ln v  - a
  \ln r -\ln v' \right)^2}{4b \ln r} \right]\;,
  \label{eq:yakhot_K62}
\end{equation}
which reduces to a Gaussian for $r=1$, as demanded.

In analogy to the K62 phenomenology, Castaing's model of a multiplicative energy cascade is based on local fluctuations of the energy dissipation rate, which follow a log-normal distribution. However, in contrast to K62 which predicts scaling of the structure functions, Castaing's formula
\begin{equation}
  f_1(v,r)= \frac{1}{2\pi \lambda(r)} \int \frac{\textrm{d}s}{s^2}\exp \left[ -\frac{\ln^2(s/s_0(r))}{2\lambda^2(r)} \right]e^{-v^2/2s^2}\;,
  \label{eq:castaing}
\end{equation}
was devised to fit experimental data via the fitting functions $s_0(r)$ and $\lambda(r)$ and does not necessarily imply structure function scaling. Castaing's formula (\ref{eq:castaing}) is more general than Yakhot's formula
(\ref{eq:yakhot_K62}), which can be recovered in substituting $s=v/v'$
and choosing $s_0(r)= r^a$ and $\lambda(r)= \sqrt{2 b \ln r}$.
\externalbibliography{yes}
\bibliography{burgers_km.bib}

\end{document}